\documentclass[11pt,a4paper]{article}
\pdfoutput=1
\usepackage{jheppub}
\usepackage[utf8]{inputenc}

\usepackage{graphicx}
\usepackage{amsmath}
\usepackage{amssymb}
\usepackage{braket} 
\usepackage{hyperref}
\usepackage{frontespizio}
\usepackage{comment}
\usepackage{epsfig}
\usepackage{float}
\usepackage{color}
\usepackage[title]{appendix}
\usepackage{multirow}
\usepackage{slashbox}
\usepackage{tikz}
\usetikzlibrary{decorations.pathmorphing}
\usepackage{varwidth}
\usepackage{tikz}
\usetikzlibrary{backgrounds,patterns,calc}
\usepackage{xparse}
\usepackage{subfigure}
\usepackage{caption}
\usetikzlibrary{decorations.pathmorphing}

\tikzset{zigzag/.style={decorate, decoration=zigzag}}

\newcommand{\bea}{\begin{eqnarray}}
\newcommand{\eea}{\end{eqnarray}}
\newcommand{\be}{\begin{equation}}
\newcommand{\ee}{\end{equation}}
\newcommand{\ba}{\begin{align}}
\newcommand{\ea}{\end{align}}

\usepackage{subfigure}

\title{Quantum Transitions Between Minkowski and\\ de Sitter Spacetimes}

\author[1]{\small Senarath P. de Alwis}
\author[2]{\small, Francesco Muia}
\author[2,3]{\small, Veronica Pasquarella} 
\author[2,4]{\small and Fernando Quevedo}

\affiliation[1]{\small \it Physics Department, University of Colorado, Boulder, CO 80309 USA}
\affiliation[2]{\small \it ICTP, Strada Costiera 11, Trieste 34151, Italy}
\affiliation[3]{\small \it Dipartimento di Fisica, Universita'  di Trieste, Strada Costiera 11, 34151 Trieste, Italy}
\affiliation[4]{\small \it DAMTP, Centre for Mathematical Sciences, Wilberforce Road, Cambridge, CB3 0WA, UK}

\emailAdd{dealwiss@colorado.edu}
\emailAdd{fmuia@ictp.it}
\emailAdd{veronica-pasquarella@virgilio.it}
\emailAdd{fq201@damtp.cam.ac.uk}

\abstract{Quantum transitions among de Sitter and Minkowski spacetimes through bubble nucleation are revisited using the Hamiltonian formalism. We interpret tunnelling probabilities as relative probabilities: the ratio of the squared wave functionals $\mathcal{P}=\frac{|\Psi_{\mathcal{N}}|^2}{|\Psi_{\mathcal{B}}|^2}$, with $\Psi_{\mathcal{B,N}}$ solutions of the Wheeler-DeWitt equation corresponding to the spacetimes $\mathcal{N}$ and $\mathcal{B}$,  gives the probability of nucleating the state $\mathcal{N}$ relative to the probability of having the state $\mathcal{B}$. We find that the transition amplitude from de Sitter to de Sitter for both up- and down-tunnelling agrees with the original result based on Euclidean instanton methods. Expanding on the work of  Fischler, Morgan and Polchinski we find that the Minkowski to de Sitter transition is possible as in the original Euclidean approach of Farhi, Guth and Guven. We further generalise existing calculations by computing the wave function away from the turning points for the classical motion of the wall in de Sitter to de Sitter transitions. We address several challenges for the viability of the Minkowski to de Sitter transition, including consistency with detailed balance and AdS/CFT. This sets this transition on firmer grounds but opens further questions. Our arguments also validate the Coleman-De Luccia formulae in the presence of gravity since it has no issues involving negative eigenmodes and other ambiguities of the Euclidean approach. We briefly discuss the implications of our results for early universe cosmology and the string landscape.
\\
\\
\\
\\
{\centerline{\it Dedicated to the memory of Joe Polchinski}}}

\arxivnumber{}

\date{\small\today}

\begin{document} 
\maketitle

\section{Introduction}
\label{sec:Introduction}
Quantum transitions among different maximally symmetric gravitational vacuum states have been thoroughly studied for almost 40 years starting from the original work of Coleman and De Luccia (CDL)~\cite{Coleman:1980aw} in which, arguing from analogy with the flat space case that makes use of Euclidean instanton methods, they gave an expression for the probability amplitude for decay from de Sitter (dS) space to Minkowski ($\mathcal{M}$) space and from the latter to Anti-de Sitter (AdS) space. In~\cite{Brown:1988kg} Brown and Teitelboim (BT),  by considering explicitly the role of a nucleated brane, were able to discuss further tunnelling possibilities such as up as well as down-tunnelling among dS spaces with different cosmological constants  as well as transitions from dS to AdS. However,
the transition from Minkowski spacetime to dS was not allowed in their formalism. This possibility was later considered by Farhi, Guth and Guven (FGG)~\cite{Farhi:1989yr}, inspired by the possibility of creating an inflationary universe `in the laboratory'\footnote{Please note that the creation of baby universes from Minkowski should not be thought as an instability of Minkowski spacetime since the original Minkowski spacetime remains after nucleating the dS bubble.} (see also Blau, Guth and Guendelman (BGG)~\cite{Blau:1986cw} and references therein). Their analysis starts from an eternal Schwarzschild black hole ($\mathcal S$), and involves a Euclidean instanton that mediates the transition. 
\begin{figure}[h!] 
\begin{center} 
\includegraphics[scale=0.6]{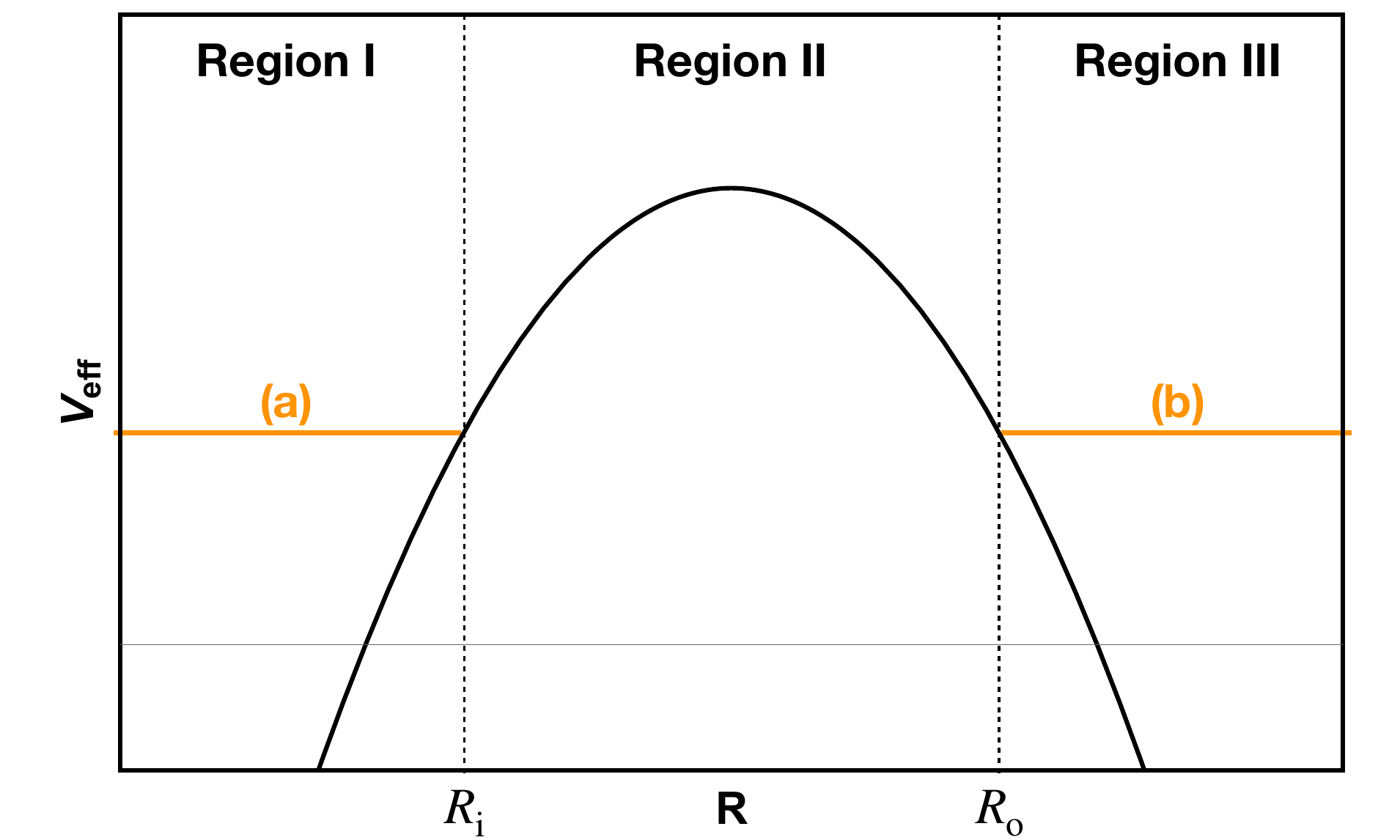} 
\caption{\footnotesize{Pictorial representation of the effective potential associated to a Schwarzschild to dS transition, see also Fig.~\ref{fig:PotentialSdS}. Region I and III are the classically allowed regions for the motion of the bubble wall, while Region II is the classically forbidden region. The horizontal lines correspond to different wall trajectories and $R_{\rm i}$ and $R_{\rm o}$ (the subscripts `i' and `o' stand for `inner' and `outer') correspond to two classical turning points of the wall trajectory. Type (a) is a bubble that can classically expand until $R = R_{\rm i}$ and then collapse to a singularity. Type (b) contracts from spatial infinity, reflects off the second turning  point and then expands back to infinity. In the quantum version classical trajectory (a) can tunnel to (b). In the dS to dS transitions the first turning point disappears, see Fig.~\ref{fig:PotentialdStodS}.\label{fig:EffectivePotential}}}
\end{center} 
\end{figure}

\begin{figure}[h!] 
\begin{center} 
\includegraphics[scale=0.7]{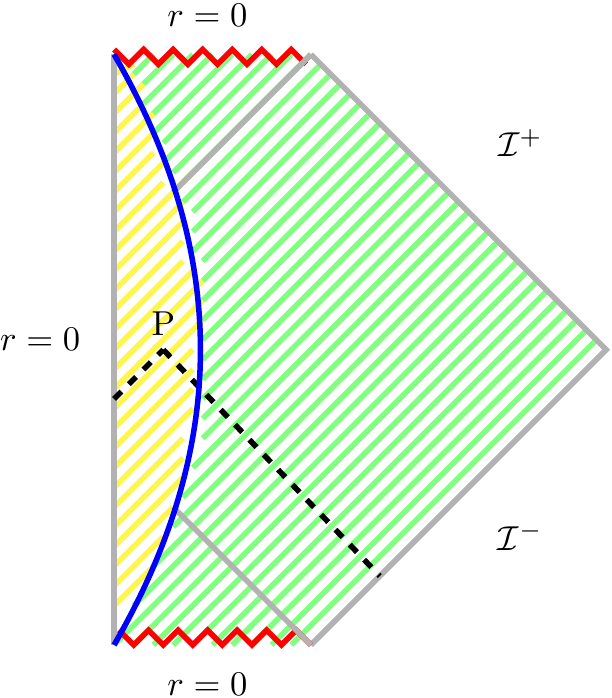} 
\qquad
\includegraphics[scale=0.7]{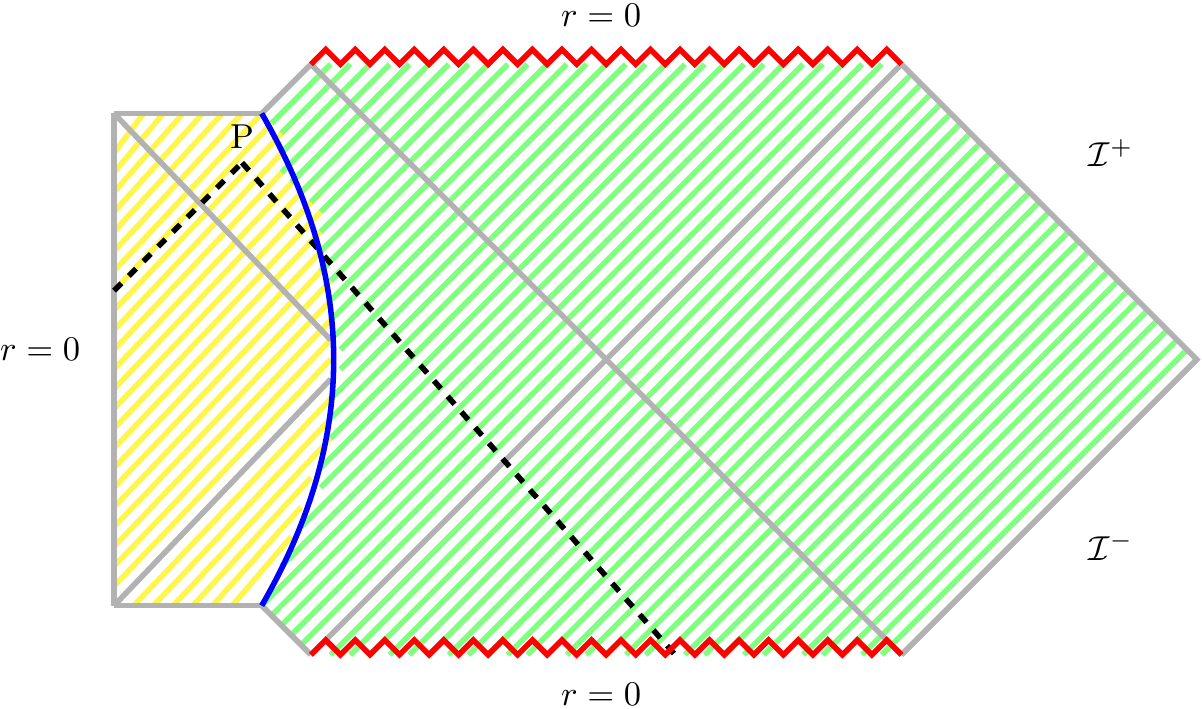} 
\caption{\footnotesize{Composite Penrose diagrams for both (a) (left) and (b) (right) wall trajectories. In each case the wall trajectory (in blue) separates the exterior Schwarzschild geometry (green) and the interior dS (yellow). The dotted black lines in both cases illustrate that the (b) trajectory represents an anti-trapped surface and is subject to Penrose theorem, whereas the (a) trajectory is not and can in principle be created from a non-singular geometry.}\label{fig:Trajectories}}
\end{center} 
\end{figure}

\begin{figure}[h!] 
\begin{center} 
\includegraphics[scale=0.7]{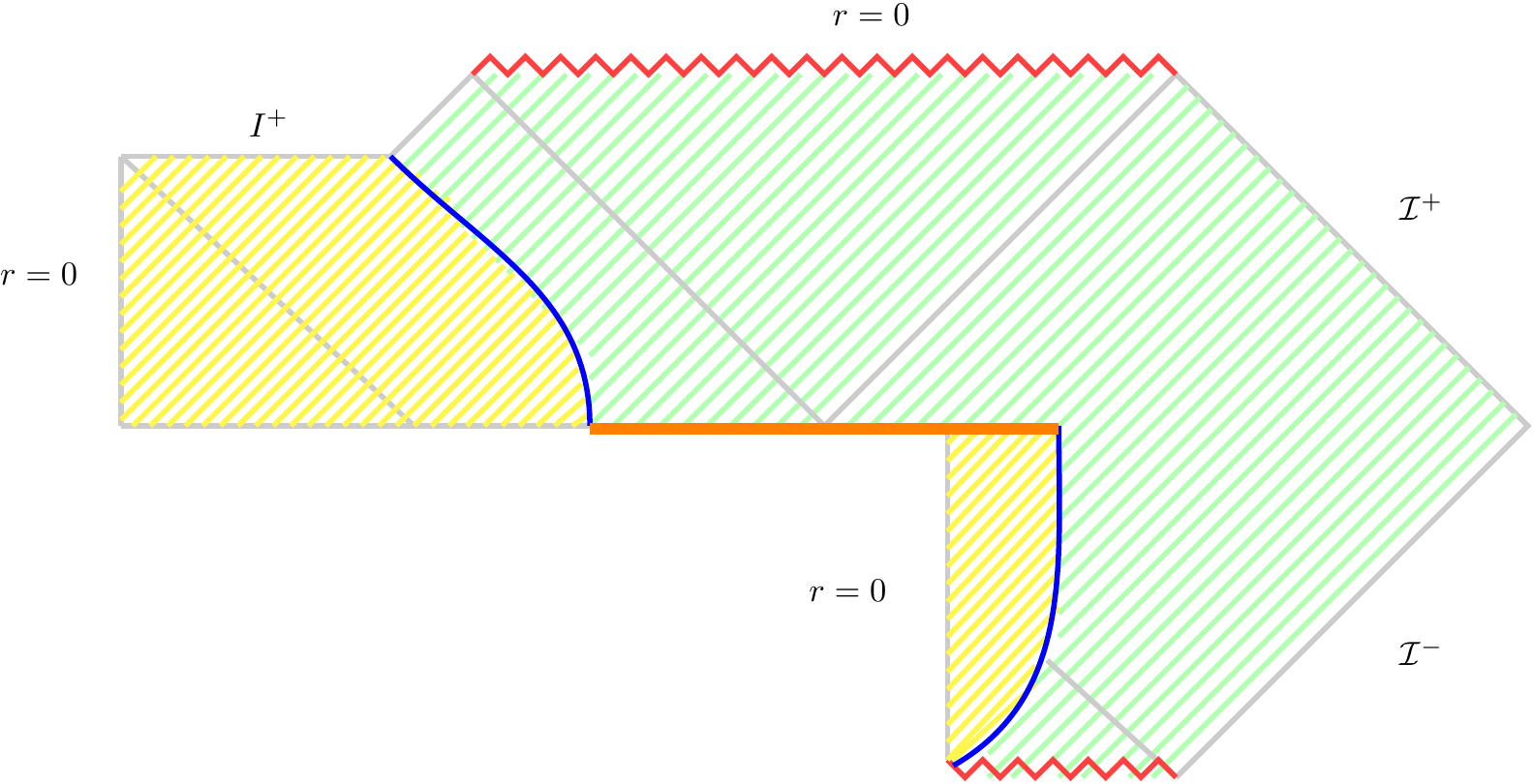}  
\caption{\footnotesize{The Penrose diagram including \textit{i)} the combined classical trajectory (in blue) starting with the (a) trajectory at $r=0$ that reaches a turning point, \textit{ii)} the corresponding tunnelling through the wormhole (horizontal line in orange) to an expanding bubble in trajectory (b) of the same energy, \textit{iii)} its further evolution towards infinite radius. The effective spacetime corresponds to patching the two shaded regions. The green shaded area on the right corresponds to the relevant part of the Schwarzschild spacetime and the dashed yellow area on the left to the corresponding part of the dS spacetime.\label{fig:CompositePenrose}}}
\end{center} 
\end{figure}

In the BGG discussion the trajectories of the bubble wall with respect to the effective potential were classified into five main types, according to the value of the mass $M$ of the black hole. We have omitted all but the ones relevant for our discussion since we are ultimately interested in the $M\rightarrow 0$ limit. In Fig.~\ref{fig:EffectivePotential} the trajectory (a) corresponds to  a  bubble coming out of the white hole singularity, bouncing off the turning point $R_{\rm i}$ and then collapsing to the black hole. Trajectory (b) represents a wall coming in from infinity, reflecting off the second turning point and then expanding back to infinity. Trajectory (a) by itself does not allow for an ever expanding universe. Trajectory (b) on the other hand allows for a continuously expanding universe but suffers from the Penrose theorem in the sense that the wall surface is an anti-trapped surface and cannot escape a singularity (see Fig.~\ref{fig:Trajectories}). Selecting a point $\textrm{P}$ on the left hand side of the wall trajectory, i.e. within the dS patch, any pair of orthogonal ingoing geodesics either hit the singularity or past asymptotic infinity behind the horizon of the observer on the right hand side. However, FGG argued that tunnelling between these trajectories can result in the spontaneous nucleation of an expanding bubble at the second turning point $R_{\rm o}$, see Fig.~\ref{fig:CompositePenrose}. The important feature of the given setup enabling type (a) trajectories to be buildable is the  choice of the range of the mass parameter $M$, such that one of the ingoing geodesics of a two-surface avoids the initial singularity. Since trajectory (a) does not represent an anti-trapped surface, it is in principle `buildable' in a laboratory\footnote{In the sense that it is at least possible to `communicate' with the wall of the bubble from the exterior of the bubble itself, without relying on a black hole singularity.} and through tunnelling allows the nucleation of a bubble in the expanding region (b) that classically would have been `unbuildable'. We note for future reference that in the limit where the black hole mass goes to zero the first turning point disappears while the second one where the bubble is nucleated, remains. This situation is similar to the tunnelling from `nothing' discussion in quantum cosmology~\cite{Vilenkin:1982de, Hartle:1983ai, Vilenkin:1984wp}. As we will see later the situation for `tunnelling' between two dS spaces is similar in that there is only one turning point: there is no initial classical motion of the bubble.

These results have been subject to criticism in the literature and up to now they are not generally accepted by the community mainly because the instanton involved in the FGG tunnelling process has some peculiar properties, as acknowledged by the authors themselves. However, these results were confirmed by an explicit calculation in the Hamiltonian formalism by Fischler, Morgan and Polchinski (FMP)~\cite{Fischler:1989se,Fischler:1990pk}. 

The importance of this issue acquires a new dimension in the context of the string theory landscape in which a huge number of vacua with essentially all possible values of the cosmological constant are allowed and it is important to understand the dynamics of this landscape. Furthermore, in the proposal of Bousso and Polchinski to address the cosmological constant problem~\cite{Bousso:2000xa}, it is necessary that the whole landscape can be populated by quantum transitions. This has been questioned in~\cite{Johnson:2008vn, Aguirre:2009tp} where it is argued that even though transitions among dS vacua are expected, the existence of a runaway solution towards decompactification or vanishing string coupling prevents the transition from being completed. This would then prevent the landscape from being populated. If, however (in the context of string theory), transitions from 10D Minkowski to 4D dS (times a compact space) were allowed, there would be no obstacle to populating the landscape (see also~\cite{Brown:2011ry}). As a first step towards this objective we believe it is important to revisit the original proposals for a Minkowski to dS transition. 

In this paper we will  address this question directly by considering 
 the nucleation of baby universes  within the four dimensional
context. We will first review the Hamiltonian argument given by FMP~\cite{Fischler:1989se, Fischler:1990pk} in
support of the claim of Guth and collaborators on the creation of
baby universes from behind the horizon of a black hole configuration~\cite{Blau:1986cw,Farhi:1986ty,Farhi:1989yr}.
While these calculations were somewhat incomplete, since a certain
boundary term was not explicitly worked out (see Sec.~\ref{sec:dSdStransitions}), in the corresponding case of transitions from dS to dS spaces this can indeed be done and explicit formulae obtained as in~\cite{Bachlechner:2016mtp}.

Next we compare these calculations with the vacuum transition probabilities obtained by CDL~\cite{Coleman:1980aw} and BT~\cite{Brown:1988kg} using Euclidean instanton methods. In fact the latter paper (which is the one used by Bousso and Polchinski~\cite{Bousso:2000xa}) is more closely related to the current investigation since it involves the nucleation of  a brane as in the string theory case. We find that while CDL/BT gives zero transition probability for up-tunnelling from flat space, the FMP calculation (in agreement with  the calculation of FGG) gives a non-vanishing probability for this. We explicitly compute this amplitude in two independent ways depending on the way we describe Minkowski space: we consider the zero cosmological constant limit of dS and the zero mass limit of the Schwarzschild solution. In the latter case we get a non-vanishing result but in the former case we find a vanishing transition amplitude. 

We explain this discrepancy by arguing that it is due to the use of different relative probabilities. We interpret the CDL/BT expression as coming from the (absolute value squared of the) ratio of the Wheeler-DeWitt (WDW) wave functions for the nucleated spacetime configuration $\mathcal{N}$ to the background spacetime configuration $\mathcal{B}$
\begin{equation}
\label{eq:RelativeProbability}
\mathcal{P} = \frac{\left|\Psi_{\mathcal{N}}\right|^2}{\left|\Psi_{\mathcal{B}}\right|^2} \,,
\end{equation}
where for instance in the particular case of dS to dS transitions the nucleated spacetime $\mathcal{N}$ is given by two dS spaces joined together by a brane, while $\mathcal{B}$ is a dS bubble, see Sec.~\ref{sec:dSdStransitions}.

The vanishing of the up-tunnelling probability in this case, in the limit in which the cosmological constant of the background dS vanishes (i.e. in the Minkowski limit for the background configuration), is entirely due to the blow up of the Hartle-Hawking (HH) wave function (in the denominator) in such a limit. This reflects the fact that the HH wave function is effectively the exponential of the horizon entropy of dS space, which becomes infinite in the Minkowski limit. On the other hand in the FMP case (i.e. for Schwarzschild to dS transitions, in the limit of $M \rightarrow 0$), while the numerator is the same in the Minkowski limit, the denominator is essentially unity. Thus both calculations are consistent with detailed balance, however with different interpretations of entropy. In the first case with the entropy of Minkowski being  given as the log of the dimension of the Hilbert space that can be built on  the space (and therefore is infinite), while in the second case it is the limit of the horizon entropy of a black hole which goes to zero in the Minkowski limit. This is consistent with the fact that in the latter case which is asymptotically flat, a notion of boundary energy exists which is not the case for dS space. A ratio of the two relative probabilities for up-tunnelling to down-tunnelling is proportional to the ratio of the  exponentials of the entropies of the two initial states. Thus we recover detailed balance for Minkowski to dS transitions and indeed find that there is no inconsistency in having up-tunnelling from Minkowski space.

Please note that in the context of Euclidean instanton calculations \textit{à la} CDL, the expression in Eq.~\eqref{eq:RelativeProbability} is usually interpreted as the tunnelling rate per unit volume and unit time of the spacetime $\mathcal{N}$ from a spacetime $\mathcal{B}$. For this reason we will loosely refer to $\mathcal{P}$ in Eq.~\eqref{eq:RelativeProbability} as the probability of tunnelling from the background spacetime $\mathcal{B}$ to the nucleated spacetime $\mathcal{N}$, while strictly speaking we are computing the relative probability of nucleating $\mathcal{N}$ against $\mathcal{B}$ from `nothing'. Of course these computations raise the usual problem related to the role of time in quantum gravity: can these transitions be interpreted as a tunnelling from one spacetime configuration to another one that happens in time? In the case of the FGG/FMP calculation, where the transition is from a Schwarzschild geometry to a dS one, the answer seems to be yes: the tunnelling takes place once the buildable bubble classically expands to its maximum radius. At that point in time it tunnels through a wormhole, disappearing from region I of the black hole geometry and appearing in region III. The relative probability can be interpreted as the probability of getting a bubble of the type (b) (from a bubble of type (a) at its maximum radius) normalised by the probability of remaining with a bubble of the type (a) that just re-collapses to the black hole singularity after having classically reached its maximum radius. In the case of dS to dS (or Minkowski to dS) transitions the interpretation is less clear, due to the absence of an initial classical bubble. We will comment more on this point in due course, see Sec.~\ref{sec:TunnellingProbability}.

We should also mention that the FMP calculation is an exercise in Hamiltonian dynamics (in contrast to the Euclidean calculations of FGG as well as those of CDL and BT), and so should not violate unitarity. There has been a claim in the literature~\cite{Freivogel:2005qh} that it does violate unitarity in the sense that it appears to take a pure state into a mixed state using AdS/CFT arguments. We revisit these arguments and conclude that there should not be an inconsistency since both states can be pure\footnote{We thank Steve Shenker for pointing out this argument.}. Also, even though the instanton calculation of FGG involves a singular configuration, the FMP Hamiltonian calculation has no such problem. Thus we believe that the latter should be taken seriously and as we pointed out above the thermodynamic issues it presents should be addressed as above.

In this paper we also generalise the FMP argument away from the turning point of the tunnelling trajectory. This enables us to 
show the relation to the well-known arguments due to HH and Vilenkin for different wave functions  for creating dS spaces from `nothing'\footnote{For recent discussions of the different proposals for the `wave function of the universe' in terms of solutions to the WDW equation see~\cite{deAlwis:2018sec} and~\cite{Halliwell:2018ejl}.}. Indeed in the case of  the dS to dS transitions there is no initial turning point very much like the case of tunnelling from `nothing'. Also this general discussion resolves the issue related to the sign of the exponent of the wave function (mentioned for instance by Bachlechner~\cite{Bachlechner:2016mtp}) in that the usual (CDL/BT) tunnelling arguments are recovered from the general solution to the WDW equation by picking the dominant term in both the numerator and the denominator of the ratio defining the relative probability.

In the next Section we start by reviewing the Hamiltonian formalism used by FMP. We introduce the transition amplitudes in terms of relative probabilities and discuss the different cases of transition among dS and Minkowski spacetimes. We finish the Section with a summary comparing the value of the different amplitudes for up- and down-tunnelling. In Sec.~\ref{sec:Wave-function-away} we generalise the formalism by computing the wave functions away from the turning points that allows us to properly study the wave function in the regions under and outside the barrier. This is also relevant since the pre-factor of the semi-classical wave function usually blows up at the turning points. The issue of the dominant components of the wave functions contributing to the transition amplitude are addressed. Sec.~\ref{sec:Comparison} is dedicated to comparison with other approaches to the Minkowski to dS transition. We address several concerns that have been raised over the years questioning the validity of the FGG proposal. We conclude that the FGG proposal survives the different challenges and it is robust. In particular we address explicitly the consistency with AdS/CFT and detailed balance. We present our conclusions in Sec.~\ref{sec:Conclusions}.

\section{ Vacuum transitions in the Hamiltonian formalism}
\label{sec:4DSummary}

\subsection{Summary of the Hamiltonian formalism}
\label{sec:SummaryHamiltonian}

We consider spherically symmetric configurations so that the metric in four dimensions take the form
\begin{equation}
\label{eq:4DMetric}
ds^{2} = -N_{t}^{2}(t,r)dt^{2}+L^{2}(t,r)(dr+N_{r}dt)^{2}+R^{2}(t,r)d\Omega_{2}^{2} \,,
\end{equation}
where as usual, $d \Omega_2^2 = d \theta^2 + \sin^2 \theta \, d \phi^2$, $\theta$ and $\phi$ being the angular coordinates on the two-sphere. In the case of a single wall separating two domains, the total action is
\begin{align}
\label{eq:4DAction}
S_{\rm tot} &= \frac{1}{16\pi G}\int_{{\cal M}}d^{4}x\sqrt{g} \,{\cal R}+\frac{1}{8\pi G}\int_{\partial{\cal M}}d^{3}y\sqrt{h} \, K+S_{\rm mat}+S_{\rm W} \equiv \nonumber \\
&\equiv S_{\rm EH} + S_K + S_{\rm mat} + S_{\rm W} \,,
\end{align}
where $\partial \mathcal{M}$ is a time-like boundary, $K$ is the extrinsic curvature and $h$ is the metric induced on the boundary. 
$S_{K}$ is the Gibbons-Hawking boundary term that needs to be added to $S_{\rm EH}$ in order to recover the correct field equation of motion when applying the variational principle. In the above metric the bulk Einstein-Hilbert action becomes\footnote{In the entire paper we denote $x'=\frac{d}{dr}x$ and $\dot{x}=\frac{d}{dt}x$.}
\begin{align}
S_{\rm EH} & = \frac{1}{2G}\int drdt\left[\frac{2}{N_{t}}(N_{r}LR)'(\dot{R}-N_{r}R')-\frac{2}{N_{t}}\partial_{t}(LR)(\dot{R}-N_{r}R')+ \right.\nonumber \\
&\left. +\frac{2}{L}(N_{t}R)'R' +\frac{N_{t}}{L}(L^{2}-R^{\prime 2})+\frac{L}{N_{t}}(\dot{R}-N_{r}R')^{2}\right] \,.
\label{eq:SEH}
\end{align}
The canonically conjugate variables to $L$, $R$ and the Hamiltonian
and momentum of the gravity theory are
\begin{align}
\label{eq:piLR}
\pi_{L} & = \frac{N_{r}R'-\dot{R}}{GN_{t}}R,\,\qquad \pi_{R}=\frac{(N_{r}LR)'-\partial_{t}(LR)}{GN_{t}} \,, \\
\label{eq:Hg}
{\cal H}_{g} & = \frac{GL\pi_{L}^{2}}{2R^{2}}-\frac{G}{R}\pi_{L}\pi_{R}+\frac{1}{2G}\left[\left(\frac{2RR'}{L}\right)^{'}-\frac{R'^{2}}{L}-L\right] \,, \\
\label{eq:Pg}
P_{g} & = R'\pi_{R}-L\pi_{L}' \,.
\end{align}
We assume that the spherical brane is located at $r = \hat{r}$. The induced metric\footnote{We choose the gauge $\sigma^{0} = t$, $\sigma^{1} = \theta$, $\sigma^{2} = \phi$.} can be written as
\begin{equation}
h_{ij} = g_{\mu\nu}\frac{\partial x^{\mu}}{\partial\sigma^{i}}\frac{\partial x^{\nu}}{\partial\sigma^{j}}\,, \qquad h_{_{00}}=-N_{t}^{2}+L^{2}(N_{r}+\dot{\hat{r}})^{2} \,,
\end{equation}
and then the determinant takes the simple form
\begin{equation}
\sqrt{h} = 4\pi\hat{R}^{2}\sqrt{h_{00}} \,,
\end{equation}
where the $ \,\, \hat{} \,\,$ denotes that the function $R(r)$ has been evaluated at $r = \hat{r}$. Finally, the domain wall action is
\begin{equation}
\label{eq:SDW}
S_{{\rm W}}=-4\pi\sigma\int dtdr \, \delta(r-\hat{r})[N_{t}^{2}-L^{2}(N_{r}+\dot{\hat{r}})^{2}]^{1/2} \,,
\end{equation}
where $\sigma$ is the tension of the wall, while the matter action is
\begin{equation}
\label{eq:Sm}
S_{{\rm mat}} = - 4\pi\int dtdr \, L N_{t} R^{2} \, \rho(r)\,, \qquad \rho = \Lambda_{{\rm O}} \,\theta(r - \hat{r}) + \Lambda_{{\rm I}}\, \theta(\hat{r}-r) \,,
\end{equation}
i.e. it just includes a cosmological constant term which takes different values on the two sides of the wall\footnote{Here and in the following we denote by a subscript $\text{I}$ the \textit{internal} region such that $r < \hat{r}$, while we denote by a subscript $\text{O}$ the \textit{outer} region such that $r > \hat{r}$.}. The Hamiltonian and momentum constraints are 
\begin{align}
\label{eq:H}
{\cal H} & = {\cal H}_{g}+4\pi LR^{2}\rho(r)+\delta(r-\hat{r})E=0 \,,\\
\label{eq:P}
P & = P_{g}-\delta(r-\hat{r})\hat{p}=0 \,,
\end{align}
where 
\begin{equation}
\label{eq:E and m}
E = \sqrt{\frac{\hat{p}^{2}}{\hat{L}^{2}}+m^{2}}\,, \qquad m=4\pi\sigma\hat{R}^{2} \,, \qquad \hat{p}=\partial{\cal L}/\partial\dot{\hat{r}} \,,
\end{equation}
and the Lagrangian can be read from Eq.~\eqref{eq:4DAction}. Away from the domain wall (i.e. $r\ne\hat{r}$) we have from the second
constraint,
\begin{eqnarray}
\mbox{\ensuremath{\pi}}_{R} & = & \frac{L}{R'}\pi'_{L}.\label{eq:Pcons}
\end{eqnarray}
Inserting Eq.~\eqref{eq:Pcons} in Eq.~\eqref{eq:H} (for $r\ne\hat{r}$)
we get 
\begin{equation}
\frac{d}{dr}\left(\frac{\pi_{L}^{2}}{2R}\right)=\frac{1}{2G^{2}}\frac{d}{dr}\left[R\left(\frac{R'}{L}\right)^{2}-R+\frac{8\pi}{3}G\rho R^{3}\right] \,,
\end{equation}
that translates into the solution
\begin{align}
\label{eq:pi_L}
\pi_{L} & = \eta\frac{R}{G}\left[\frac{R'^{2}}{L^{2}}-A_{\alpha}\right]^{1/2}\,, \quad  \alpha={\rm O}, \,{\rm I}\,, \quad \eta = \pm 1 \,,\\
\label{eq:Aalpha}
A_{\alpha} & = 1-\frac{2GM_{\alpha}}{R}-H_{\alpha}^{2}R^{2}\,, \qquad H_{\alpha}^{2}=\frac{8\pi G}{3}\Lambda_{\alpha} \,,
\end{align}
where $M_{\alpha}$ is an integration constant. This of course corresponds to the general solution to the spherically symmetric metric ansatz, i.e Schwarzschild-dS ($\mathcal{S}$dS). If the constant $M_{\alpha} = 0$, $\Lambda_{\alpha} \ne 0$, we have a pure dS solution and if $\Lambda_{\alpha} = 0$, $M_{\alpha} \ne 0$ we have a Schwarzschild black hole. In the static coordinate system with $R$ as one of the coordinates, the spherically symmetric $\mathcal{S}$dS metric takes the static form:
\begin{equation}
ds_\alpha^2=-A_\alpha(R)\, d\tau^2+A_\alpha^{-1}(R)\, dR^2 + R^2\, d\Omega_2^2 \,.
\end{equation}

\subsubsection*{Constraints and dynamics of the wall}
The constraints on the domain wall are imposed by integrating
Eq.~\eqref{eq:H} and Eq.~\eqref{eq:P} from $\hat{r}-\epsilon$ to $\hat{r}+\epsilon$ leading to
\begin{align}
\label{eq:I1}
\frac{\hat{R}}{\hat{L}}(R'(\hat{r}+\epsilon)-R'(\hat{r}-\epsilon)) & = -GE \,,\\
\label{eq:I2}
\pi_{L}(\hat{r}+\epsilon)-\pi_{L}(\hat{r}-\epsilon) & = \frac{\hat{p}}{\hat{L}}=0 \,,
\end{align}
where to get the last equality we have transformed to the rest frame
of the wall so that $\hat{p}=0$ and $E = m = 4 \pi \hat{R}^{2} \sigma$. We note for future reference that in the limit $\kappa\rightarrow 0$ , $A_{\rm I} = A_{\rm O}$, i.e. there is not change in the geometry in the absence of the wall.  Combining Eq.~\eqref{eq:I2} with Eq.~\eqref{eq:pi_L} and then using Eq.~\eqref{eq:I1} gives 
\begin{equation}
\label{eq:R'pm}
\frac{R'(\hat{r}\pm\epsilon)}{\hat{L}}=\frac{1}{2\kappa\hat{R}}\left(\hat{A}_{{\rm I}}-\hat{A}_{{\rm O}}\right)\mp\frac{\kappa}{2}\hat{R} \,,
\end{equation}
where we have defined 
\begin{equation}
\kappa\equiv4\pi\sigma G=\frac{Gm}{\hat{R}^{2}}.
\end{equation}
Using Eq.~\eqref{eq:R'pm} in Eq.~\eqref{eq:pi_L} and Eq.~\eqref{eq:piLR} (in the gauge $N_{r}=0$ and $N_t = 1$) one can write an equation of motion for the wall:
\begin{equation}
\label{eq:EofM}
\dot{\hat{R}}^{2}+V = -1 \,,
\end{equation}
where, depending on the solution chosen in Eq.~\eqref{eq:R'pm}, the potential takes the form
\begin{align}
\label{eq:V}
V & = -\frac{1}{(2\kappa\hat{R})^{2}}\left((\hat{A}_{{\rm I}}-\hat{A}_{{\rm O}})-\kappa^{2}\hat{R}^{2}\right)^{2}+(\hat{A}_{{\rm O}}-1) = \\
\label{eq:Valt}
 & = -\frac{1}{(2\kappa\hat{R})^{2}}\left((\hat{A}_{{\rm I}}-\hat{A}_{{\rm O}})+\kappa^{2}\hat{R}^{2}\right)^{2}+(\hat{A}_{{\rm I}}-1) \,.
\end{align}
So the equation $V=-1$ gives the turning points for the classical
motion of the wall, i.e. the points at $R = R_{\rm i}$ and $R = R_{\rm o}$ in Fig.~\ref{fig:EffectivePotential}. At a turning point we see from Eq.~\eqref{eq:V} that $\hat{A}_{{\rm O}}>0$ and from Eq.~\eqref{eq:Valt} that $\hat{A}_{{\rm I}}>0$. The classical turning points for the geometry occur at $\pi_{L}=0$, i.e. (see Eq.~\eqref{eq:pi_L}) at
\begin{equation}
\label{eq:TurningPoints}
\frac{R'^{2}}{L^{2}} = A(R) = 1-\frac{2MG}{R}-H^{2}R^{2} \,.
\end{equation}
When $r=\hat{r}$ these are the turning points for the wall, i.e.
the solutions of $V = -1$.

\subsection{Wheeler-DeWitt equation and tunnelling probability}
\label{sec:TunnellingProbability}

The problem of computing the tunnelling probability in semiclassical gravity is similar to the tunnelling problem of a barrier potential in usual quantum mechanics. In this context, instead of the Schr\"oedinger equation we need to solve the WDW equation
\begin{equation}
\label{eq:WDW}
\mathcal{H} \Psi = 0 \,,
\end{equation}
where the wave functional $\Psi$ is a functional of the geometry and a function of the brane position $\hat r$. We employ the WKB approximation, for which the wave function $\Psi$ can be written at leading order in the semi-classical approximation as
\begin{equation}
\label{eq:GeneralWF}
\Psi = a e^{I} + b e^{- I} \,,
\end{equation}
where, given a configuration with action $S$, we have denoted the combination $iS = I$ and the action $S$ is evaluated on a classical solution. In the following we will refer to both $S$ and $I$ as the `action'. The two solutions are given by $\eta = \pm 1$ corresponding to the two solutions for $\pi_L$, see Eq.~\eqref{eq:pi_L}.

The problem is qualitatively very different depending on whether the background geometry includes a black hole or not. If it does, the problem is similar to the usual quantum mechanical tunnelling through a potential barrier: as shown in Fig.~\eqref{fig:EffectivePotential}, there are two classically allowed (I and III) and one classically forbidden regions (II). Classically, the wall expands (or contracts) up to a classical turning point and then re-collapses (or re-expands). Quantum mechanically, it can tunnel under the barrier and resurface after the second turning point. The WDW equation has two independent solutions as in Eq.~\eqref{eq:GeneralWF} in each of these regions, that need to be matched at the classical turning points. Some of the coefficients $a$ and $b$ (there are two coefficients for each region) can be fixed by imposing boundary conditions. For instance, requiring that there is only an outgoing wave in region III (that amounts to requiring that the nucleated universe is expanding and not contracting, as it happens in the Vilenkin proposal for the creation of a dS bubble from `nothing') gives a relation between the two coefficients of the under the barrier wave function. On the contrary, if there is no boundary condition, in general one of the two solutions (positive or negative exponent) in Eq.~\eqref{eq:GeneralWF} will dominate in each region, depending on the sign of the action $I$. 
The tunnelling  probability in this case can be computed as the ratio between the squared wave function associated to the expanding bubble in the classical Region III and the squared wave function of the expanding bubble in the classical Region I. This can be approximated (up to subtleties that will be addressed in Sec.~\ref{sec:Wave-function-away}) as the ratio
\begin{equation}
\label{eq:TunnellingProbabilitySdS}
\mathcal{P} = \frac{\left|\Psi(R_{\rm o})\right|^2}{\left|\Psi(R_{\rm i})\right|^2} \,,
\end{equation}

\noindent In general we interpret the ratio
\begin{equation}
\label{eq:GeneralDefProbability}
\mathcal{P}(\mathcal{B} \rightarrow \mathcal{N}) = \left|\frac{\Psi_{\mathcal{N}}}{\Psi_{\mathcal{B}}}\right|^2 \,,
\end{equation}
as the relative probability of finding the system in the `nucleated' state $\mathcal{N}$ versus the `background' state $\mathcal{B}$, see Sec.~\ref{sec:Wave-function-away} for more details. Notice that the two states $\mathcal{B}$ and $\mathcal{N}$ do not always have the meaning of `initial' and `final' states (see Sec.~\ref{sec:Introduction}): the transition can be clearly interpreted as happening in time if there is an initial classical motion of the bubble wall. In such a case, the configuration with a type (a) bubble at its maximum radius (i.e. evaluated at the first turning point $R_{\rm i}$, see Fig.~\ref{fig:EffectivePotential}), plays the role of the background geometry. In the cases in which there is no initial classical motion of the wall the interpretation is less clear. For this reason we will refer to the state $\mathcal{B}$ as the `background' spacetime, instead of initial spacetime. Since we compute a  relative probability, it does not have to be smaller than one, and we avoid the problem of the normalization of wave functionals. Observe that in the context of Euclidean instanton computations \textit{à la} CDL, the relative probability in Eq.~\eqref{eq:GeneralDefProbability} is interpreted as tunnelling rate per unit volume and time, see e.g.~\cite{Weinberg:2012pjx} for details.
\begin{equation}
\mathcal{P}(\mathcal{B} \rightarrow \mathcal{N}) \equiv \Gamma_{\mathcal{B} \rightarrow \mathcal{N}} \,.
\end{equation}

In the typical transitions that we will consider in the next Sections, the background configuration is given by some spacetime state $A$, while the nucleated state corresponds to the spacetime $A$ joined to another spacetime $B$ through a wall $\text{W}$. We will denote such a configuration by $A/B\oplus \text{W}$. Then the relative probability for being in the configuration $A/B\oplus \text{W}$ versus being in the background state $A$ is
\begin{equation}
\mathcal{P}(A\rightarrow A/B\oplus \text{W}) = \frac{|\Psi(A/B\oplus \text{W})|^{2}}{|\Psi(A)|^{2}} \,. \label{eq:Psi}
\end{equation}

Note that the denominator represents the probability of creating the state $A$ out of `nothing' (`nothing', whose wave function is simply $1$, corresponds to the background spacetime in the `nothing' to $A$ transition). If the state $A$ is a dS bubble, the denominator corresponds to the standard HH or Vilenkin wave function, depending on the choice of the sign in the exponent of the wave function. In Sec.~\ref{sec:Wave-function-away} we will argue that both in the case of dS to dS transitions and in the zero mass limit of the Schwarzschild to dS transitions the numerator has a similar interpretation.

If Eq.~\eqref{eq:GeneralDefProbability} is dominated by one term in the numerator and the denominator (see Sec.~\ref{sec:Wave-function-away} for more details), it simplifies to an expression of the form
\begin{equation}
\label{eq:ApproxProbability}
\mathcal{P}(\mathcal{B} \rightarrow \mathcal{N}) \simeq \text{exp} \bigg[2 \, \text{Re}\left(I_{\rm tot}(\mathcal{N}) - I(\mathcal{B})\right)\bigg] \,,
\end{equation}
where $I_{\rm tot}(\mathcal{N})$ denotes the total action evaluated on the nucleated configuration\footnote{In order to simplify the notation, in the following we will suppress the explicit dependence on $\mathcal{N}$: only over-barred quantities will refer to the background configuration.}, while $I(\mathcal{B}) \equiv \overline{I}$ denotes the total action evaluated on the background configuration. Notice that, if there is an initial classical motion of the wall of the bubble (i.e. if there is a first turning point), the background configuration corresponds to the spacetime evaluated at the first turning point of the wall. Otherwise, the background must be specified on a case by case basis.

\subsubsection*{Computation of the classical action}

We note that the dynamical variables are $R, L, \hat{r}$ and their conjugates $\pi_{R}$, $\pi_{L}$, $\hat{p}$. On a classical trajectory ending at some given values of $R$, $L$, $\hat{r}$ we have in Hamilton-Jacobi theory\footnote{Note that ${\cal L}$ depends on $R'$ since $\pi_{L}$ does, see Eq.~\eqref{eq:pi_L}.}
\begin{equation}
S_{\rm tot} = S_{\rm tot} [R, L, \hat{r}] = \int \, dr \, {\cal L}(L, R, R', \hat{r}) \,,
\end{equation}
where
\begin{equation}
\label{eq:SVariation}
\delta S_{\rm tot} = \int dr \left(\pi_{L} \, \delta L + \pi_{R} \, \delta R \right) +\hat{p} \, \delta\hat{r}+\frac{\partial{\cal L}}{\partial R'}\, \delta R\bigg|_{\hat{r} - \epsilon}^{\hat{r} + \epsilon} \,.
\end{equation}
 The last (boundary) term needs to be removed from the bulk
action in order to have a well-defined functional derivative with respect to $R$~\cite{Fischler:1990pk}. The integration to get the classical action may be done along
any path ending at the given values. In particular one may choose
the path that keeps $R$, $\hat{r}$ fixed as the integrability conditions
will guarantee that the other functional partial derivatives are satisfied. With this choice we can neglect the second and third terms in Eq.~\eqref{eq:SVariation} and we can write schematically
\begin{equation}
\label{eq:Scl}
I_{\rm tot} = I_{{\rm B}} + I_{{\rm b}} \,,
\end{equation}
where $I_{\rm B}$ is the \textit{bulk} action arising from the integration of the first term in Eq.~\eqref{eq:SVariation}, while $I_{\rm b}$ is the \textit{boundary} action coming from the integration of the last term in Eq.~\eqref{eq:SVariation}. The two terms can be written in detail as
\begin{align}
\label{eq:SclassicalBulk}
I_{{\rm B}} & = \frac{\eta}{G}\int_{0}^{\hat{r}-\epsilon}drR\left[ \sqrt{A_{{\rm I}}L^{2}-R'^{2}}-R'\cos^{-1}\left(\frac{R'}{L\sqrt{A_{{\rm I}}}}\right)\right] + \int_{\hat{r}+\epsilon}^{\pi}dr \,\left[{\rm I}\rightarrow{\rm O}\right] \,, \\
\label{eq:SclassicalBoundary}
I_{{\rm b}} & = \frac{\eta}{G}\int\delta\hat{R} \, \hat{R}\cos^{-1}\left(\frac{R'}{L\sqrt{\hat{A}}}\right)\bigg|_{\hat{r}-\epsilon}^{\hat{r}+\epsilon} \,,
\end{align}
with the inverse cosine defined be be between $0$ and $\pi$, while $\eta = \pm 1$, as from Eq.~\eqref{eq:pi_L} and $[\text{I} \rightarrow \text{O}]$ means that the integrand is the same as in the first term of Eq.~\eqref{eq:SclassicalBulk}, with the subscript I substituted by O. FMP did not evaluate the boundary term in Eq.~\eqref{eq:SclassicalBoundary} since the integral
cannot be done analytically for $M \ne 0$. However the most extreme
form of the puzzle that we encounter comes from the limiting case
$M\rightarrow0$, i.e. dS bubble nucleation from flat space.
In this case there is no inner turning point and it is easy to calculate the boundary integral. 

Since we want to compute the relative probability for the nucleation of the spacetime $\mathcal{N}$, the action $I_{\rm tot}$ must be computed at the (second) turning point. We will denote all quantities evaluated at the turning point(s) by the subscript `tp'. In the cases with two turning points, the subscript tp will denote the difference between the quantity evaluated at the second turning point and the same quantity evaluated at the first turning point, incorporating then the background subtraction, see Sec.~\ref{sec:StodS}.

If Eq.~\eqref{eq:ApproxProbability} holds, the transition probability is given by the real part of Eq.~\eqref{eq:SclassicalBulk} and Eq.~\eqref{eq:SclassicalBoundary}. In order to compute the tunnelling probability we need to evaluate the classical action in Eq.~\eqref{eq:SclassicalBulk} and Eq.~\eqref{eq:SclassicalBoundary} in each different case as we do in the next Sections.

\subsection{de Sitter to de Sitter transitions}
\label{sec:dSdStransitions}

\begin{figure}[h!] 
\begin{center} 
\includegraphics[scale=0.6]{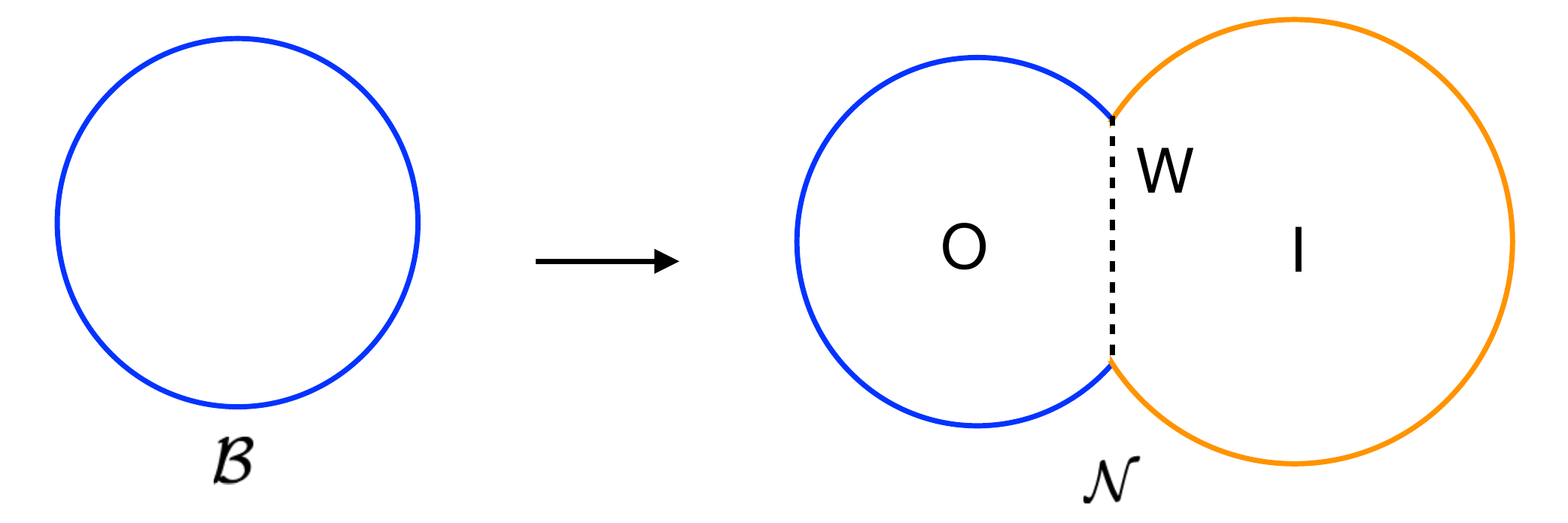} 
\caption{\footnotesize{Pictorial representation of the background spacetime $\mathcal{B}$ and the nucleated spacetime $\mathcal{N}$. The letters O and I represent the outer and inner regions respectively, while W represents the wall that separates the two regions.\label{fig:dSdSdSW}}}
\end{center} 
\end{figure}

\begin{figure}[h!] 
\begin{center} 
\includegraphics[scale=0.6]{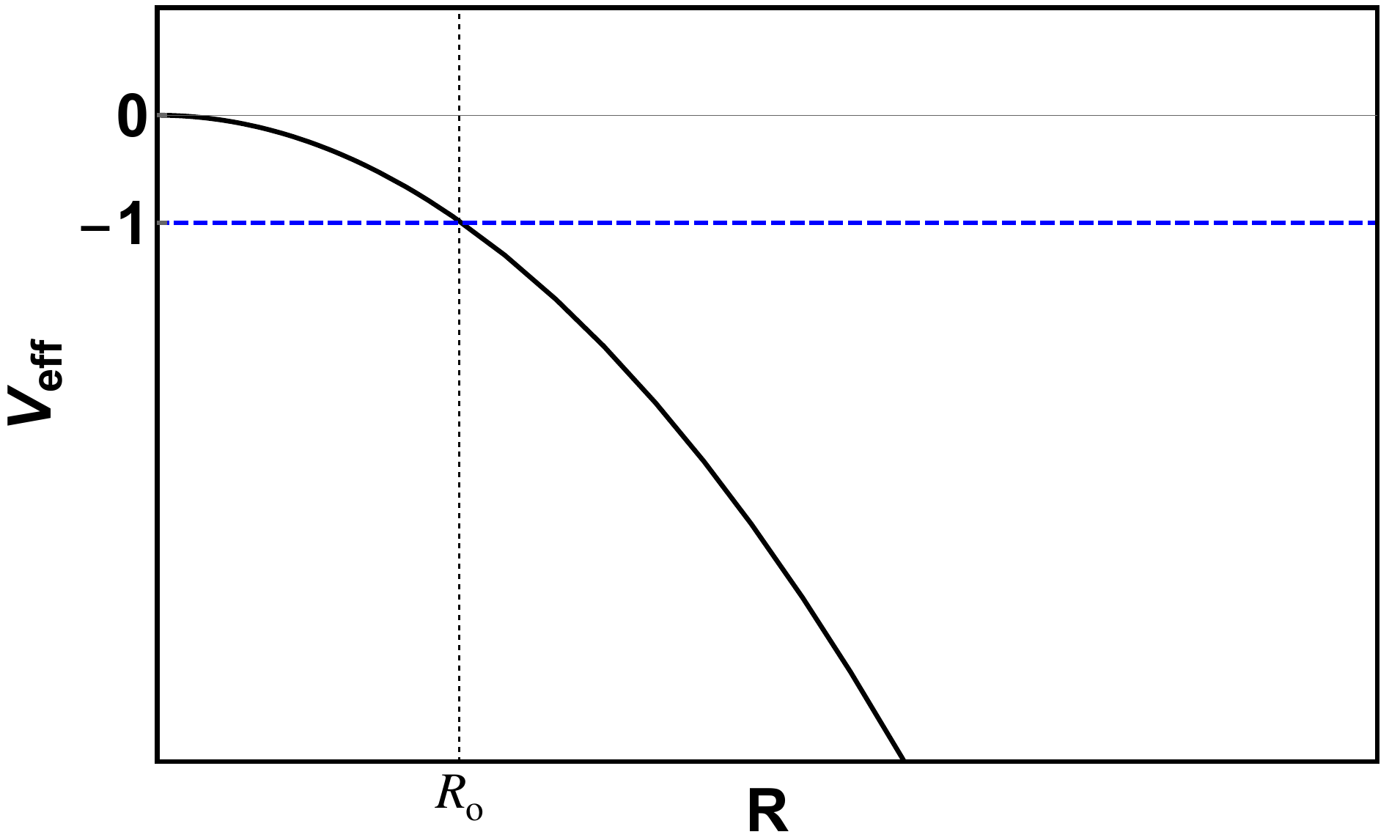} 
\caption{\footnotesize{Effective potential for dS to dS transitions. Notice there is only one turning point.\label{fig:PotentialdStodS}}}
\end{center} 
\end{figure}

In this Section we are interested in the relative probability of nucleating a configuration with two dS spaces joined at a wall versus the probability of having a single dS space, see Fig.~\ref{fig:dSdSdSW}:
\begin{equation}
\mathcal{P}(\text{dS} \rightarrow \text{dS}/\text{dS} \oplus \text{W}) = \frac{|\Psi(\text{dS}/\text{dS} \oplus \text{W})|^{2}}{|\Psi(\text{dS})|^{2}} \,.
\end{equation}
It is possible to calculate the general case of dS to dS transitions (with Hubble constants $H_{{\rm O}}$ and $H_{\rm I}$) using
\begin{eqnarray}
A_{{\rm O}} & = & 1-H_{{\rm O}}^{2}R^{2}\,,\qquad A_{{\rm I}}=1-H_{{\rm
I}}^{2}R^{2}\,,\label{eq:AOIHH}\\
V & = & -\frac{1}{4\kappa^{2}}\hat{R}^{2}\left[(H_{{\rm O}}^{2}-H_{{\rm
I}}^{2})^{2}+2\kappa^{2}(H_{{\rm O}}^{2}+H_{{\rm
I}}^{2})+\kappa^{4}\right]\,,\label{eq:VHH}\\
R_{\rm o}^{2} & = & \frac{4\kappa^{2}}{(H_{{\rm O}}^{2}-H_{{\rm
I}}^{2})^{2}+2\kappa^{2}(H_{{\rm O}}^{2}+H_{{\rm
I}}^{2})+\kappa^{4}}\,,\label{eq:RoHH}
\end{eqnarray}
where $R_{\rm o}$ is the turning point (i.e. solution of $V=-1$) There is no initial turning point in this case since the potential $V\propto -R^2$ which has only one turning point (see Fig.~\ref{fig:PotentialdStodS}). So effectively the integration in Eq.~\eqref{eq:ApproxProbability}
starts from $R(0)=0$ analogous to the tunnelling from `nothing' case studied by HH and Vilenkin. The matching conditions are given by
\begin{equation}
\frac{\hat{R}'_{\pm}}{L}=\frac{1}{2\kappa\hat{R}}(H_{{\rm O}}^{2}-H_{{\rm
I}}^{2}\mp\kappa^{2})\hat{R}^{2}\equiv c_{\pm}\hat{R} \,.\label{eq:R'HH}
\end{equation}

\noindent The boundary action in Eq.~\eqref{eq:SclassicalBoundary} then becomes\footnote{We have used the definite
integral
\begin{equation}
\int_{0}^{1/\sqrt{1+a}}dx \, x\cos^{-1}\frac{x}{\sqrt{1-ax^{2}}}=\frac{\pi}{4a}\left(1-\frac{1}{\sqrt{1+a}}\right) \,.
\end{equation}
Also note that $\cos^{-1}(-x)=\pi-\cos^{-1}x$.\label{fn:DefInt}}

\begin{align}
I_{{\rm b}}\bigg|_{\rm tp} & = -\frac{\eta}{G}\int_{0}^{R_{\rm o}}\delta\hat{R}\hat{R}\left[
\cos^{-1}\left(\frac{R'_{+}}{L\sqrt{\hat{A}}_{{\rm
O}}}\right)-\cos^{-1}\left(\frac{R'_{-}}{L\sqrt{\hat{A}}_{{\rm
I}}}\right)\right] =  \label{eq:SbHH}\\
 & = -\frac{\eta}{G}\frac{\pi}{4}R_{\rm o}^{2}\left[\frac{\epsilon(R_{+}')}{1+|c_{+}|R_{\rm o}}-\frac{\epsilon(R_{-}')}{1+|c_{-}|R_{\rm o}}+2\left(\theta(-\hat{R}_{+}')-\theta(-\hat{R}_{-}')\right)\right] \,.\label{eq:SbHH2}
\end{align} 
In this case, for instance, the subscript `tp' amounts to evaluating the integral in Eq.~\eqref{eq:SbHH2} between 0 and $R_{\rm o}$. After some algebra this becomes 
\begin{eqnarray}\notag
I_{{\rm b}}\bigg|_{\rm tp} = \frac{\eta}{G} &&\left[\frac{\pi}{2}R_{\rm o}^{2} \left(\theta(-\hat R^{\prime}_{-})-\theta(-\hat R^{\prime}_{+})\right)+\frac{\pi}{4H_{\rm I}^{2}}\ \epsilon(\hat R^{\prime}_{-})-\frac{\pi}{4H_{\rm O}^{2}}\ \epsilon(\hat R^{\prime}_{+}) + \right. \\
&& \,\,\, \left.- \pi \frac{(H_{\rm O}^{2}-H_{\rm I}^{2})^{2}+\kappa^{2}(H_{\rm O}^{2}+H_{\rm I}^{2})}{8\kappa H_{\rm O}^{2} H_{\rm I}^{2}} R_{\rm o} \right] \,. 
\label{eq:wall} 
\end{eqnarray}
  
\begin{figure}[h!] 
\begin{center} 
\includegraphics[scale=0.6]{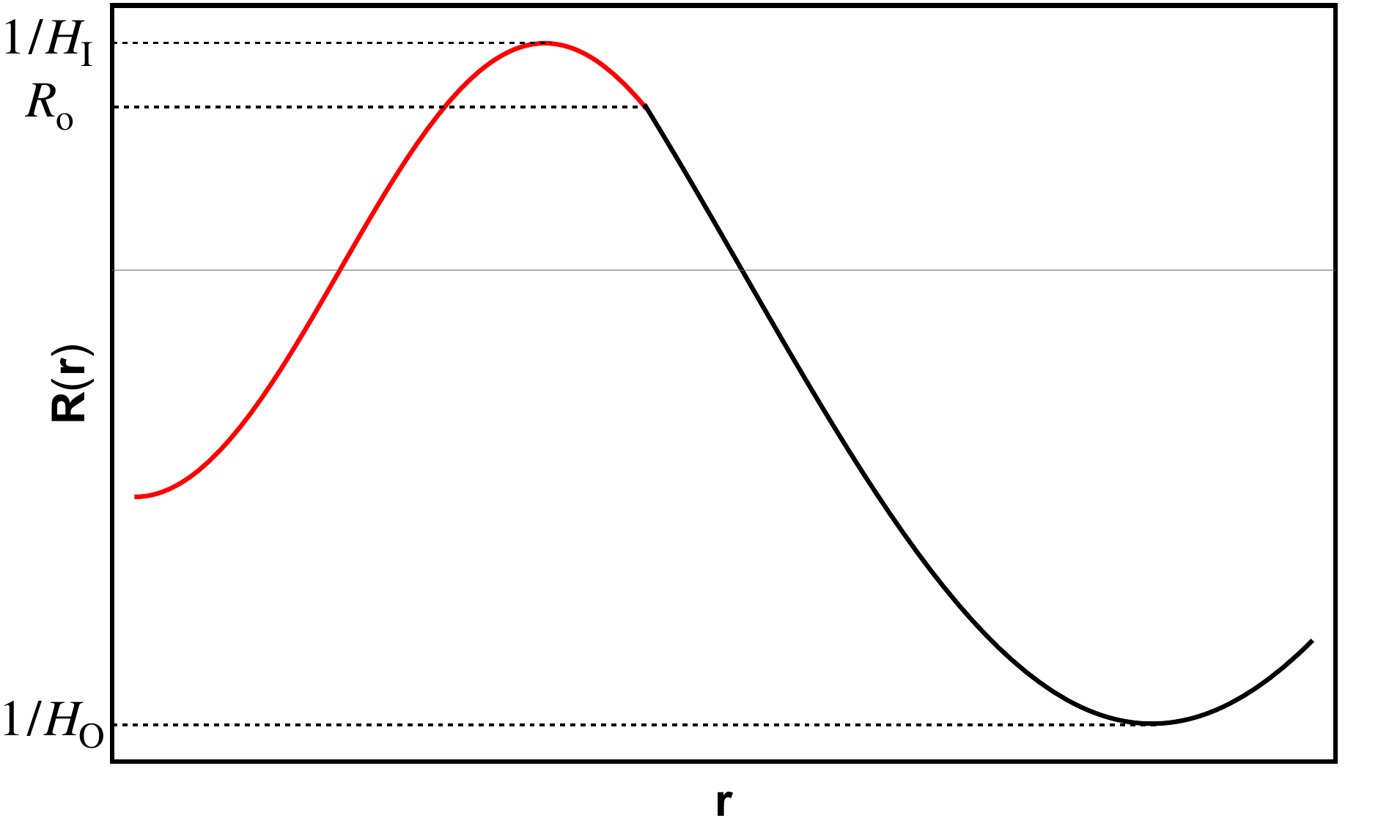} 
\caption{\footnotesize{This plot shows $R(r)$ for both dS configurations and it can be obtained by solving Eq.~\eqref{eq:TurningPoints} on both sides. The red line describes the profile of the vacuum $H_{\rm I}$, nucleated inside the background $H_{\rm O}$. The two are joined at the wall, where $r = \hat r$. We have highlighted the locations of the two horizons $1/H_{\rm O}$ and $1/H_{\rm I}$. This plot is needed in order to determine the relevant contributions to the bulk integrals on either side of the wall. The two vacua are joined in such a way that $R^{\prime}_{\pm} < 0$ at the turning point $R_{\rm o}$, which is needed in order to get a nontrivial contribution from the second term featuring in the bulk integrals in Eq.~\eqref{eq:SclassicalBulk}.}}
\end{center} 
\end{figure} 
\vspace{0.2cm}
The bulk action (see Eq.~\eqref{eq:SclassicalBulk}) for dS to dS transitions gets a contribution from each dS vacuum. It can be most easily obtained using the explicit parametrization for the nucleated spacetime $\mathcal{N}$:
\begin{equation}
\begin{split}
\text{I}:& \qquad R = a_{\rm I} \sin r \,, \, \quad L = a_{\rm I} \,, \\
\text{O}:&\qquad R = a_{\rm O}\sin r \,, \quad L = a_{\rm O} \,,
\label{eq:aR}
\end{split}
\end{equation}
where $a_{\rm I,O} = H_{\rm I, O}^{-1}$. The integral in Eq.~\eqref{eq:SclassicalBulk} can then be done explicitly and the result is
\begin{equation}
I_{\rm B} = \frac{\eta\pi}{2G}\left[(\theta(-\hat{R}'_{+})-\theta(-\hat{R}'_{-}))\hat{R}_{{\rm o}}^{2}+\theta(-\hat R^{\prime}_{-}) H_{{\rm I}}^{-2}+H_{\rm O}^{-2} \left[1 - \theta(-\hat R^{\prime}_{+})\right]\right]\,.
\label{eq:SBHH}
\end{equation}
Please note that the first term in Eq.~\eqref{eq:SBHH} cancels the last term within the square brackets in Eq.~\eqref{eq:SbHH2} so adding the bulk and the boundary contributions we get
\begin{equation}
I_{\rm tot}\bigg|_{\rm tp} = -\frac{\eta \pi}{G} \left[\frac{\left[(H_{\rm O}^{2} - H_{\rm I}^{2})^{2}+\kappa^{2} (H_{\rm O}^{2} + H_{\rm I}^{2})\right] R_{{\rm o}}}{8 \kappa H_{\rm O}^{2} H_{\rm I}^{2}}-\frac{1}{4} \left(H_{\rm I}^{-2} + H_{\rm O}^{-2}\right)\right] \,.\label{eq:HH}
\end{equation}
It is easy to check that this expression is positive (negative) for $\eta = +1(-1)$ and $H_{I} \neq 0$ and it is symmetric under the interchange ${\rm O}\leftrightarrow {\rm I}$. The above equation corresponds to Eq.~(3.41) in~\cite{Bachlechner:2016mtp}, except that we set the numbers $N_{A,B}$ to zero, as will be explained in detail in Sec.~\ref{sec:Comparison}. Taking the flat space limit in the outer region $H_{\rm O} \rightarrow 0$ in Eq.~\eqref{eq:HH} we get a non-zero numerator of the expression in Eq.~\eqref{eq:GeneralDefProbability} corresponding to the finite action 

\begin{equation}
I_{\rm tot}\bigg|_{\rm tp} = \frac{\eta\pi}{2G}\frac{H_{{\rm I}}^{2}+2\kappa^{2}}{(H_{{\rm I}}^{2}+\kappa^{2})^{2}} \,.\label{eq:0toH}
\end{equation}

Note that since the expression in Eq.~\eqref{eq:HH} is symmetric under the interchange  ${\rm I}\leftrightarrow{\rm O}$ the action for down-tunnelling to Minkowski is exactly the same with ${\rm I}\rightarrow{\rm O}$ in the Eq.~\eqref{eq:0toH}. Observe that since the $M\rightarrow 0$ limit of $A=1-2MG/R$ is the same as the $H\rightarrow 0$ limit of $A=1-H^2R^2$ ($A \rightarrow 1$) the Minkowski limit of the black hole to dS action is the same as the dS to dS action where one of the dS's goes to the Minkowski limit. What is different in the two cases is the expression for the background action (which is zero in the black hole case and is the HH or the Vilenkin wave function for the (initial) dS case (see below). 

Recall now that in dS to dS transitions there is no initial classical trajectory. This is exactly like the tunnelling from `nothing' discussion in quantum cosmology in which solutions of the WDW equation are compared at the final (and only) turning point with the solution to be at scale factor zero (i.e. `nothing'). Thus these expressions should be interpreted as giving the probability to find the space of two dS (or dS and Minkowski) spaces mediated by a wall compared to have `nothing'. We will discuss this further in Sec.~\ref{sec:Wave-function-away}.

In this case, the latter configuration is the background, whose action, denoted by $\bar I$, is given by 
\begin{equation} 
\overline{I} = \frac{\eta}{G}\int_{0}^{\pi} dr \, \left[R \sqrt{1-H_{\rm O}^{2}R^{2}-R^{\prime}_{+}} - R R^{\prime} \cos^{-1}\left(\frac{R^{\prime}_{+}}{L \sqrt{1 - H_{\rm O}^{2}R^2}}\right)\right] \,.
\end{equation} 
\begin{equation} 
\Rightarrow \quad \overline{I} = \frac{\eta \pi}{2G H_{\rm O}^{2}} \,. 
\label{eq:bkgr}  
\end{equation} 
which (for $\eta=+1$) gives the HH wave function (Vilenkin's tunnelling wave function corresponds to $\eta=-1$). Adding Eq.~\eqref{eq:SBHH} to Eq.~\eqref{eq:wall} and subtracting Eq.~\eqref{eq:bkgr} we get
\begin{equation}
I_{\rm tot}\bigg|_{\rm tp} - \overline{I} = -\frac{\eta\pi}{G}\left[\frac{\left[(H_{{\rm
O}}^{2}-H_{{\rm I}}^{2})^{2}+\kappa^{2}(H_{{\rm O}}^{2}+H_{{\rm
I}}^{2})\right] R_{\rm o}}{8\kappa H_{{\rm O}}^{2}H_{{\rm
I}}^{2}}-\frac{1}{4}\left(H_{{\rm I}}^{-2}-H_{{\rm O}}^{-2}\right) \right] \,,
\label{eq:iStr}
\end{equation}
which is exactly what BT obtained. It is easy to check that the quantity in Eq.~\eqref{eq:iStr} is negative (positive) for $\eta = + 1$ ($\eta = - 1$). Observe that, when taking the limit $H_{\rm O} \rightarrow 0$, the relative probability is exponentially suppressed (choosing $\eta = +1$), implying nucleation of a dS vacuum is exponentially suppressed with respect to the nucleation of a Minkowski spacetime. This is in contrast to the limit $H_{\rm O}\rightarrow 0$ of Eq.~\eqref{eq:HH}. Thus this suppression is coming entirely from the blow up of the HH wave function. We will discuss this further in Sec.~\ref{sec:Wave-function-away} where we will explain how the sign $\eta$ is determined. 

\subsection{Schwarzschild to de Sitter transitions}
\label{sec:StodS}

\begin{figure}[h!] 
\begin{center} 
\includegraphics[scale=0.6]{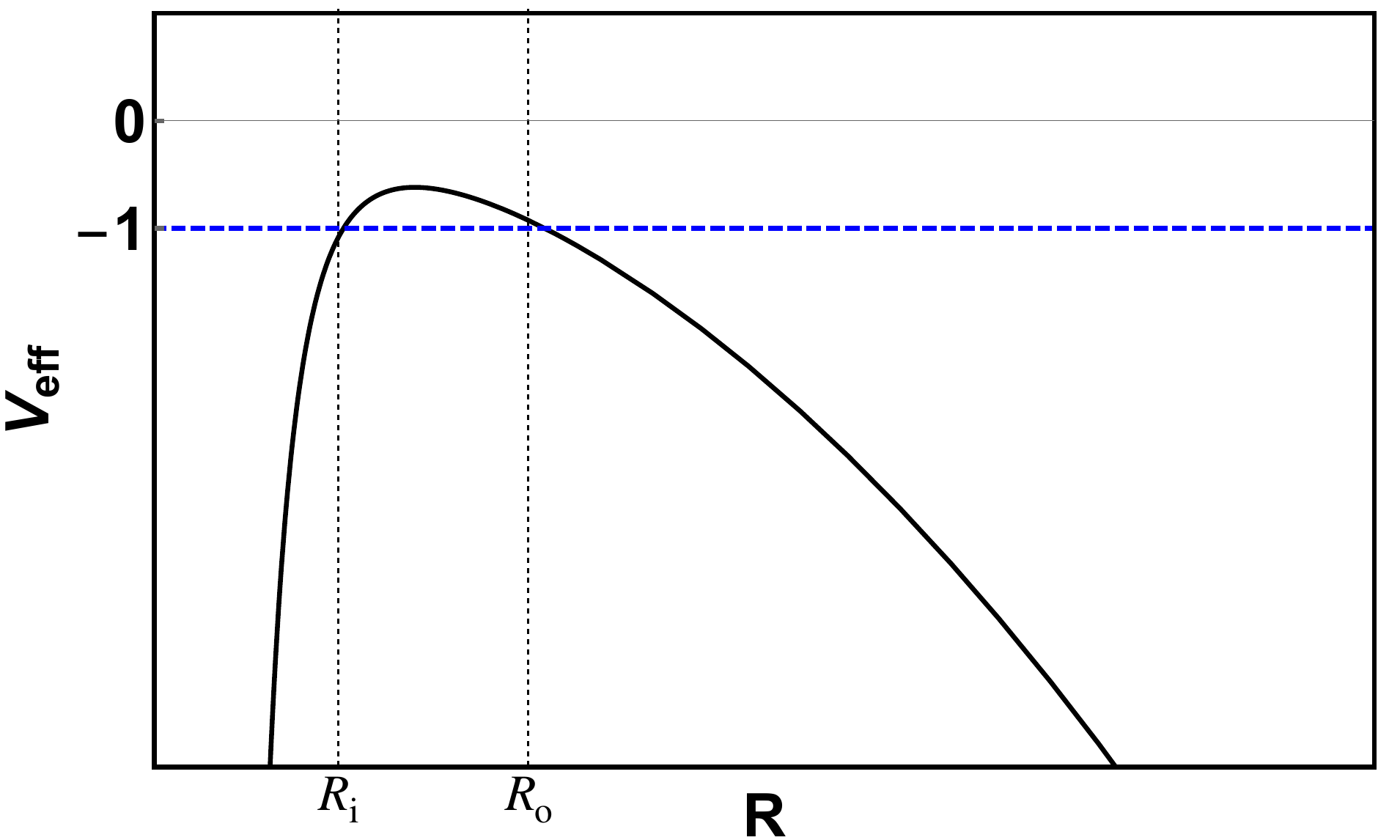} 
\caption{\footnotesize{Effective potential for Schwarzschild to dS transitions. Notice there are two turning points. \label{fig:PotentialSdS}}}
\end{center} 
\end{figure}

The case  investigated by FGG/FMP corresponds to s Schwarzschild to dS transition, namely one where
\begin{equation}
A_{{\rm O}} = 1 - \frac{2GM_{{\rm O}}}{R}\,, \qquad A_{{\rm I}}=1-H_{{\rm I}}^{2}R^{2}.
\end{equation}
In this case, Eq.~\eqref{eq:R'pm} becomes
\begin{equation}
\frac{R'(\hat{r}\pm\epsilon)}{\hat{L}}=\frac{1}{2\kappa\hat{R}}\left(-\hat{R}^{2}(H^{2}\pm\kappa^{2})+\frac{2GM}{\hat{R}}\right)\,.\label{eq:R'+-}
\end{equation}
The motion of the wall is then given by Eq.~\eqref{eq:E and m} with
\begin{eqnarray}
V & = & -\frac{1}{(2\kappa\hat{R})^{2}}\left((H_{{\rm
I}}^{2}+\kappa^{2})\hat{R}^{2}-\frac{2GM_{{\rm
O}}}{\hat{R}}\right)^{2}-\frac{2GM_{{\rm O}}}{\hat{R}} = \label{eq:VBGG}\\
 & = &
-\frac{1}{(2\kappa\hat{R})^{2}}\left((H_{\rm I}^{2}-\kappa^{2})\hat{R}^{2}-\frac{2GM_{{\rm
O}}}{\hat{R}}\right)^{2}-H_{{\rm I}}^{2}\hat{R}^{2} \,.\label{eq:VBGGalt}
\end{eqnarray}
Note that while $\Lambda_{{\rm O}}$ (and hence $H_{{\rm O}}$) is
a parameter in the Lagrangian (see Eq.~\eqref{eq:Sm}), $M$ is an integration constant (see Eq.~\eqref{eq:pi_L}). Note also that the turning points are defined by $V = - 1$ (see Fig.~\ref{fig:PotentialSdS}) and are ordered such that (see discussion after Eq.~\eqref{eq:Valt}),
\begin{equation}
R_{\rm S}\equiv 2M_{{\rm O}}G\le\hat{R}_{{\rm i}} < \hat{R}_{{\rm o}} \le R_{\rm D} \equiv H_{{\rm I}}^{-1} \,.
\end{equation}

For $R<R_{{\rm i}}$ there is a classical solution where the bubble
emerges from the black hole singularity grows to $R_{{\rm i}}$ and then collapses to the singularity. For $R>R_{{\rm o}}$ the bubble emerges from infinity contracts to $R_{{\rm o}}$ and then expands to infinity\footnote{For more details see~\cite{Blau:1986cw}.}.
Also the character of the turning point changes when the integration
constant is such that the points where $\hat{R}'_{\pm}=0$ coincide
with a turning point, i.e.
\begin{equation}
\hat{R}'_{+} = 0 \quad \Rightarrow \quad \hat{R}=R_{\mathcal S}\equiv 2MG \,; \quad V  = -1\quad{\rm
for\quad}M= M_{\rm S}\equiv\frac{1}{2G\sqrt{(H_{{\rm
I}}^{2}+\kappa^{2})\ }}\,,\label{eq:Ms}
\end{equation}
\begin{equation}
\hat{R}'_{-} = 0 \quad \Rightarrow \quad \hat{R} = R_{\rm dS}\equiv\frac{1}{H_{{\rm
I}}} \,; \quad V = -1 \quad{\rm for\quad} M =M_{\rm D}\equiv \frac{H_{{\rm
I}}^{2}-\kappa^{2}}{2GH_{{\rm I}}^{3}}\,,
\end{equation}
Note that $M_{\rm D} < M_{\rm S}$. FMP evaluated the difference in the bulk actions using Eq.~\eqref{eq:SclassicalBulk} and find the following results for different values of the integration constant\footnote{Note that the sign of $R'$ is determined by Eq.~\eqref{eq:R'pm} at $r=\hat{r}$ and then preserves this sign for $r\ne\hat{r}$ until $R$ crosses a horizon. It is useful to look at the figures  in~\cite{Blau:1986cw} (particularly Fig. 6) in deriving this. Also $\beta$ defined there is equal to $\hat{R}'$ in FMP notation used here.}
\begin{eqnarray}
I_{{\rm B}}\bigg|_{\rm tp} \equiv I_{{\rm B}}\bigg|_{R_{{\rm i}}}^{R_{\rm o}} =\begin{cases}
\frac{\eta\pi}{2G}(R_{\rm o}^{2}-R_{{\rm
i}}^{2})\,, \qquad & M > M_{\rm S} \,,\label{eq:SB1}\\
\frac{\eta\pi}{2G}(R_{\rm o}^{2}-R_{\mathcal S}^{2})\,, \qquad & M_{\rm S} > M > M_{\rm D} \,,\label{eq:SB2}\\
\frac{\eta\pi}{2G}(R_{\rm dS}^{2}-R_{\mathcal S}^{2})\,, \qquad & M_{\rm D} > M \,.\label{eq:SB3}
 \end{cases}
\end{eqnarray}
In this case, the spacetime configuration at the first turning point represents the background configuration of the transition. FMP did not compute the boundary term since, in contrast to the dS to dS considered before, the integral cannot be done analytically. We will consider in detail the last case in Eq.~\eqref{eq:SB3}, which is the case of interest in order to take the limit $M \rightarrow 0$. As we will show in the next Section, the background action for the black hole (unlike the HH wave function for dS) does not blow up in the Minkowski limit and therefore the transition does not vanish even in this limit. We also note for future reference that the last line in the above equation is simply the difference of (half) the horizon entropies of dS and Schwarzschild.

\subsection{Schwarzschild to de Sitter transitions (and viceversa) in the limit $M \rightarrow 0$}

In the zero mass limit of Schwarzschild to dS transitions (and viceversa) in contrast to the general case the wall integral can be performed analytically. 

\subsubsection*{\underline{Down-tunnelling to Minkowski: de Sitter to Minkowski transitions}}
The total action (before subtracting the background action) is given by our previous expression in Eq.~\eqref{eq:0toH} with $H_{\rm I}\rightarrow H$ (see also the discussion below Eq.~\eqref{eq:0toH}). Therefore the total action after background subtraction is:
\begin{equation}
I_{\rm tot} \bigg|_{\rm tp}-\bar I = - \frac{\eta\pi}{G} \frac{\kappa^{4}}{2H^{2}(H^{2}+\kappa^{2})^{2}}  \,.
\label{eq:13} 
\end{equation} 
This (with $\eta=+1$) is the standard result for CDL as can be also verified by starting with Eq.~\eqref{eq:iStr} and taking the limit $H_{\rm I} \rightarrow 0$. Therefore we conclude that both Minkowski limits: the $H_{\rm I} \rightarrow 0$ limit of dS to dS and the $M \rightarrow 0$ of Schwarzschild to dS give the same result for CDL down-tunnelling.

\subsubsection*{\underline{Up-tunnelling from Minkowski: Minkowski to de Sitter transitions}} 

The total action for up-tunnelling from Minkowski is given again by Eq.~\eqref{eq:0toH} with $H_{\rm I} \rightarrow H$. The background action is now the limit of the black hole action when $M\rightarrow 0$: the background action vanishes. Thus for two spaces $A$ and $B$ we denote the relative probability for being in the configuration $A/B\oplus \text{W}$ (i.e. the two  spaces joined by a wall) versus being in the initial state $A$ as $\mathcal{P}(A\rightarrow A/B\oplus \text{W})$. The transition probabilities for up- and down-tunnelling therefore read
\begin{equation}
\mathcal{P}(\mathcal{M} \rightarrow \mathcal{M}/{\rm{dS}} \oplus \text{W}) = \exp\left[\frac{\eta \pi}{G H^2} \left(1 -  \frac{\kappa^4}{(H^2 + \kappa^2)^2} \right) \right]  
\label{eq:P1}
\end{equation}
and
\begin{equation}
\mathcal{P} ({\rm{dS}}\rightarrow {\rm{dS}}/\mathcal{M} \oplus \text{W}) = \exp \left[\frac{\eta \pi}{G H^2} \left(- \frac{\kappa^4}{(H^2 + \kappa^2)^2} \right) \right]
\label{eq:P2}
\end{equation}
respectively. The ratio between the two transitions in Eq.~\eqref{eq:P1} and Eq.~\eqref{eq:P2} is thus
\begin{equation}
\frac{\mathcal{P} (\mathcal{M} \rightarrow \mathcal{M}/\rm{dS} \oplus \text{W})}{\mathcal{P} (\rm{dS} \rightarrow \rm{dS}/\mathcal{M} \oplus \text{W})} = \exp \left[\eta\frac{\pi}{G} \frac{1}{H^2} \right] \,,
\end{equation}
which is the ratio of the exponentials of the entropy of dS to the exponential of the entropy of Minkowski (taken as the $M \rightarrow 0$ limit of a Schwarzschild black hole entropy, i.e. zero), implying that detailed balance is correctly recovered even in the black hole case in the limit $M \rightarrow 0$.

Now depending on the sign $\eta$ we have two different puzzles. Take the brane tension $\kappa$ to be vanishingly small  $\kappa\ll H$ and consider first the case $\eta=+1$. Then Eq.~\eqref{eq:P1} (which is the relative probability for tunnelling from `nothing' to the composite of Minkowski and dS joined at the brane compared to remaining in the Minkowski ground state) goes over to the HH wave function for tunnelling from `nothing' to dS space. Now if we take smaller and smaller values of $H$ we get the well-known divergence of the HH wave function for zero cosmological constant. This appears to mean that it is infinitely more probable to be in a dS space with cosmological constant tending to zero than to be in the vacuum state of Minkowski. Of course this is again a reflection of the fact that the HH wave function is the exponential of the horizon entropy. In some sense this ratio (in the limit) then is the probability of being in a random state of the Hilbert space built on the Minkowski vacuum relative to being in the ground state.

On the other hand with $\eta=-1$ the same ratio gives the probability of being in dS after tunnelling from `nothing' according to Vilenkin compared to the probability of being in the Minkowski vacuum. This ratio, in the limit of the cosmological constant of the dS space going to zero, goes to zero. It is unclear how to interpret this. In Sec.~\ref{sec:Wave-function-away} we will argue that the dominant contribution to these relative probabilities corresponds to choosing $\eta=+1$ i.e. with the HH wave function for dS.

\subsection{Schwarzschild-de Sitter to Schwarzschild-de Sitter transitions}

For completeness let us briefly consider also the most general case of $\mathcal{S}$dS to $\mathcal{S}$dS. Taking two metric functions of the same kind
\be
A_{\rm I} = 1 - \frac{2GM_{\rm I}}{R} - H_{\rm I}^2 R^2 \,, \qquad
A_{\rm O} = 1 - \frac{2GM_{\rm O}}{R} - H_{\rm O}^2 R^2 \,,
\ee
at the turning points the bulk action reads 
\begin{equation}
\label{eq:BulkActionGeneric}
I_{\rm B} \bigg|_{\rm tp} = \frac{\eta \pi}{2G} \left[\theta(-\hat R^{\prime}_{-})\left(R_{\rm{I},2}^{2}-R_{\rm o}^2\right) + \theta(-\hat R^{\prime}_{+})\left(R_{\rm o}^2 - R_{\rm{O},1}^{2}\right) + R_{\rm{O},2}^{2} - R_{\rm{O},1}^{2}\right] \,,
\end{equation} 
In this setup, the metric function has two positive and one negative roots. The former actually correspond to an event and a cosmological horizon. We will refer to them as $R_{\rm{O},1} < R_{\rm{O},2}$ and $R_{\rm{I},1} < R_{\rm{I},2}$ for the outer and inner vacua, respectively.

The boundary action is given by
\begin{align}
I_{\rm b} = &- \frac{\eta}{G}\int dR R \cos^{-1} \left(\frac{\frac{2G}{R}(M_{\rm O} - M_{\rm I}) + R^2(H_{\rm O}^2 - H_{\rm I}^2 - \kappa^2)}{2 \kappa R \sqrt{1 - \frac{2GM_{\rm O}}{R} - H_{\rm O}^2 R^2}}\right) + \nonumber \\
&+ \frac{\eta}{G}\int dR R \cos^{-1} \left(\frac{\frac{2G}{R}(M_{\rm O} - M_{\rm I}) + R^2 (H_{\rm O}^2 - H_{\rm I}^2 + \kappa^2)}{2 \kappa R \sqrt{1 - \frac{2GM_{\rm I}}{R} - H_{\rm I}^2 R^2}}\right) \,.
\end{align}
Since the junction conditions now read
\begin{equation}
\frac{R^{\prime}_{\pm}}{L} = \frac{1}{2 \kappa R}\left(R^2 (H_{\rm O}^2 - H_{\rm I}^2 \mp \kappa^2) + \frac{2G(M_{\rm O} - M_{\rm I})}{R}\right) \,,
\end{equation}
and the background action is
\begin{equation} 
\overline{I} = \frac{\eta \pi}{2G} \left[R_{\rm{O},2}^{2} - R_{\rm{O},1}^{2}\right] \,.
\end{equation} 
\begin{figure}[h!]
\begin{center} 
\includegraphics[scale=0.6]{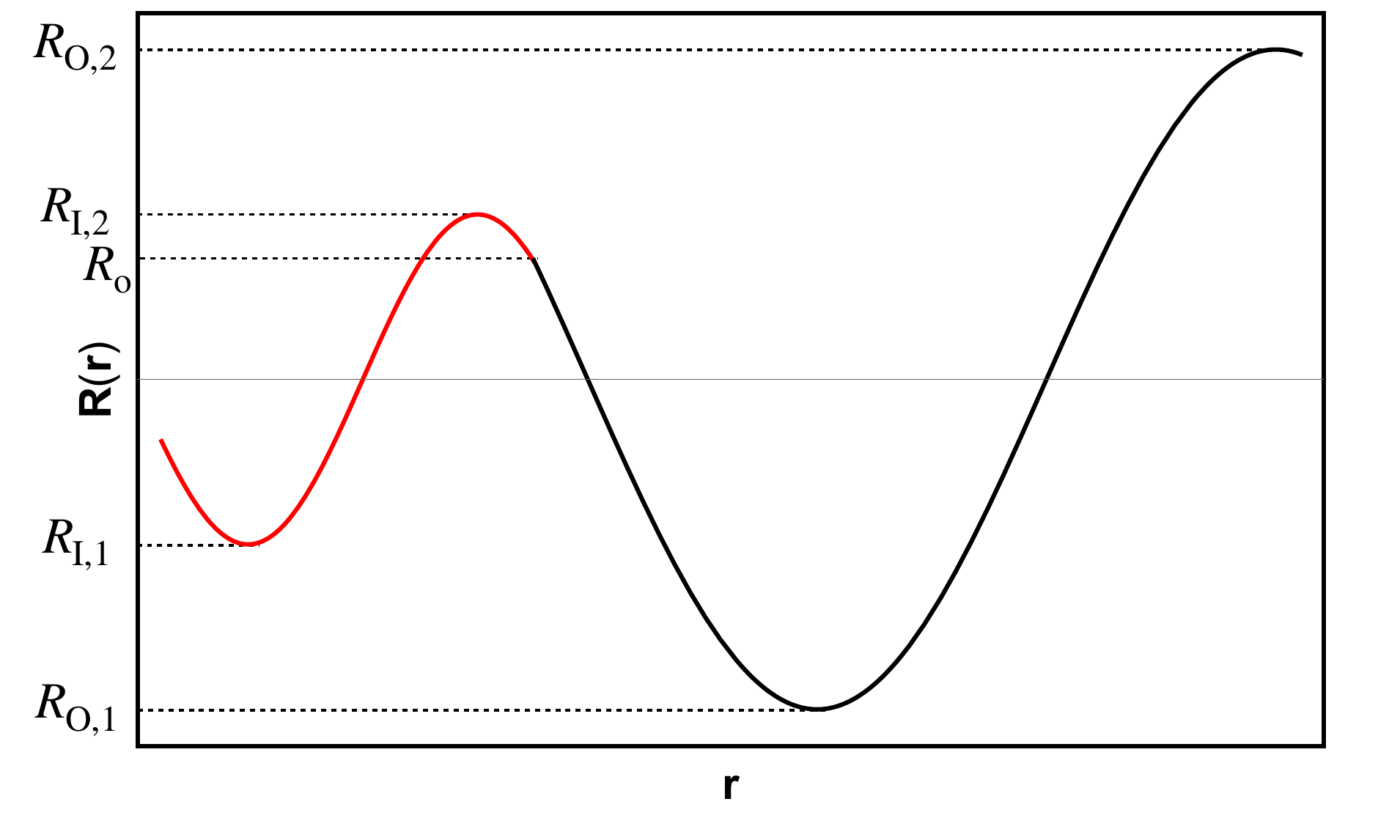} 
\caption{\footnotesize{Still relying upon the same arguments as for the dS to dS case, the plot of $R(r)$ on either side is what enables to determine the nontrivial contributions to the bulk action. Indeed, in absence of any disconnected spacetimes, $R^{\prime}_{\pm} < 0$ only in between $1/H_{\rm I}$ and $2GM_{\rm O}$.}}
\end{center} 
\end{figure}

\begin{figure}[h!]  
\centering
\centerline{\includegraphics[scale=1.1, trim = 0cm 0.5cm 0cm 0cm, clip]{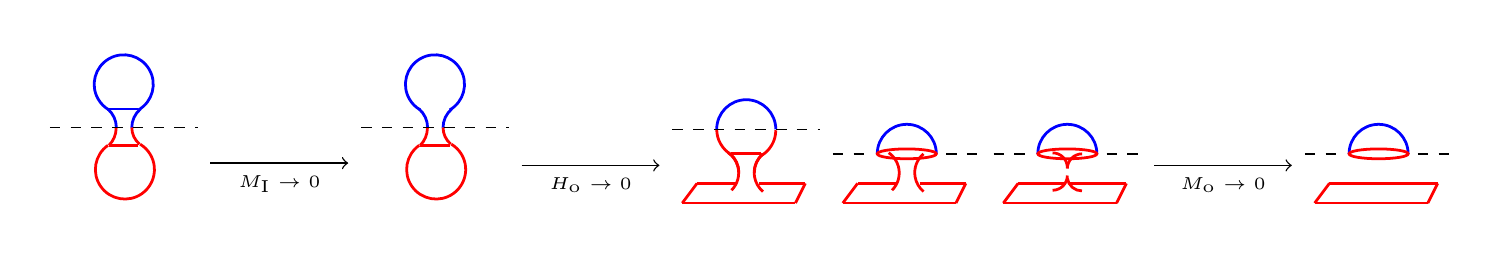}} 
\caption{\footnotesize{This case, obtained by first taking $H_{\textrm{O}}\rightarrow 0$ and then $M_{\textrm{I}}\rightarrow 0$, is an example leading to an up-tunnelling Minkowski to dS transition. In particular, notice that such particular choice of the order of the limits reduces ${\cal S}$dS to ${\cal S}$dS transitions to the Schwarzschild to dS case also dealt with by BGG. \label{fig:MIMO}}}
\end{figure}

\begin{figure}[h!]
\centering
\centerline{\includegraphics[scale=1.2, trim = 0cm 0.5cm 0cm 0cm, clip]{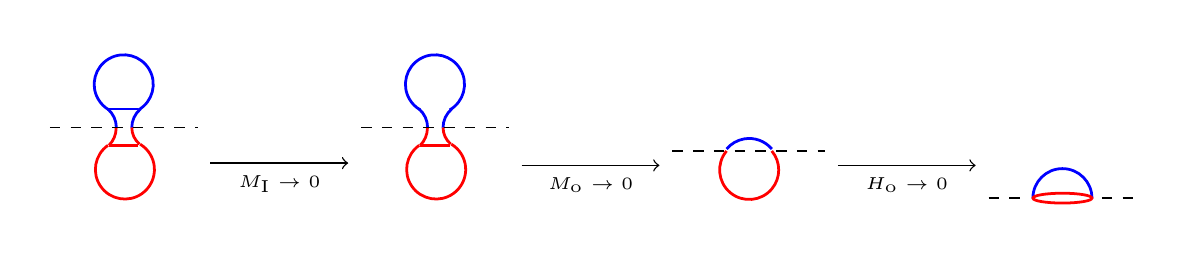}}
\caption{\footnotesize{In this case instead, the $H_{\textrm{O}} \rightarrow 0$ and $M_{\textrm{O}} \rightarrow 0$ limits commute, reducing the more general $\mathcal{S}$dS to $\mathcal{S}$dS case to the ususal BT calculation, which prohibits up-tunnelling from Minkowski. \label{fig:MIHO}}}
\end{figure}

\noindent In this case the classical turning points are determined by solving the following system of equations
\begin{equation}
V_{\rm eff}^{\rm out} = -1 \quad \Rightarrow \quad - \frac{1}{(2 \kappa R)^2} \left(\frac{2G}{R}(M_{\rm O} - M_{\rm I}) + R^2 (H_{\rm I}^2 - H_{\rm O}^2) \right)^2 + H_{\rm O}^2 R^2 +\frac{2GM_{\rm O}}{R} = -1 \,, \nonumber
\end{equation}
\begin{equation}
V_{\rm eff}^{\rm in} = -1 \quad  \Rightarrow \quad -\frac{1}{(2 \kappa R)^2} \left(\frac{2G}{R}(M_{\rm O} - M_{\rm I}) + R^2 (H_{\rm I}^2 - H_{\rm O}^2)\right)^2 + H_{\rm I}^2 R^2 +\frac{2GM_{\rm I}}{R} = -1\,. \nonumber
\end{equation}
These are polynomials of degree 8 in $R$, the solutions dependening on the different parameters. In general, the  $\mathcal{S}$dS to $\mathcal{S}$dS transitions are characterised by four parameters, namely $H_{\rm I,O}$ and $M_{\rm I,O}$ and the particular case Minkowski to dS can be extracted by taking $H_{\rm O}, M_{\rm I,O} \rightarrow 0$. According to the particular order with which these three limits are taken, a non-commutative behaviour emerges. Of the six possible cases, four of them allow transitions as in the Schwarzschild to de Sitter case and two do not allow Minkowski to dS transitions as in the dS to dS case. Examples of both cases are depicted in Fig.~\ref{fig:MIMO} and Fig.~\ref{fig:MIHO}, the dashed black line denotes the location of the wall. 


\vspace{0.5cm}
Before taking the $M_{\textrm{O}}\rightarrow 0$ limit, Fig.~\ref{fig:MIMO} corresponds to BGG's sequence for Schwarzschild to dS transitions. In their article, the authors argue that, selecting constant time slices intersecting the black hole event horizon, the corresponding three-geometry is split up into two regions, lying on the left and right hand sides of the maximally-extended Schwarzschild configuration, respectively. The former is joined to the newly nucleated dS by the wall, whereas the latter is the relic vacuum prior to the transition. When ultimately taking the limit of vanishing mass parameter, the right hand side amounts to the Minkowski flat spacetime in the last step of the sequence in Fig.~\ref{fig:MIMO}.

\subsection{Summary of vacuum decays}

\begin{figure}[h!] 
\begin{center} 
\includegraphics[scale=0.6]{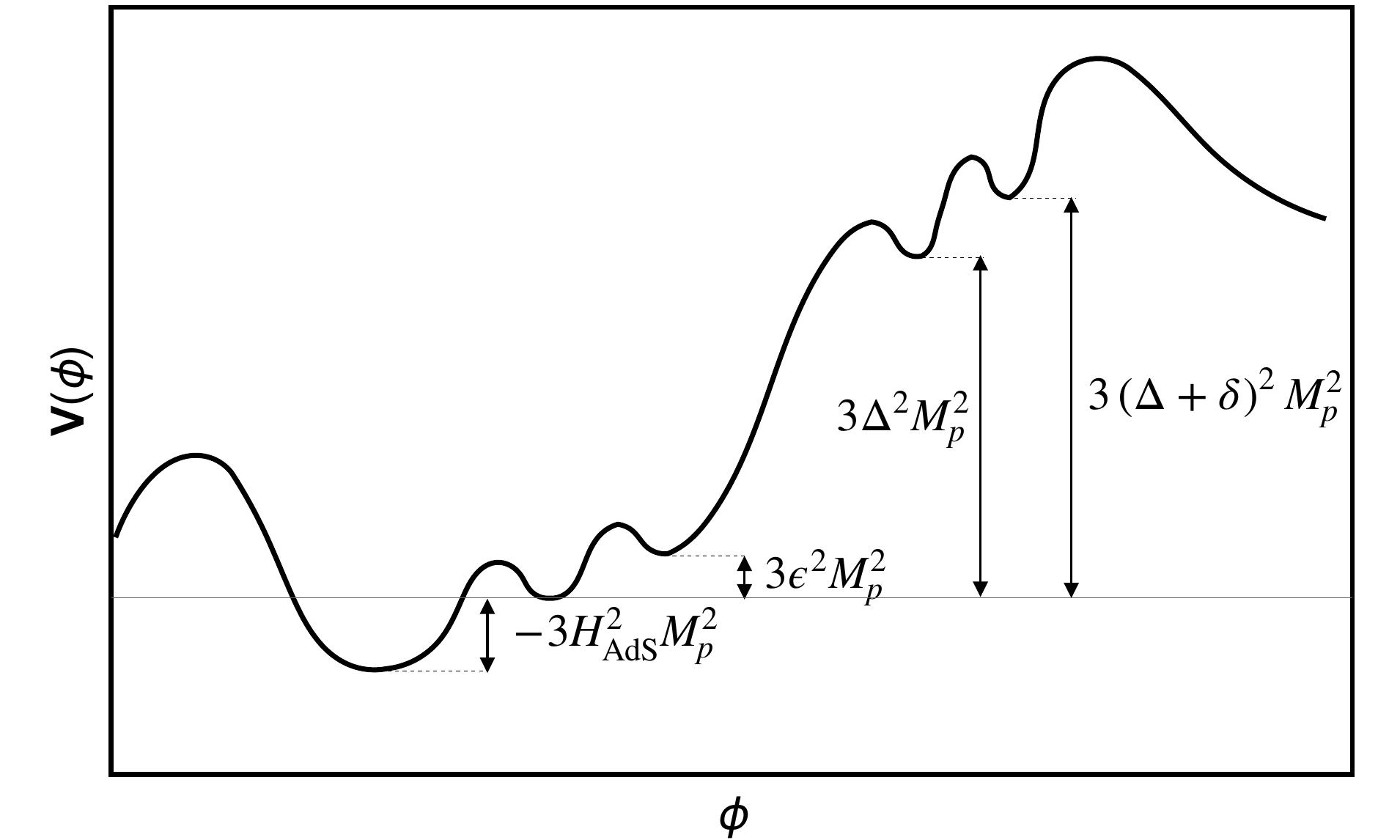}
\caption{\footnotesize{Potential that mimics a simplified situation occurring in the landscape, with various minima at different (positive, vanishing and negative) values of the cosmological constant.\label{fig:LandscapePotential}}}
\end{center} 
\end{figure}
\noindent Let us summarise our general results regarding transitions between different vacua. We consider for both up-tunnelling and down-tunnelling processes, different ranges for the energy gap between the nucleated and background configurations, see Fig.~\ref{fig:LandscapePotential}. To provide a qualitative idea of the leading-order behaviour of such transition amplitudes, we consider four different values of the cosmological constant $\Lambda = 3 H_i^2 M_p^2$ and $H_i = $ 0, $\epsilon$, $\Delta$, $\Delta + \delta$ with $0 < \epsilon$, $\delta \ll \Delta$. For completeness we add to the discussion an AdS vacuum with vacuum energy $- 3 H_{\rm AdS}^2 M_p^2$. We will use Eq.~\eqref{eq:iStr} to compute transitions between dS vacua, Eq.~\eqref{eq:P1} to compute up-tunnelling transitions from Minkowski and the following expressions, obtained by CDL, for the down-tunnelling dS to AdS
\begin{equation}
I_{\rm tot}\bigg|_{\rm tp} = - \frac{\eta \pi}{G} \left[\frac{\left(H_{\rm O}^2 - H_{\rm AdS}^2\right)^2 + \kappa^2 \left(H_{\rm O}^2 + H_{\rm AdS}^2\right)}{8 \kappa H_{\rm O}^2 H_{\rm AdS}^2} \tilde{R}_{\rm o} + \frac{1}{4 H_{\rm O}^2}\right] \,,
\end{equation}
where the turning point in this case is given by
\begin{equation}
\tilde{R}_{\rm o} = \frac{2 \kappa}{\sqrt{\left(\left(H_{\rm AdS}^2 + H_{\rm O}^2 + \kappa^2\right)^2 - 4 H_{\rm AdS}^2 \kappa^2 \right)}} \,.
\end{equation}
We do not consider up-tunnelling AdS to Minkowski/dS since it is prohibited: it can be checked that it is not possible to find a solution for the turning point radius $R_{\rm o}$. The $ij$-th entry in Tab.~\ref{tab:TransitionsSummary} gives $2 \left(I_{\rm tot} - \overline{I}\right)_{ij} = \ln\left(\mathcal{P}_{ij}\right)$, where $\mathcal{P}_{ij}$ refers to $\Lambda_{i}\rightarrow \Lambda_{j}$ transitions. Given the positivity of the action (see e.g. Eq.~\eqref{eq:HH} and Eq.~\eqref{eq:bkgr}), we take $\eta = +1$ in this Section, assuming that the raising solution under the barrier dominates: all the table entries would change sign with the opposite choice, and the preferred transitions would change accordingly. We define $\omega \equiv \epsilon/\kappa$, $\xi = \Delta/H_{\rm AdS}$, $\zeta = \kappa/H_{\rm AdS}$, $\iota = \epsilon/H_{\rm AdS}$ and
\begin{table}[h!]
\centering
\begin{tabular}{|*{6}{c|}}\hline
\backslashbox{$H_{\rm O}$}{$H_{\rm I}$} & \makebox[2em]{$H_{\rm AdS}$}
&\makebox[2em]{$0$}&\makebox[2em]{$\epsilon$}&\makebox[2em]{$\Delta$}
&\makebox[2em]{$\Delta + \delta$} \\ \hline
$0$ & $\Omega_0^{\rm AdS}$ & $\times$ & $\Omega_0^{\epsilon}$ & $\Omega_0^\Delta$ & $\Omega_0^\Delta \left(1 - g_1 \frac{\delta}{\Delta}\right)$ \\ \hline
$\epsilon$ & $\Omega_\epsilon^{\rm AdS}$ & $\Omega_{\epsilon}^0$ & $\times$ & $\Omega_\epsilon^\Delta$ & $\Omega_\epsilon^\Delta \left(1 + g_2 \frac{\delta \epsilon^2}{\Delta^3}\right)$ \\ \hline
$\Delta$ & $\Omega_{\Delta}^{\rm AdS}$ & $\Omega_\Delta^0$ & $\Omega_\Delta^0 \left(1 + g_3 \frac{\epsilon^2}{\Delta^2}\right)$ & $\times$ & $\Omega_\Delta^{\Delta + \delta}$ \\ \hline
$\Delta + \delta$ & $\Omega_{\Delta}^{\rm AdS} \left(1 + g_4 \frac{\delta}{\Delta}\right)$ & $\Omega_{\Delta+\delta}^0 $  & $\Omega_{\Delta + \delta}^0 \left(1 + g_3 \frac{\epsilon^2}{\Delta^2}\right)$ & $\Omega_{\Delta + \delta}^\Delta$ & $\times$ \\ \hline
\end{tabular}
\caption{Minkowski/dS to AdS/Minkowski/dS transitions: this table shows the leading contribution to $2 \left(I_{\rm tot} - \overline{I}\right)_{ij} = \ln\left(\mathcal{P}_{ij}\right)$ with $i, j = H_{\rm AdS}$, $0$, $\epsilon$, $\Delta$, $\Delta + \delta$ and $0 < \epsilon$, $\delta \ll \Delta$. \label{tab:TransitionsSummary}}
\end{table}
\vspace{0.2cm}

\begin{equation}
\Omega_0^{\rm AdS} = - \frac{\pi}{2 G H_{\rm AdS}^{2}} \frac{\kappa^{4}}{(H_{\rm AdS}^{2} - \kappa^{2})^{2}} \,, \nonumber
\end{equation}
\begin{equation}
\Omega_0^\epsilon = \frac{\pi}{G \kappa^2} \frac{2 + \omega^2}{\left(1 + \omega^2\right)^2} \,, \qquad\Omega_{\epsilon}^0 = - \frac{\pi}{G \epsilon^2} \frac{1}{\left(1 + \omega^2\right)^2} \,, \nonumber
\end{equation}
\begin{equation}
\Omega_0^\Delta = \frac{\pi}{G \Delta^2} \frac{1 + 2 \lambda^2}{(1 + \lambda^2)^2} \,, \qquad \Omega_\Delta^0 = - \frac{\pi}{G \Delta^2} \frac{\lambda^4}{\left(1 + \lambda^2\right)^2} \,,  \nonumber
\end{equation}
\begin{equation}
\Omega_\epsilon^\Delta = - \frac{\pi}{G \epsilon^2} \left(1 - \frac{\epsilon^2}{\Delta^2} \frac{1 + 2 \lambda^2}{(1 + \lambda^2)^2}\right) \,, \nonumber
\end{equation}
\begin{equation}
\Omega_{\Delta + \delta}^0 = \Omega_\Delta^0 \left(1 - \frac{\delta}{\Delta} \frac{6 + 2 \lambda^2}{1 + \lambda^2} + \frac{\delta^2}{\Delta^2} \frac{3 (7 + 4 \lambda^2 + \lambda^4)}{(1+ \lambda^2)^2} \right) \,, \nonumber
\end{equation}
\begin{equation}
\Omega_{\Delta + \delta}^\Delta = - \frac{\pi}{G \Delta^2} \frac{\lambda}{\sqrt{4 + \lambda^2}} \left(1 - \frac{\delta}{\Delta} \frac{\left(\lambda (\lambda + \sqrt{4 + \lambda^2}) + \left(6 + \frac{4}{\lambda} \sqrt{4 + \lambda^2}\right)\right)}{4 + \lambda^2}\right) \,, \nonumber
\end{equation}
\begin{equation}
\Omega_\Delta^{\Delta + \delta} = - \frac{\pi}{G \Delta^2} \frac{\lambda}{\sqrt{4 + \lambda^2}} \left(1 - \frac{\delta}{\Delta} \frac{\left(\lambda (\lambda - \sqrt{4 + \lambda^2}) + \left(6 - \frac{4}{\lambda} \sqrt{4 + \lambda^2}\right)\right)}{4 + \lambda^2}\right) \,, \nonumber
\end{equation}
\begin{equation}
\Omega_{\epsilon}^{\rm AdS} = - \frac{\pi}{2 G \epsilon^2} \frac{\left(1 - \iota^2\right)^2 + \zeta^2 \left(1 + \iota^2\right) + \sqrt{\Theta}}{\sqrt{\Theta}} \,, \nonumber
\end{equation}
\begin{equation}
\Omega_{\Delta}^{\rm AdS} = - \frac{\pi}{2 G \Delta^2} \frac{1 + \xi^2 \left(\xi^2 + \zeta^2 - 2\right) + \zeta^2 + \sqrt{\Xi}}{\sqrt{\Xi}} \,, \nonumber
\end{equation}
\begin{equation}
g_1 = \frac{2 + 6 \lambda^2}{1 + 3 \lambda^2 + 2 \lambda^4} \,, \qquad g_2 = \frac{2 + 6 \lambda^2}{\left(1 + \lambda^2\right)^3} \,, \qquad g_3 = \frac{3}{(1+\lambda^2)^2} \,, \nonumber
\end{equation}
\begin{footnotesize}
\begin{equation}
g_4 = - \frac{2 \left(1 + 3 \xi^2 \left(1 - \frac{\xi^2}{3} - \xi^4 + \xi^2 \zeta^2 + \zeta^4 + 2 \zeta^2\right) + \zeta^2 \left(\zeta^4 - \zeta^2 - 1\right) + \sqrt{\Xi} \left(1 + 2 \xi^2  - 2 \zeta^2 + (\xi^2 + \zeta^2)^2\right) \right)}{\Xi \left(1 + \xi^2 \left(\xi^2 + \zeta^2 - 2\right) + \zeta^2 + \sqrt{\Xi} \right)} \,, \nonumber
\end{equation}
\end{footnotesize}
\begin{equation}
\Theta = \left(1 + \iota^2 + \zeta^2 \right)^2 - 4 \zeta^2 \,, \nonumber
\end{equation}
\begin{equation}
\Xi = 1 + 2 \left(\xi^2 - \zeta^2\right) + \left(\xi^2 + \zeta^2\right)^2 \,. \nonumber
\end{equation}

\noindent Notice that the functions $g_1$, $g_2$, $g_3$ are non-negative and they all tend to $0$ in the limit $\lambda \rightarrow \infty$. From Tab.~\ref{tab:TransitionsSummary} it is possible to read which transitions are preferred:
\begin{itemize}
\item $\mathcal{P}(0 \rightarrow \epsilon)$ vs $\mathcal{P}(0 \rightarrow \Delta)$.\\
Please note that, since $\lambda = \frac{\kappa}{\Delta} = \frac{1}{\omega} \frac{\epsilon}{\Delta}$, then $\lambda \ll 1$ implies $\frac{\epsilon}{\Delta} \ll \omega$. On the other hand, $\lambda \gg 1$ implies that $\omega \ll \frac{\epsilon}{\Delta} \ll 1$, i.e. $\omega$ is bound to be small. $\lambda \omega = \frac{\epsilon}{\Delta} \ll 1$ is always bound to be small. Then
\begin{align}
\label{eq:0epsilonvs0Delta}
\frac{\mathcal{P}\left(0 \rightarrow \epsilon\right)}{\mathcal{P}\left(0 \rightarrow \Delta\right)} &\simeq \exp\left[\frac{\pi}{G \kappa^2} \left(\frac{2 + \omega^2}{\left(1 + \omega^2\right)^2} - \frac{\lambda^2 (1 + 2 \lambda^2)}{(1 + \lambda^2)^2}\right)\right] \simeq \nonumber \\
&\simeq \begin{cases}
\frac{\pi}{G \kappa^2} \,\mathcal{O}(1) \quad &\lambda \ll 1 \,, \, \frac{\epsilon}{\Delta} \ll \omega \lesssim \mathcal{O}(1) \,, \\
\frac{\pi}{G \kappa^2} \frac{1}{\omega^2} \left(1 - \frac{\epsilon^2}{\Delta^2} \right) \quad &\lambda \ll 1 \,, \, \omega \gg 1 \,, \\
\frac{3 \pi}{G \kappa^2} \frac{1}{\lambda^2}  \left(1 - \frac{\epsilon^2}{\Delta^2} \right) \quad &\lambda \gg 1 \,, \, \omega \ll \frac{\epsilon}{\Delta} \ll 1 \,.
\end{cases}
\end{align}
Given that $\frac{\epsilon}{\Delta} \ll 1$, the ratio is always positive and the $0 \rightarrow \epsilon$ transition is preferred over the $0 \rightarrow \Delta$ transition.

\item $\mathcal{P}(0 \rightarrow \Delta)$ vs $\mathcal{P}(0 \rightarrow \Delta + \delta)$.

In this case the comparison is simple
\begin{equation}
\label{eq:0Deltavs0Deltadelta}
\frac{\mathcal{P}\left(0 \rightarrow \Delta\right)}{\mathcal{P}\left(0 \rightarrow \Delta + \delta\right)} \simeq \exp\left[g_1 \, \Omega_0^\Delta \, \frac{\delta}{\Delta} \right] \,,
\end{equation}
and given that $g_1 > 0$, the transition $0 \rightarrow \Delta$ is preferred over the transition $0 \rightarrow \Delta + \delta$.

\item $\mathcal{P}(\epsilon \rightarrow 0)$ vs $\mathcal{P}(\epsilon \rightarrow \Delta)$.\\
$\mathcal{P}\left(\epsilon \rightarrow 0\right)$ is exact in $\epsilon$, so
\begin{align}
\frac{\mathcal{P}\left(\epsilon \rightarrow 0\right)}{\mathcal{P}\left(\epsilon \rightarrow\Delta\right)} & \simeq \exp\left[\frac{\pi}{G \epsilon^2} \left(\frac{\omega^2(2 + \omega^2)}{(1 + \omega^2)^2} - \frac{\epsilon^2}{\Delta^2} \frac{1 + 2 \lambda^2}{(1 + \lambda^2)^2}\right)\right] \simeq \nonumber\\
& \simeq \begin{cases} \frac{\pi}{G \epsilon^2} \, \mathcal{O}(1) \quad &\lambda \ll 1 \,, \, \frac{\epsilon}{\Delta} \ll \omega \lesssim \mathcal{O}(1) \,, \\
\frac{\pi}{G \epsilon^2} \left(1 - \frac{\epsilon^2}{\Delta^2}\right)  \quad &\lambda \ll 1 \,, \, \omega \gg 1 \,, \\
\frac{9 \pi}{G \kappa^2} \frac{1}{\lambda^2} \quad &\lambda \gg 1 \,, \, \omega \ll \frac{\epsilon}{\Delta} \ll 1 \,.
\end{cases}
\end{align}
Hence, down-tunnelling towards a Minkowski vacuum is always preferred.

\item $\mathcal{P}(\epsilon \rightarrow \Delta)$ vs $\mathcal{P}(\epsilon \rightarrow \Delta + \delta)$.\\
In this case the comparison is simple:
\begin{align}
\frac{\mathcal{P}\left(\epsilon \rightarrow \Delta\right)}{\mathcal{P}\left(\epsilon \rightarrow \Delta + \delta\right)} & \simeq \exp\left[g_2 \, \left|\Omega_{\epsilon}^{\Delta}\right| \, \frac{\delta \epsilon^2}{\Delta^3}\right] \,,
\end{align}
where we used that $\Omega_{\epsilon}^{\Delta} < 0$. Since $g_2 > 0$ the transition $0 \rightarrow \Delta$ is always preferred over the transition $0 \rightarrow \Delta + \delta$.

\item $\mathcal{P}(\Delta \rightarrow 0)$ vs $\mathcal{P}(\Delta \rightarrow \Delta + \delta)$.
\begin{align}
\frac{\mathcal{P}\left(\Delta \rightarrow 0\right)}{\mathcal{P}\left(\Delta \rightarrow \Delta + \delta\right)} &\simeq \exp\left[\frac{\pi}{G \Delta^2} \left(\frac{\lambda}{\sqrt{4 + \lambda^2}} - \frac{\lambda^4}{(1 + \lambda^2)^2} + \frac{\delta}{\Delta} \alpha_1\right)\right] \simeq \nonumber \\
&\simeq \begin{cases} \frac{\pi}{G \Delta^2} \left(\frac{\lambda}{2} + \frac{\delta}{\Delta}\right) \quad &\lambda \ll 1 \,, \\
\frac{3 \pi}{G \kappa^2} \frac{1}{\lambda^2} \left(1 + 2 \frac{\delta}{\Delta}\right) \quad &\lambda \gg 1 \,,
\end{cases}
\label{eq:PD0DDd}
\end{align}
where
\begin{equation}
\alpha_1 = \left(\frac{4 \sqrt{4 + \lambda^2} - \lambda \left(6 + \lambda^2 - \lambda \sqrt{4 + \lambda^2}\right)}{\left(4 + \lambda^2\right)^{3/2}}\right) \,.
\end{equation}
Hence, from Eq.~\eqref{eq:PD0DDd} we infer that a large down-tunnelling to a Minkowski vacuum is preferred over a small up-tunnelling in both limits.

\item $\mathcal{P}(\Delta \rightarrow \epsilon)$ vs $\mathcal{P}(\Delta \rightarrow \Delta + \delta)$.
\begin{align}
\frac{\mathcal{P}\left(\Delta \rightarrow 0\right)}{\mathcal{P}\left(\Delta \rightarrow \Delta + \delta\right)} &\simeq \exp\left[\frac{\pi}{G \Delta^2} \left(\frac{\lambda}{\sqrt{4 + \lambda^2}} - \frac{\lambda^4}{(1 + \lambda^2)^2} + \frac{\delta}{\Delta} \alpha_1 + \frac{\epsilon^2}{\Delta^2} \alpha_2\right)\right] \,, \nonumber
\label{eq:PDeDDd}
\end{align}
where
\begin{equation}
\alpha_2 = - \frac{3 \lambda^4}{\left(1 + \lambda^2\right)^4} \,. \nonumber
\end{equation}
The term proportional to $\alpha_2$ introduces $\mathcal{O}\left(\frac{\epsilon^2}{\Delta^2}\right)$ corrections to the result in Eq.~\eqref{eq:PD0DDd} so that the qualitative behaviour does not change.

\item $\mathcal{P}(\Delta + \delta \rightarrow \epsilon)$ vs $\mathcal{P}(\Delta + \delta \rightarrow \Delta)$.
\begin{align}
\frac{\mathcal{P}\left(\Delta + \delta \rightarrow \epsilon\right)}{\mathcal{P}\left(\Delta + \delta \rightarrow\Delta\right)} &\simeq \exp\left[\frac{\pi}{G \Delta^2} \left(\frac{\lambda}{\sqrt{4 + \lambda^2}} - \frac{\lambda^4}{(1 + \lambda^2)^2} + \frac{\delta}{\Delta} \gamma_1 - \frac{\delta^2}{\Delta^2} \gamma_2 - \frac{\epsilon^2}{\Delta^2} \gamma_3\right)\right] \simeq \nonumber \\
&\simeq \begin{cases} \frac{\pi}{G \Delta^2} \left(\frac{\lambda}{2} - \frac{\delta}{\Delta}\right) \quad &\lambda \ll 1 \,, \\
- \frac{3 \pi}{G \Delta^2} \frac{\delta^2}{\Delta^2} \quad &\lambda \gg 1 \,,
\end{cases}
\label{eq:PDdDDde}
\end{align}
where
\begin{equation}
\begin{gathered}
\gamma_1 = 1 - \frac{\lambda (6 + \lambda^2)}{(4 + \lambda^2)^{3/2}} - \frac{6}{(1 + \lambda^2)^2} + \frac{4}{(1 + \lambda^2)^3} \,, \\
\gamma_2 = \frac{3 \lambda^4 (7 + 4 \lambda^2 + \lambda^4)}{(1 + \lambda^2)^4} \,, \\
\gamma_3 = \frac{3 \lambda^4}{(1 + \lambda^2)^4} \,.
\end{gathered}
\end{equation}
From Eq.~\eqref{eq:PDdDDde} we see that in the limit $\lambda \ll 1$ the sign of the ratio depends on the sign of the bracket, while in the limit $\lambda \gg 1$ it is always negative, i.e. the small down-tunnelling $\Delta + \delta \rightarrow \Delta$ is preferred over the large down-tunnelling $\Delta + \delta \rightarrow \epsilon$. The result is qualitatively similar to the case $\mathcal{P}(\Delta + \delta \rightarrow 0)$ vs $\mathcal{P}(\Delta + \delta \rightarrow \Delta)$ where the $\epsilon$-dependent correction proportional to $\gamma_3$ would disappear.

\item $\mathcal{P}(0 \rightarrow\epsilon)$ vs $\mathcal{P}(0 \rightarrow H_{\rm AdS})$.
\begin{align}
\frac{\mathcal{P}\left(0 \rightarrow \epsilon\right)}{\mathcal{P}\left(0 \rightarrow H_{\rm AdS}\right)} &\simeq \exp\left[\frac{\pi}{G} \left(\frac{1}{\kappa^2} \frac{2 + \omega^2}{\left(1 + \omega^2\right)^2} + \frac{1}{2 H_{\rm AdS}^2} \frac{\zeta^4}{\left(1 - \zeta^2\right)^2}\right)\right] \,,
\end{align}
which is always positive: the up-tunnelling transition towards a dS vacuum $0 \rightarrow \epsilon$ is always preferred over the down-tunnelling transition towards an AdS vacuum $0 \rightarrow H_{\rm AdS}$. The comparison between the transitions $0 \rightarrow \Delta (+ \delta)$ and $0 \rightarrow H_{\rm AdS}$ is analogous.

\item $\mathcal{P}(\epsilon \rightarrow 0)$ vs $\mathcal{P}(\epsilon \rightarrow H_{\rm AdS})$.\\
We consider only the case in which $H_{\rm AdS} \simeq \Delta \gg \epsilon$, i.e. $\iota \ll 1$. In this case, if $\omega \ll 1$ then $\iota \ll \zeta$ and $\zeta$ can be either much larger or much smaller than one. However, if $\omega \gg 1$ then $\iota \gg \zeta$ and so $\zeta \ll 1$.
\begin{align}
\frac{\mathcal{P}\left(\epsilon \rightarrow 0\right)}{\mathcal{P}\left(\epsilon \rightarrow H_{\rm AdS}\right)} &\simeq \exp\left[- \frac{\pi}{G \epsilon^2} \left(\frac{1}{\left(1 + \omega^2\right)^2} - \frac{\left(1 - \iota^2\right)^2 + \zeta^2 \left(1 + \iota^2\right) + \sqrt{\Theta}}{2 \sqrt{\Theta}}\right) \right] = \, \nonumber \\
& = \frac{\pi}{G \epsilon^2} \begin{cases}
2 \omega^2 + \zeta^2 - \frac{3}{2} \iota^2 \,, \qquad &\omega \ll 1 \,, \iota \ll 1 \,, \zeta \ll 1 \,, \\
\frac{1 + \zeta^2 + \sqrt{(1 + \zeta^2)^2 - 4 \zeta^2}}{2 \sqrt{(1 + \zeta^2)^2 - 4 \zeta^2}} - 1\,, \qquad & \omega \ll 1 \,, \iota \ll 1 \,, \zeta \gtrsim 1 \,, \\
1 - \frac{3}{2} \iota^2 \,, \qquad & \omega \gg 1 \,, \zeta \ll \iota \ll 1 \,. \\
\end{cases}
\end{align}
The down-tunnelling to AdS $\epsilon \rightarrow H_{\rm AdS}$ can be preferred over the down-tunnelling to Minkowski in the cases $\omega, \iota, \zeta \ll 1$ and $\omega, \iota \ll 1$, $\zeta \gtrsim 1$, depending on the sign of the bracket. The cases of down-tunnelling transitions from a larger initial cosmological constant are analogous.
\end{itemize}

\section{Wave function away from the turning point\label{sec:Wave-function-away}}

Let us consider again the dS to dS transitions. So far we have evaluated the semi-classical wave function only at the turning points. This simplifies the evaluation of the integral in Eq.~\eqref{eq:SclassicalBulk} considerably since at the turning point geometry $AL^{2} = R'^{2}$ so that the first term is zero and the second integral is zero or $\pi$. However, one should really study the general solutions of the WDW equation with appropriate boundary conditions. In fact in the familiar quantum mechanics tunnelling problem one has two solutions inside the barrier as well as two in each classically allowed region, and they should be matched at the turning point(s). In order to implement the same procedure for dS to dS transitions we need to parametrise the dS solutions in both regions and discuss the matching between outside and inside the barrier. At the end of the day, however, we will see that effectively the probability of tunnelling is given by the ratio of the dominant terms in the numerator and denominator (unless for some reason the corresponding integration constant(s) are set to zero). 

Away from the turning  points one cannot evaluate the integral in Eq.~\eqref{eq:SclassicalBulk} without knowing the functional dependencies $R = R(r)$, $L = L(r)$. As for the boundary integral, it is defined at $r=\hat{r}\pm\epsilon$, but we evaluated it (in the dS $\rightarrow$ dS case) only at the turning point. Away from a turning point to evaluate both the wall and the bulk integrals we need the result 
\begin{eqnarray*}
\int^{u} dx x \, \cos^{-1}\frac{x}{\sqrt{1 - a x^{2}}} & = & \frac{1}{2} \left[u^{2} \cos^{-1}\left(\frac{u}{\sqrt{1-a u^{2}}}\right) +\right.\\
 & + & \frac{1}{a}\left.\left( \sin^{-1}\frac{u}{\sqrt{1-au^{2}}}+\frac{1}{\sqrt{1+a}}\sin^{-1}\left(-\sqrt{1+a} u \right) \right)\right] + {\rm const.} \,.
\end{eqnarray*}
At the turning point (i.e. at $u=1/\sqrt{1+a}$) and taking $0$ as the lower limit of integration, this gives the result quoted in footnote~\ref{fn:DefInt}\footnote{In evaluating that integral we substituted $x=|c|\hat{R}$ where $c_{\mp}=\frac{1}{2\kappa}(H_{{\rm O}}^{2}-H_{{\rm I}}^{2}\pm\kappa^{2})$ with appropriate changes for $\hat{R}'<0$ as in footnote~\ref{fn:DefInt}.}.

Let us now work out the result of doing the wall integral away from
the turning point. In the two regions, inside and outside the wall, we parametrise
\begin{equation}
\begin{split}
\text{I}:& \qquad R = a_{\rm I} \sin r \,, \, \quad L = a_{\rm I} \,, \\
\text{O}:&\qquad R = a_{\rm O}\sin r \,, \quad L = a_{\rm O} \,.
\label{eq:a0aIR}
\end{split}
\end{equation}
The wall is located at $r = \hat{r}_{-}$ and at $r = \hat{r}_{+}$ in the inside and outside coordinate systems respectively. Its proper radius is then 
\begin{equation}
\hat{R} = a_{\rm O} \sin\hat{r}_{+} = a_{\rm I} \sin\hat{r}_{-} \,,\label{eq:Rhata0aI}
\end{equation}
 and the matching conditions in Eq.~\eqref{eq:R'HH} become 
\begin{equation}
\frac{\hat{R}'_{\pm}}{L} = \cos\hat{r}_{\pm} = \frac{1}{2 \kappa}(H_{\rm O}^{2} - H_{\rm I}^{2} \mp \kappa^{2}) \hat{R} \equiv c_{\pm} \hat{R} \,.\label{eq:Rhatprime}
\end{equation}
We also need the useful relations 
\begin{equation}
\label{eq:R0Rhat}
\begin{split}
R_{\rm o}^{-2} & = c_{-}^{2} + H_{\rm I}^{2} = c_{+}^{2} + H_{\rm O}^{2} \,,\\
1 - H_{\rm I,O}^{2} \, a_{\rm I,O}^{2} & = \frac{1 - \hat{R}^{2}/R_{\rm o}^{2}}{1 - c_{-,+}^{2} \hat R^{2}} \,.
\end{split}
\end{equation}
Note that the last relation implies that at the turning point $\hat{R}= R_{\rm o}$ we have $a_{\rm O,I} = H_{\rm O,I}^{-1}$.

As usual, the total classical action can be written as in Eq.~\eqref{eq:Scl}, i.e. $I_{\rm tot} = I_{\rm B} + I_{\rm b}$, where each term can be decomposed as
\begin{equation}
\label{eq:IDecomposition}
I_{\rm B} = I_{\rm B}^- + I_{\rm B}^+ \,, \qquad \text{and} \qquad I_{\rm b} = I_{\rm b}^- - I_{\rm b}^+ \,,
\end{equation}
where the $+/-$ subscripts refer to the outside and inside regions respectively. The general solution to the WDW equation is given in Eq.~\eqref{eq:GeneralWF}, where the two signs of the exponentials correspond to the choice of $\eta$ in Eq.~\eqref{eq:SclassicalBulk} and Eq.~\eqref{eq:SclassicalBoundary}. In this Subsection we consider the solution with $\eta = +1$, which is equivalent to setting $b = 0$ in Eq.~\eqref{eq:GeneralWF}. We will come back to this choice in Sec.~\ref{sec:SuperpositionWF}.

The boundary integral at arbitrary $\hat{R}$ is (with $c = c_{+}$, $H = H_{\rm O}$ or $c = c_{-}$, $H = H_{\rm I}$)
\begin{align}
I_{\rm b}^{\hat{R}} = \,\,& \frac{1}{G} \,\int^{\hat{R}}d\hat{R}\cos^{-1}\left(\frac{c\hat{R}}{(1 - H^{2}\hat{R}^{2})^{1/2}}\right) =\nonumber \\
 = \,\,& \frac{\hat{R}^{2}}{2 G}\cos^{-1}\left(\frac{c\hat{R}}{(1 - H^{2}\hat{R}^{2})^{1/2}}\right)+\nonumber \\
 & + \frac{1}{2 G H^{2}}\left[ {\rm \sin}^{-1}\left(\frac{c \hat{R}}{(1 - H^{2}\hat{R}^{2})^{1/2}}\right) - |c| R_{\rm o}\sin^{-1}\left(\frac{\epsilon(c)\hat{R}}{R_{\rm o}}\right)\right] \,. \label{eq:IhatW}
\end{align}
Note that $I_{\rm b}^{\hat{R}}\bigg|_{\rm tp} = (I_{\rm b}^{-} - I_{\rm b}^{+})\bigg|_{\hat{R} = R_{\rm o}} - (I_{\rm b}^{-} - I_{\rm b}^{+})\bigg|_{\hat{R} = 0}$, i.e. the boundary contribution at the turning point minus its background contribution (the latter is of course zero), is exactly the expression given in Eq.~\eqref{eq:SbHH2}.

The bulk contribution is given by the following integrals obtained
by using Eq.~\eqref{eq:a0aIR} in Eq.~\eqref{eq:SclassicalBulk}:
\begin{small}
\begin{align*}
I_{\rm B}^{-}(\hat{r}_{-}) & = \frac{1}{G} \, \int_{0}^{\hat{r}_{-}}dr\, a_{\rm I}\sin r \left[\sqrt{1 - H_{\rm I}^{2} a_{\rm I}^{2}} \, a_{\rm I} \sin r -a_{\rm I} \cos r \cos^{-1} \left(\frac{\cos r}{\sqrt{1 - H_{\rm I}^{2}a_{\rm I}^{2} \sin^{2} r}} \right) \right] \,, \\
I_{\rm B}^{+}(\hat{r}_{+}) & = \frac{1}{G} \, \int_{\hat{r}_{+}}^{\pi_{-}} dr\, a_{\rm O} \sin r \left[\sqrt{1 - H_{\rm O}^{2} a_{\rm O}^{2}} \, a_{\rm O} \sin r - a_{\rm O} \cos r \cos^{-1} \left(\frac{\cos r}{\sqrt{1 - H_{\rm O}^{2} a_{\rm O}^{2} \sin^{2} r}}\right) \right] \,.
\end{align*}
\end{small}
These integrals can be evaluated, giving  
\begin{align}
I_{\rm B}^{-}(\hat{r}_{-}) = \,& \frac{a_{\rm I}^{2}}{2G} \left[\sqrt{1 - H_{\rm I}^{2} a_{\rm I}^{2}}(\hat{r}_{-} - \sin\hat{r}_{-} \cos\hat{r}_{-}) + \cos^{2}\hat{r}_{-} \cos^{-1} \left(\frac{\cos\hat{r}_{-}}{\sqrt{1 - H_{\rm I}^{2} a_{\rm I}^{2} \sin^{2}\hat{r}_{-}}}\right)\right]\nonumber - \\
 & -\frac{(1 - H_{\rm I}^{2} a_{\rm I}^{2})}{2 G H_{\rm I}^{2}}\left[\sin^{-1}\left(\frac{\cos\hat{r}_{-}}{\sqrt{1 - H_{\rm I}^{2} a_{\rm I}^{2} \sin^{2}\hat{r}_{-}}}\right) - \frac{\pi}{2} \right] -\frac{\sqrt{1 - H_{\rm I}^{2} a_{\rm I}^{2}}}{2 G H_{\rm I}^{2}} \hat{r}_{-} \,,\label{eq:IB-}
\end{align}
and 
\begin{small}
\begin{align}
I_{\rm B}^{+}(\hat{r}_{+}) = \, & \frac{a_{\rm O}^{2}}{2G} \left[\sqrt{1 - H_{\rm O}^{2} a_{\rm O}^{2}}(\pi - \hat{r}_{+} + \sin\hat{r}_{+} \cos\hat{r}_{+}) + \pi -\cos^{2}\hat{r}_{+} \cos^{-1}\left(\frac{\cos\hat{r}_{+}}{\sqrt{1 - H_{\rm O}^{2} a_{\rm O}^{2}\sin^{2}\hat{r}_{+}}}\right)\right] +\nonumber \\
 & +\frac{(1 - H_{\rm O}^{2} a_{\rm O}^{2})}{2 G H_{\rm O}^{2}}\left[\sin^{-1}\left(\frac{\cos\hat{r}_{+}}{\sqrt{1 - H_{\rm O}^{2}a_{\rm O}^{2} \sin^{2}\hat{r}_{+}}}\right) + \frac{\pi}{2}\right] +\frac{\sqrt{1 - H_{\rm O}^{2} a_{\rm O}^{2}}}{2 G H_{\rm O}^{2}}(\hat{r}_{+} - \pi) \,.\label{eq:IB+}
\end{align}
\end{small}

The total action away from the turning point (but still under the barrier) is

\begin{small}
\begin{equation}
\begin{gathered}
\label{eq:B1i}
I_{\rm tot} = \, \frac{\pi}{4H_{\rm I}^{2}}\left[1-\epsilon(\hat R^{\prime}_{-}) \frac{2}{\pi} \left(\cos^{-1}\left(\frac{\hat R}{R_{\rm o}}\sqrt{1 - H_{\rm I}^{2} R_{\rm O}^{2}}\right)\right)\left(\frac{R_{\rm o}^{2} - \hat R^{2}}{R_{\rm o}^{2} - \hat R^{2}(1 - H_{\rm I}^{2} R_{\rm o}^{2})}\right)^{3/2}\right] - \\  
- \frac{\pi}{4 H_{\rm O}^{2}} \left[1 - \epsilon(\hat R^{\prime}_{+}) \left(2 + \frac{2}{\pi} \cos^{-1}\left(\frac{\hat R}{R_{\rm o}}\sqrt{1-H_{\rm O}^{2} R_{\rm o}^{2}}\right)\right)\left(\frac{R_{\rm o}^{2}-\hat R^{2}}{R_{\rm o}^{2} - \hat R^{2}(1 - H_{\rm O}^{2} R_{\rm o}^{2})}\right)^{3/2} \right] + \\ 
+ \frac{\hat R^{3}}{2 R_{\rm o}} \sqrt{R_{\rm o}^{2} - \hat R^{2}}  \left[\frac{(H_{\rm O}^{2} - H_{\rm I}^{2} + \kappa^{2})}{\sqrt{1 - c_{-}^{2}\hat R^{2}}} - \frac{(H_{\rm O}^{2} - H_{\rm I}^{2}-\kappa^{2})}{\sqrt{1 - c_{+}^{2}\hat R^{2}}}\right] - \\
- \left[\frac{H_{\rm O}^{2} - H_{\rm I}^{2} + \kappa^{2}}{4 \kappa H_{\rm I}^{2}} - \frac{H_{\rm O}^{2} - H_{\rm I}^{2} - \kappa^{2}}{4 \kappa H_{\rm O}^{2}}\right] R_{\rm o} \sin^{-1} \left(\frac{\hat R}{R_{\rm o}}\right) \,.
\end{gathered}
\end{equation}
\end{small}

The background action is obtained by setting $\hat{r}_{\pm}=0$ in the above expressions (corresponding to having the complete dS space with Hubble parameter $H_{\rm O}$), giving us
\begin{equation}
\overline{I} = -\frac{\pi}{2 G H_{\rm O}^{2}}\left[(1 - H_{\rm O}^{2} a_{\rm O}^{2})^{3/2} - 1 \right] \,,\label{eq:HHV}
\end{equation}
which gives the HH (under the barrier) wave function when
substituted into Eq.~\eqref{eq:GeneralWF} with $b=0$ and gives the Vilenkin version when $b=2ia$.

Let us first check the turning point bulk contribution.
This is easily obtained from Eq.~\eqref{eq:IB-} and Eq.~\eqref{eq:IB+} by putting $a_{\rm I} = H_{\rm I}^{-1}$, $a_{\rm O} = H_{\rm O}^{-1}$ and noting that $\cos\hat{r}_{\pm}/|\cos\hat{r}_{\pm}|=\epsilon(c_{\pm})=\epsilon(\hat{R}'_{\pm})$. We get
\begin{equation}
I_{\rm B}(\hat{r})\bigg|_{{\rm tp}} = \frac{\pi}{2G}R_{\rm o}^{2}(\theta(-R_{+}^{'}) - \theta(-R_{-}^{'})) + \frac{\pi}{2 G H_{\rm I}^{2}} \theta(-R'_{-}) + \frac{\pi}{2 G H_{\rm O}^{2}}(1 - \theta(-R'_{+})) \,,
\end{equation}
which is the same as Eq.~\eqref{eq:SBHH}. If we wish to compute the probability relative to that of having the initial dS space then as before we would subtract the background contribution in Eq.~\eqref{eq:bkgr} (that can be read also from Eq.~\eqref{eq:HHV} at $a_{\rm O} = H^{-1}_{\rm O}$) to get 
\begin{equation}
I_{\rm B}(\hat{r})\bigg|_{{\rm tp}} - \overline{I} = \frac{\pi}{2 G}R_{\rm o}^{2}(\theta(-R_{+}^{'}) - \theta(-R_{-}^{'})) + \frac{\pi}{2 G H_{\rm I}^{2}} \theta(-R'_{-}) - \frac{\pi}{2 G H_{\rm O}^{2}} \theta(-R'_{+}) \,.
\end{equation}
When added to Eq.~\eqref{eq:SbHH2} (still using $\eta = +1$) this is precisely the same as the amplitude corresponding to the BT/CDL expression, i.e. Eq.~\eqref{eq:iStr}. Accordingly the tunnelling probability is computed as in Eq.~\eqref{eq:ApproxProbability}.

\subsection{General solution}
\label{sec:SuperpositionWF}

The result obtained in the previous Section however depends on just picking one branch of the wave function in Eq.~\eqref{eq:Psi}, namely the $\eta = +1$ one, by setting $b = 0$ in Eq.~\eqref{eq:Psi}. In fact, it has been argued in~\cite{Bachlechner:2016mtp} that one should choose one or the other branch depending on the sign of the right hand side of Eq.~\eqref{eq:iStr}. However, there are several problems with this. Firstly, let us point out that there could be a problem with working at the turning point itself since the fluctuation determinant may be singular there (as is the case in the corresponding mini-superspace calculation of HH and Vilenkin~\cite{Hartle:1983ai, Vilenkin:1984wp} for the quantum creation of inflating universes from `nothing'). In fact the semi-classical calculation breaks down at the turning point: it is a leading asymptotic expression to the Airy function, valid only far away from the turning point. Secondly, the choice of $b = 0$ or $a = 0$ in Eq.~\eqref{eq:GeneralWF} corresponds to a particular choice of the boundary conditions.  

In particular, in the well-known HH no-boundary proposal for the wave function of the universe it is in fact the rising term which is kept - i.e. the solution given in Eq.~\eqref{eq:HHV} - whereas for the `tunnelling' proposal of Vilenkin one has a linear combination of the rising and falling solutions (i.e. $a = 2ib$ in Eq.~\eqref{eq:GeneralWF}) so that in the classical regime one only has outgoing (corresponding to an expanding universe) waves.

In order to elucidate this issue it is useful to analytically continue
the solutions $I_{\rm b}$, $I_{\rm B}^{\pm}$ to the `classical' region $\hat{R} > R_{\rm o}$ (so that $a_{\rm I} > H_{\rm I}$, $a_{\rm O} > H_{\rm O}$). We find\footnote{We use the formulae $\sin^{-1}z=-i\ln\left(iz+\sqrt{1-z^{2}}\right),\,\cos^{-1}z=-i\ln\left(z+\sqrt{1-z^{2}}\right)$ to continue the functions to the whole complex plane. We have in particular $\sin^{-1}\left(\pm ix\right)=-i\ln(\mp x+\sqrt{1+x^{2}})$, which is pure imaginary and $\cos^{-1}\left(\pm ix\right)=\frac{\pi}{2}-i\ln(\pm x+\sqrt{1+x^{2}})$ where the second term is pure imaginary. We also have for $|x|>1$, $\sin^{-1}x=\epsilon(x)\frac{\pi}{2}-i\ln\left(x+\sqrt{x^{2}-1}\right)$ where the second term is pure imaginary, and $\cos^{-1}x=\theta(-x)\pi-i\ln\left(|x|-\epsilon(x)\sqrt{x^{2}-1}\right)$ where again the second term is pure imaginary.} for the real part of the action
\begin{align}
\text{Re}\left(I_{\rm tot} (\hat{R}) \right) & = \frac{\pi}{2 G}\left(\frac{\theta(- R'_{-})}{H_{\rm I}^{2}} + \frac{1 - \theta(-R'_{+})}{H_{\rm O}^{2}}\right) + \nonumber \\
 & +\frac{\pi}{4H_{\rm I}^{2}} \left(-R_{\rm o}| c_{-}| + 1\right)\epsilon(R'_{-}) - \label{eq:realIRhat} \\
 & -\frac{\pi}{4H_{\rm O}^{2}} \left(-R_{\rm o}| c_{+}| + 1\right)\epsilon(R'_{+}) \nonumber \,.
\end{align}
Here the first line comes from the bulk $(I_{\rm B}^{-} + I_{\rm B}^{+})$ and the last two lines from the boundary $(I_{\rm b}^{-} - I_{\rm b}^{+}$). For future reference let us rewrite this as
\begin{equation}
\text{Re}\left(I(\hat{R})\right) = \frac{\pi}{2G}\left(\frac{1}{2H_{\rm I}^{2}} + \frac{1}{2H_{\rm O}^{2}}\right) + \frac{\pi}{4}R_{\rm o}\left(\frac{c_{+}}{H_{\rm O}^{2}} - \frac{c_{-}}{H_{\rm I}^{2}}\right) \,.\label{eq:realIRhat2}
\end{equation}
We note here that this expression is symmetric between the `outside' and `inside', (since this labeling is a matter of convention), as observed before for the turning point expression in Eq.~\eqref{eq:HH}, and in fact the two are the same. 

As expected these are exactly the values of the under the barrier actions at the turning point but now there is an imaginary contribution which gives in the wave function, an oscillatory contribution corresponding to an expanding or contracting geometry depending on the sign of $\eta$. 

However, the usual tunnelling calculation assumes that the coefficient $a$  in Eq.~\eqref{eq:GeneralWF} is zero, but this choice is the opposite of what gives the HH wave function, which as we pointed out before is the rising (as a function of $R$) contribution. So if we are to follow the logic of the arguments for either the HH or Vilenkin wave functions, we should at least {\it ab initio} keep both terms in the wave function.

The above discussion implies that we have for the inside the barrier (hence the subscript `ins') wave function
\begin{equation}
\Psi^{{\rm ins}}_{{\cal N}}(\hat{R}) = a \, e^{I_{{\rm ins}}} + b \, e^{-I_{{\rm ins}}} \,,\label{eq:Psiin}
\end{equation}
 where $I_{{\rm ins}}$ is given by Eq.~\eqref{eq:B1i} evaluated at $\hat{R} < R_{\rm o}$ i.e. $a_I<H_I,~a_{\rm O}<H_{\rm O}$. This wave function is to be compared with the wave function for the background $\Psi^{{\rm ins}}_{{\cal B}}$ which is given by the same expression with the action being given now by Eq.~\eqref{eq:HHV}. 

Actually the situation is more complicated than that since as we argued
before one should really evaluate these semi-classical wave functions
away from the turning points, and in any case one should compare the
probabilities for the emergence in the classical region, where one
would have an expanding/contracting universe(s). However in analogy
with the mini-superspace case, where there is an exact solution -
namely the Airy function - one expects that the wave function in the
classsical region to be a linear combination of the under the barrier
wave functions. In other words one would have (we denote the analytically continued action `outside' the barrier by $I_{{\rm out}}$, and analogously for the relative probability $\mathcal{P}_{\rm out}$ and the wave function $\Psi^{\rm out}_{\mathcal{B}/\mathcal{N}}$) 
\begin{equation}
\Psi_{\mathcal{N}}^{\rm out}(\hat{R})=(a+ib/2)e^{\left[\text{Re}\left(I_{{\rm out}}(\hat{R})\right)+ i \text{Im} \left(I_{{\rm out}}(\hat{R})\right)\right]} + (a-ib/2) e^{-\left[\text{Re}\left(I_{{\rm out}}(\hat{R})\right) + i \text{Im} \left(I_{{\rm out}}(\hat{R})\right)\right]} \,,
\end{equation}
with  the same combination for the wave function for the
background\footnote{Recall that  in the quantum mechanics problem the wave function near the turning point can be approximated by the Airy function since the potential locally can be approximated by a linear function.}. Note that in the latter case one has an exact solution
(it is of course the usual HH/Vilenkin construction). 
Then the relative probability is 
\begin{align}
\label{eq:Pout}
& \mathcal{P}_{{\rm {\rm out}}}(\mathcal{B} \rightarrow \mathcal{N}) = \frac{|\Psi^{{\rm out}}_{{\cal N}}|^{2}}{|\Psi^{{\rm out}}_{{\cal B}}|^{2}} =  \\
& = \frac{|a+i\frac{b}{2}|^{2} e^{2 \text{Re}\left(I_{\rm out}(\hat{R})\right)} + |a-i\frac{b}{2}|^{2}e^{-2 \text{Re}\left(I_{{\rm out}}(\hat{R})\right)}+2\text{Re}\left[(a+i\frac{b}{2})(a^{*}-i\frac{b^{*}}{2})\right]e^{2i\text{Im} \left(I_{\rm out}(\hat{R})\right)}}{|a+i\frac{b}{2}|^{2}e^{2 \text{Re}\left(I_{{\rm out}}(0)\right)} + |a-i\frac{b}{2}|^{2} e^{-2 \text{Re}\left(I_{{\rm out}}(0)\right)} + 2 \text{Re}\left((a+i\frac{b}{2})(a^{*}-i\frac{b^{*}}{2})\right) e^{2i \text{Im} \left(I_{{\rm out}}(0)\right)}} \,.\nonumber
\end{align}
Now we observe that $\text{Re}\left(I_{{\rm out}}(\hat{R})\right)$ is positive as is easily seen from Eq.~\eqref{eq:HH} or Eq.~\eqref{eq:realIRhat2}. Hence the choice of $\eta = 1$ that we operated in the previous Sections is justified: the first term in the numerator dominates and one gets the same result as in the Euclidean argument of BT/CDL and corresponds to a generalization of the HH argument for tunnelling from `nothing'. On the other hand if we insist on an expanding configuration in the classical region (as opposed to a superposition of an expanding and a contracting geometry) one must set $a + ib/2 = 0$. This would then give a result that is different from the Euclidean calculation but would be a generalised version of Vilenkin's tunnelling from `nothing' wave function. 

\subsection{Detailed balance}

Let us now revisit the question of detailed balance. Let us take two
dS spaces $A$ and $B$ with Hubble parameters such that $H_A < H_B$. Then the relative probabilty for being in the configuaration $A/B \oplus \text{W}$ versus being in the initial state $A$ is 
\begin{equation}
\mathcal{P}(A \rightarrow A/B \oplus \text{W}) = \frac{|\Psi(A/B \oplus \text{W})|^{2}}{|\Psi(A)|^{2}}\,.\label{eq:PsiABW}
\end{equation}
Here the numerator is given by the numerator of Eq.~\eqref{eq:Pout} with $\text{Re}\left(I_{{\rm out}}\right)$ given by Eq.~\eqref{eq:realIRhat2} with $\text{O} = A$ and $\text{I} = B$ i.e. 
\begin{align}
\text{Re}\left(I_{{\rm out}}(\hat{R})\right) = \,\, & \text{Re}\left(I_{{\rm out}}(A/B \oplus \text{W}) \right) = \nonumber \\
= \,\, & \frac{\pi}{2G} \left(\frac{1}{2 H_B^{2}} + \frac{1}{2 H_A^{2}}\right) + \frac{\pi}{4}R_{\rm o}\left(\frac{c_{+}}{H_{A}^{2}} - \frac{c_{-}}{H_{B}^{2}}\right)= \text{Re}\left(I(B/A \oplus \text{W})\right) \,,\label{eq:IABW}
\end{align}
where the last equality follows from the symmetry property that we
observed after Eq.~\eqref{eq:realIRhat2}. The denominator is that of
Eq.~\eqref{eq:Pout} with 
\begin{equation}
\text{Re}\left(I_{{\rm out}}(0) \right) = \frac{\pi}{2 G H_{A}^{2}} \,.
\end{equation}
On the other hand, interchanging $A$ and $B$ we have 
\begin{equation}
\mathcal{P}(B \rightarrow B/A \oplus \text{W}) = \frac{|\Psi(B/A \oplus \text{W})|^{2}}{|\Psi(B)|^{2}} =\frac{|\Psi(A/B \oplus \text{W})|^{2}}{|\Psi(B)|^{2}} \,,\label{eq:PsiBAW}
\end{equation}
where the second equality is a consequence of the last relation in
Eq.~\eqref{eq:IABW}. Since the entropy of a dS space with Hubble parameter $H$ is given by $S = \frac{\pi}{G H^{2}}$ we have 
\begin{equation}
\frac{\mathcal{P}(A \rightarrow A/B \oplus \text{W})}{\mathcal{P}(B \rightarrow B/A \oplus \text{W})} = \frac{|\Psi(B)|^{2}}{|\Psi(A)|^{2}} \simeq \frac{e^{S_{B}}}{e^{S_{A}}} \,.\label{eq:SASB}
\end{equation}
Note that the last relation is valid only as long as the coefficient
$a+ib/2$ is not set to zero so that this term dominates the denominator of Eq.~\eqref{eq:Pout}. Thus our Hamiltonian argument is in agreement with detailed balance as was the case with the Euclidean
instanton argument of CDL/BT, see Eq.~\eqref{eq:iStr}. It is important
to note though that here the detailed balance agreement is a simple
consequence of the fact that wave function for the configuration $A/B \oplus \text{W}$ cancelled in the ratio on the left hand side of the Eq.~\eqref{eq:SASB}. Thus the ratio of up to down-tunnelling is simply given by the relative probability for being in the state $B$  with large cosmological constant (and hence smaller entropy) to that for being in the state $A$ with smaller cosmological constant and hence larger entropy, and is therefore necessarily smaller than one.

On the other hand if we had used the Vilenkin/tunnelling wave function (i.e. put $a+ib/2=0$), we would not have satisfied detailed balance since the ratio on the right hand side of Eq.~\eqref{eq:SASB} would be replaced by its inverse. However one should not expect detailed balance in that case since one has only an outgoing wave (corresponding to expansion of the universe) which is not an equilibrium situation - while the HH case with essentially a real wave function corresponds to a superposition of expanding and contracting universes - essentially a standing wave.

\subsection{The flat space limit}

Let us now reconsider the flat space limit in particular the up-tunnelling transitions Minkowski to dS space. As we discussed before the probablity of up-tunnelling goes to zero simply because the entropy associated with dS space goes to infinity as $H \rightarrow 0$. On the other hand, the original calculation of FGG/FMP gave a non-zero probability for up-tunnelling to dS when the mass $M$ of the black hole cofiguration went to zero resulting in flat space. This is of course consistent with the interpretation of black hole entropy as being given by the horizon area in Planck units of the so that in the $M\rightarrow 0$ limit we get flat space with zero entropy. In effect we are assigning
two different wave functions to flat space depending on which limit
is being taken. Starting from dS, in the flat limit $H\rightarrow 0$ we had
\begin{equation}
| \Psi_{\rm dS}|^{2}\sim e^{\frac{\pi}{G H^2}} \rightarrow \infty \,,
\end{equation}
whereas starting from the black hole, in the flat space limit $R_{\mathcal{S}} = 2GM \rightarrow 0$ we get 
\begin{equation}
| \Psi_{{\mathcal{S}}}|^{2} \sim e^{\frac{\pi R_{\mathcal{S}}^{2}}{G}} \rightarrow 1 \,.
\end{equation}

\section{Comparison to other approaches}
\label{sec:Comparison}

\subsection{Hamiltonian description and disconnected spacetimes}

In the previous Sections we followed the general formalism developed in FMP and  extended their results in several directions\footnote{For an alternative approach see~\cite{Bachlechner:2018pqk, Bachlechner:2018jmq, Bachlechner:2018rht, UnpublishedBachlechner}.}. In~\cite{Bachlechner:2016mtp} Bachlechner addressed the tunnelling processes by extending the Hamiltonian formalism outlined by FMP by adding disconnected spacetimes. In effect this is tantamount to extending the integration in the two equations after Eq.~\eqref{eq:IhatW} from $0$ to $-N_{\rm I}\pi$ in the first integral and from $\pi$ to $N_{\rm O}\pi$ in the second integral. One can of course do the same with the background integrals as well. 

For the case of dS to dS transitions, the total action generalises Eq.~\eqref{eq:iStr} to
\begin{equation} 
\small{I_{\rm tot} = \frac{\eta \pi}{G} \left[-\frac{(H_{\rm O}^{2} -H_{\rm I}^{2})^{2} + \kappa^{2}(H_{\rm O}^{2} + H_{\rm I}^{2})}{8 \kappa H_{\rm O}^2 H_{\rm I}^2} R_{\rm o} + \frac{(N_{\text{O}, \mathcal{N}} - N_{\text{O}, \mathcal{B}}) - \frac{1}{2}}{2 H_{\rm O}^{2}} - \frac{(N_{\text{I}, \mathcal{N}} - N_{\text{I}, \mathcal{B}}) - \frac{1}{2}}{2 H_{\rm I}^{2}}\right]} \,.
\end{equation} 
In the flat spacetime limit $H_{\rm O} \rightarrow 0$ the transition Minkowski to dS is given by\footnote{Observe that there is a minor difference with the corresponding expression found in~\cite{Bachlechner:2016mtp}. This is due to the way we take background subtraction and the setting of $M_{\rm O}=0$ (implying $R_{\rm{O},1} = 0$ in Eq.~\eqref{eq:BulkActionGeneric}) in the $\mathcal{S}$dS to $\mathcal{S}$dS transition, before the evaluation of the bulk integral. However there is no difference on the general point that there are values of the $N_{\rm{O}, f}, N_{\rm{O},i}$ integers for which the amplitudes do not vanish.}
\be
\mathcal{P}(\mathcal{B} \rightarrow \mathcal{N}) = \exp\left[-\frac{\eta \pi}{G}\left(\frac{N_{\text{O}, \mathcal{B}} - N_{\text{O}, \mathcal{N}} +1 }{H_{\rm O}^2} - \frac{N_{\text{I}, \mathcal{B}} - N_{\text{I}, \mathcal{N}} + 1}{H_{\rm I}^2} + \frac{k^4}{H_{\rm I}^2\left(H_{\rm I}^2 + k^2\right)^2}\right)\right] \,,\label{disconnected}
\ee
which gives a finite non-vanishing amplitude as long as $N_{\text{O}, \mathcal{B}} - N_{\text{O}, \mathcal{N}} + 1 = 0$. The simplest case being $N_{\text{O}, \mathcal{B}} = 0, N_{\text{O}, \mathcal{N}} = 1$ (and $N_{\text{I}, \mathcal{B}} = N_{\text{I}, \mathcal{N}} = 0$). In this case we recover exactly the amplitude computed from the zero mass limit of Schwarzschild without introducing any disconnected spaces. Notice that as expected if the number of disconnected components is the same in the initial and final configurations their role is irrelevant.

As FMP observed, the consideration of an arbitrary number of disconnected parts may be problematic. For a fixed background, for instance $N_{\rm O, \mathcal{B}}=N_{\rm I, \mathcal{B}}=0$, the transition amplitude in the $H_{\rm O} \rightarrow 0$ limit goes like:
\be
\mathcal{P}(\mathcal{B} \rightarrow \mathcal{N}) = \exp\left[\frac{\eta\pi}{G H_{\rm O}^2}\left(N_{\rm O, \mathcal{N}}-1\right)\right] \,.
\ee
For $\eta=+1$ this vanishes for $N_{\text{O}, \mathcal{N}}=0$, gives the black hole finite result for $N_{\text{O},\mathcal{N}} = 1$ and blows-up for $N_{\text{O}, \mathcal{N}} > 1$. Since the total wave function should be a sum of all the sectors (using $N_{\text{O}, \mathcal{N}} = N$ for simplicity of notation):
\be
\Psi=\sum_{N} C_N \Psi_N \,,
\ee
it would naturally blow-up unless the coefficients $C_N = 0$ for $N > 1$. More generally, this would be needed for the $\eta = + 1$ case with opposite results for the $\eta = -1$ case. In both cases $N = N_{\text{O}, \mathcal{N}} = 1$ is singled out to give the full answer.
This potential blow-up was considered as a possible instability of HH in FMP. The problem is that there is no way to limit these disconnected sums and including them will simply make the whole formalism ambiguous. Hence we simply set the coefficients of such additional contributions to the wave function to zero as being physically meaningless. We are only interested in wave functions that can be interpreted as transitions mediated by our wall/brane and that means we just keep the connected terms. A better understanding of this situation would be desirable.

\subsection{Euclidean approaches}
Let us now compare our results with the standard Euclidean approach for tunnelling, starting from the original CDL and BT and then with the FGG approach for the Minkowski to dS transition.

\subsubsection*{BT/CDL}

The original treatments on vacuum transitions were done following the standard instanton techniques which are formulated in Euclidean space. Both CDL and BT formalisms are Euclidean. Up-tunnelling dS to dS transitions are forbidden in CDL but not in BT (and also Lee-Weinberg~\cite{Lee:1987qc}) but Minkowski to dS transitions are forbidden in both CDL and BT. Let us investigate the difference. CDL and BT give the following expression for the transition probability: $\mathcal{P} = e^{-|2 I_{\rm BT}|}$, where
\begin{equation}
\small{
I_{{\rm BT}} (\hat{R}) =\frac{\pi}{4G}\left[
\left(\frac{\epsilon(\hat{R}'_{-})}{H_{\rm I}^{2}}\left(1-H_{{\rm
I}}^{2}\hat{R}^{2}\right)^{3/2}-H_{{\rm
I}}^{-2}\right)-\left(\frac{\epsilon(\hat{R}'_{+})}{H_{{\rm
O}}^{2}}\left(1-H_{{\rm O}}^{2}\hat{R}^{2}\right)^{3/2}-H_{{\rm {\rm
O}}}^{-2}\right)+\kappa\hat{R}^{3}\right] \,. }\label{eq:BCdL}
\end{equation}
This expression is extremised (so that the probability is maximised)
at $\hat{R} = R_{\rm o}$ with the latter given by Eq.~\eqref{eq:RoHH}. If we substitute this into Eq.~\eqref{eq:BCdL} we get
\begin{equation}
I_{{\rm BT}} (\hat{R} = R_{\rm o}) = \frac{\pi}{2 G} \left[\frac{\left[(H_{{\rm O}}^{2}-H_{{\rm I}}^{2})^{2}+\kappa^{2}(H_{{\rm O}}^{2}+H_{{\rm I}}^{2})\right] R_{{\rm o}}}{4\kappa H_{{\rm O}}^{2}H_{{\rm I}}^{2}}-\frac{1}{2}\left(H_{{\rm I}}^{-2}-H_{{\rm O}}^{-2}\right)\right] \,, \label{eq:BCdL2}
\end{equation}
which is exactly the same result as in Eq.~\eqref{eq:iStr}, obtained using the Hamiltonian approach for dS to dS transitions. When setting the initial Hubble parameter to zero, the transition probability vanishes. As mentioned in the previous Subsection this result  agrees with the Hamiltonian approach in the absence of spacetimes disconnected to the wall. However, this does not prevent up-transitions: the argument for this is that FMP and FGG, instead, focused on another transition, i.e. Schwarzschild to dS and subsequently they took $M \rightarrow 0$. \footnote{Note that, in this case, the bulk action would have a term like $2 G M \theta(-\hat R^{\prime}_{+}) \rightarrow 0$ in the flat spacetime limit.} The main difference  between the two approaches is that, taking the Minkowski limit in the latter implies the vanishing of the term proportional to the black hole mass parameter; thereby, the total action still remains finite, leading to a nontrivial transition probability.

Notice that away from the turning points our general expression \eqref{eq:B1i} does not coincide with the BT expression \eqref{eq:BCdL}. However both equations are such that they reproduce the same expression \eqref{eq:BCdL2} upon minimisation and evaluation at the turning points. This is the relevant comparison.

An important non-trivial check of the validity of the Schwarzschild to dS calculation for up-tunnelling is that it exactly reproduces the CDL result when applied to down-tunnelling as we explicitly derived in the previous Section. As already discussed, the fact that the two limits $H_{\rm O}$, $M \rightarrow 0$ lead to different results is a consequence of the fact that though both background actions may be interpreted as giving the relevant entropies in the black hole limit we get zero entropy corresponding to the entropy of the vacuum state of Minkowski (interpreted as the log of the dimension of the non-degererate vacuum state) while the $H\rightarrow 0$ limit of dS gives the entropy (i.e. log of the dimension) of the entire HIlbert space that can be built on Minkowski space. We will discuss this further in the next Section.

\subsubsection*{FGG}

The original study of the Minkowski to dS transition was performed in~\cite{Farhi:1989yr} using the Euclidean formalism. A very detailed study was made of the transition probability and it was found that  the corresponding instantons are singular. Concretely the instantons correspond to Euclidean manifolds over degenerate metrics. This has cast doubt on the validity of the transition.

Here we want to emphasise that even though the validity of the use of these degenerate metrics can be questioned, one of the merits of the subsequent work of FMP was to put these results on firmer ground  by using the Hamiltonian approach, in which case there is no need to introduce degenerate metrics.

In summary, the advantages of Hamiltonian over Euclidean are 
\begin{itemize}
\item{} Both results agree but in FMP there is no need to introduce singular geometries;
\item{} Some explicit terms in the action are derived in FMP but introduced by hand in FGG;
\item{} The spacetime trajectory of the wall can be properly described in a causal diagram;
\item{} Unitarity of the process is built-in within the formalism.
\end{itemize}

However FGG, aware of the limitations of the Euclidean approach were able to add the right ingredients to obtain a non-zero amplitude and the fact that their result agrees with the Hamiltonian approach makes their assumptions more robust. We conclude that even though the original Euclidean approach for up-tunnelling from Minkowski spacetime is subject to criticism, the fact that the subsequent Hamiltonian approach gives the same results provide  strong evidence for the validity of this approach.

\subsection{Thermal/Tunnelling Approach} 

A further argument  questioning the validity of up-tunnelling from Minkowski space goes as follows\footnote{We thank Alan Guth for a discussion of this point. See for instance~\cite{GuthSusskind}.}. Starting from dS to dS, two expressions for the amplitude can be estimated in the Minkowski limit of one dS ($H_{\rm O} \rightarrow 0$). The first expression assumes detailed balance:
\begin{equation} 
\Gamma^{(1)}_{\rm up} = \Gamma_{\rm down} \exp\left[\frac{\pi}{G} \left(\frac{1}{H_{\rm I}^{2}} - \frac{1}{H_{\rm O}^{2}}\right)\right] = \Gamma_{\rm CDL} \exp\left(S_{\rm I} - S_{\rm O}\right)\,,
\label{GuthOne}\end{equation} 
where $\Gamma_{\rm up}$ is the transition rate for up tunnelling, $\Gamma_{\rm down}$ is the well known standard CDL amplitude which does not vanish in the limit $H_{\rm O} \rightarrow 0$ and $S_{\rm I}$, $S_{\rm O}$ are the two dS entropies. Of course this relation is the same as our Eq.~\eqref{eq:SASB} with 
\be
\frac{\Gamma^{(1)}_{\rm{up}}}{\Gamma_{\rm{CDL}}}=\frac{|\Psi({\rm I})|^2}{|\Psi({\rm O})|^2}=e^{S_{\rm I}-S_{\rm O}}.
\ee
The second expression assumes a non-vanishing up-tunnelling amplitude $\Gamma_{\rm tunn}$ and the $H_{\rm O}$ dependence of the total amplitude could come from a thermal amplitude for bubble nucleation given by the Boltzmann factor $e^{-E/T}$, where $T = H_{\rm O}/2\pi$ is the Gibbons-Hawking temperature and $E\propto \kappa$ the bubble energy\footnote{Note that the thermal component is symmetric under the exchange  $H_{\rm I}\leftrightarrow H_{\rm O}$ and could be used in the standard true vacuum bubble case as a thermal factor multiplying $\Gamma_{\rm CDL}$ (for a discussion of thermal effects in vacuum transition see for instance~\cite{Brown:2007sd}).}

\begin{equation} 
\Gamma^{(2)}_{\rm up} \propto \Gamma_{\rm tunn} \exp\left[-\frac{2\pi \mathcal{O}(\kappa)}{H_{\rm O}}\right] \,,
\label{GuthTwo}
\end{equation} 
where $\Gamma_{\rm tunn}$ is the transition rate for up-tunnelling in dS $\rightarrow$ dS. 

These two expressions should really describe the same process. However, when taking the limit $H_{\rm O} \rightarrow 0$, the latter vanishes much more slowly if $\Gamma_{\rm tunn}$ is assumed (as in~\cite{GuthSusskind}) to be slowly varying with $H_{\rm O}$. At this point~\cite{GuthSusskind} suggest two possible solutions to the paradox. Either $\Gamma_{\rm tunn} = 0$, or the process does not carry off any information from the original universe. Indeed, the latter would allow the evolution of the original universe to proceed unitarily with no information loss. Our findings show that in fact the first expression is satisfied but in the second expression $\Gamma_{\rm tunn} \sim e^{-1/H_{\rm O}^2} \rightarrow 0$ so the two expressions vanish at the same rate. Consistent with the first proposed solution and no Minkowski to dS transition.

Let us try to repeat the analysis of~\cite{GuthSusskind} for the Schwarzschild to dS case.

In this case \eqref{GuthOne} is replaced by
\begin{equation}
\Gamma^{(1)}_{\rm up} = \Gamma_{\rm CDL}\,  e^{S_{\rm dS} - S_{\mathcal{S}}} \,,\label{guth3}
\end{equation}
with $S_{\rm dS} \sim 1/H^2$ the de Sitter entropy and $S_{\mathcal{S}} =\pi R^2_{\mathcal S}/G= 4\pi GM^2 \rightarrow 0$ the black hole entropy. On the other hand the thermal argument for this gives (using the black hole temperature $T = 1/8\pi G M$),
\begin{equation} 
\Gamma^{(2)}_{\rm up} \sim \Gamma_{\rm tunn}\, e^{-E/T} \sim \Gamma_{\rm tunn}\, e^{-\mathcal{O}(\kappa )8\pi GM}  \,.\label{guth4}
\end{equation}
But now the tunnelling amplitude up to $\mathcal{O}(\kappa M)$ bubble terms is (see last line of Eq.~\eqref{eq:SB3} with $\eta=+1$), 
\begin{align}  
\Gamma_{\rm tunn} & \sim \exp\left[\frac{\pi}{G} \left(\frac{1}{H^{2}} -R^2_{\mathcal S} -\frac{\kappa^{4}}{H^{2}(H^{2}+\kappa^{2})^{2}} \right)\right] \,, \nonumber \\ 
\Gamma_{\rm CDL} & \sim \exp\left[-\frac{\pi}{G} \left(\frac{ \kappa^{4}}{H^{2}(H^{2}+\kappa^{2})^{2}}\right)\right] \,.
\label{eq:2e} 
\end{align}
From this follows that the two expressions for $\Gamma_{\rm up}$ \eqref{guth3} and \eqref{guth4} are the same up to terms which vanish like $M\kappa$. We then conclude that there is no contradiction in having a Minkowski to dS transition from this thermal argument. 

\subsection{AdS/CFT}

Using AdS/CFT arguments Freivogel {\it et al.}~\cite{Freivogel:2005qh} concluded that the FGG mechanism was not achievable. The argument goes as follows. As we have seen, classically Farhi and Guven had proven that it is not possible to create an inflating dS universe from non-singular initial conditions, however in FGG/FMP a non-inflationary buildable state could be created classically from non-singular initial conditions and then quantum mechanically tunnel to a dS state\footnote{Buildable and unbuildable solutions have already been defined in Sec.~\ref{sec:Introduction}.}. \cite{Freivogel:2005qh} claimed that since the original buildable state comes from a unitary evolution it is a pure state but using AdS/CFT they concluded that any inflating state has to be a mixed state and therefore it could not come from a pure state, similar to the black hole information loss paradox.  

The key point of their argument, as stressed in~\cite{Freivogel:2005qh}, lies in the definition of the mixed boundary state, and on their inevitable emergence whenever dealing with causally-disconnected spacetimes. To see this, explicitly, let us consider, at first, a pure Schwarzschild\\/AdS spacetime, having two boundaries. Placing an observer on the right hand boundary, e.g., is equivalent to tracing over the left hand side of the wedge provided by the black hole event horizon. Therefore, any pure boundary state for the initial CFT would inevitably turn into a mixed state after tracing over one of the two boundaries. 

\begin{figure}[h!]  
\begin{center} 
\includegraphics[scale=0.3]{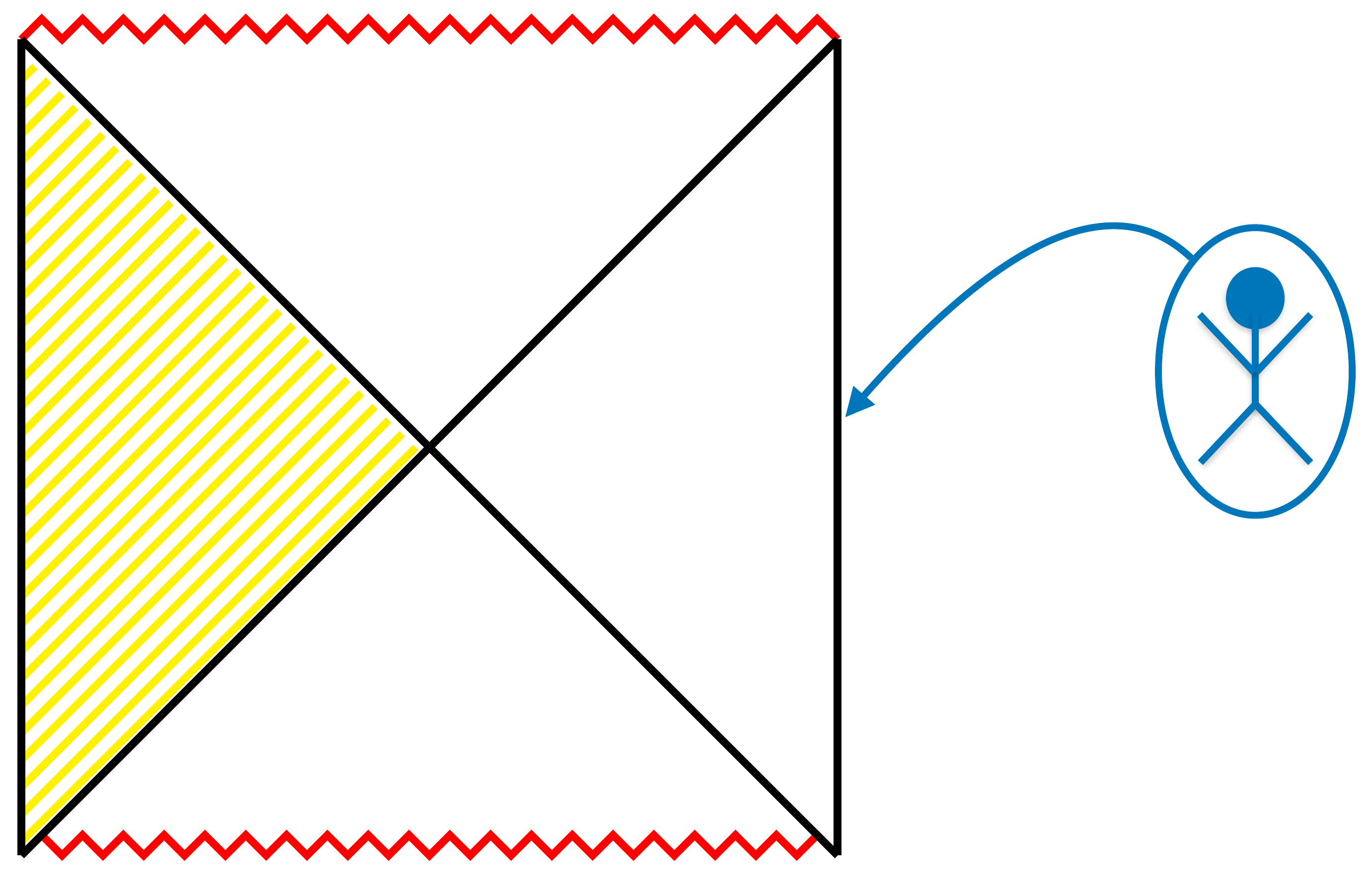} 
\caption{\footnotesize{Tracing over the left hand boundary, the boundary state on the observer's side becomes mixed.}} 
\end{center} 
\end{figure} 

Freivogel \emph{et al.} claim that the FGG process is very similar to the aforementioned setup, once having replaced the left hand side of the wedge with a portion of dS spacetime\footnote{Examples of patched up metrics are provided in the introduction.}. At this point, they argue that, the black hole separating the observer from the false vacuum has fewer degrees of freedom with respect to the ones defining the dS. By doing this an entropy puzzle is identified in the sense that the radius of the bubble is smaller than the corresponding dS radius and bigger than the black hole radius, implying that the black hole entropy must necessarily be smaller than the dS entropy (see also~\cite{Banks:2002nm}). The puzzle is how can a pure state coming up from few black hole degrees of freedom give rise to the large number of degrees of freedom needed to define the dS geometry. From this they conclude that the corresponding state living on the AdS boundary must in turn be a mixed one. 

However, we present a solution to this issue, relying upon arguments outlined in~\cite{Marolf:2008tx}. Patched-up metrics, also known as \emph{bag of gold} metrics, are characterised by having a unique boundary, which, for the case of interest, corresponds to the one on the right of the wedge, i.e. where the observer is assumed to be placed. Being there no asymptotic region on the left hand side of the causal diagram, no tracing has to be performed. The dS vacuum is still hidden behind the black hole event horizon, therefore, all the information that can reach the observer on the AdS boundary is provided by the Bekenstein-Hawking entropy, $S_{\mathcal{S}}$ , which, as also stated in~\cite{Marolf:2008tx}, is not associated  to the counting of all of the degrees of freedom of the black hole, but, rather, just to the ones that are accessible to the external observer. 

Once having clarified this new setup, it is straightforward to claim that the new dS phase can indeed be described by a pure state in the boundary CFT, thereby enabling the tunnelling in between the buildable and unbuildable trajectories. 

\begin{figure}[h!] 
\begin{center} 
\includegraphics[scale=0.7, trim = 0cm 1cm 0cm 0cm, clip]{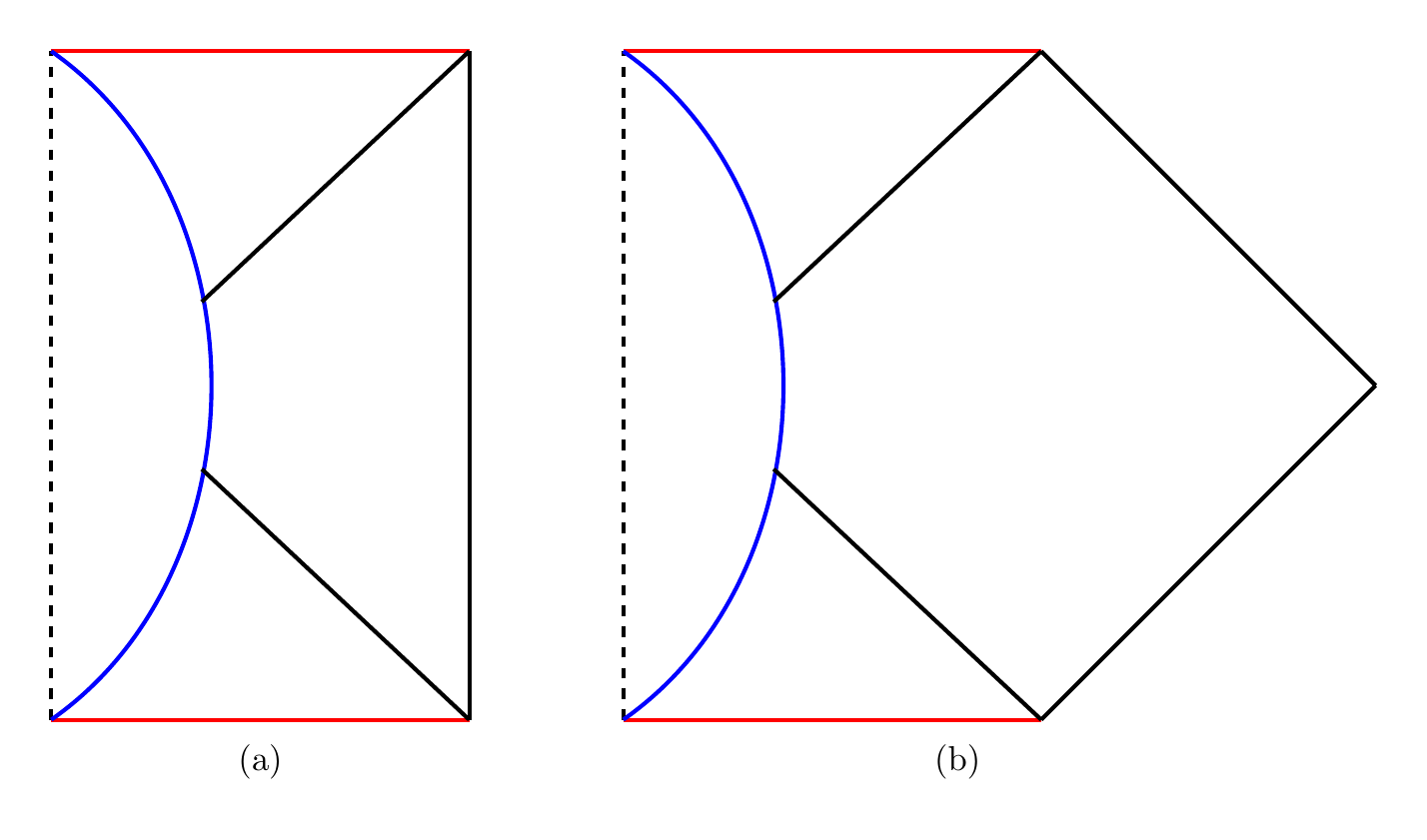} 
\caption{\footnotesize{dS/Schwarzschild-AdS and dS/Schwarzschild patched-up metrics, used for the Ryu-Takayanagi and FGG arguments, respectively.} }
\end{center} 
\end{figure}  

Developments after the appearance of~\cite{Freivogel:2005qh} suggest that the inflating geometries may actually correspond to pure states. The new ingredient is the Ryu-Takayanagi~\cite{Ryu:2006bv} definition of entanglement entropy. If we want to compute the entanglement entropy in a region $B$ inside the CFT geometry we construct a surface $\gamma$ of minimal area in the bulk such that the boundary of $\gamma$ coincides with the boundary of $B$ and the entanglement entropy is:
\be
S_{\rm HEE}=\frac{A_\gamma}{4G}
\ee 
where $A_\gamma$ is the area of the minimal surface $\gamma$.  In the FGG/FMP case the geometry is such that on a confomal Penrose diagram the left boundary corresponds to the point $r=0$ of dS space and the right boundary is one for AdS. Since $r=0$ on the left, the surface $\gamma$ has its boundary on the AdS side and is contractible, i.e. the corresponding entropy vanishes implying a pure state. Therefore this criticism of FGG/FMP does not hold since if the Ryu-Takayanagi argument holds, now we have a transition between two pure states which does not contradict any principle\footnote{We thank Steve Shenker for pointing out this argument.}. 

In this Section we prove that the holographic entanglement entropy associated to a patched up metric obtained by joining dS with Schwarzschild/AdS is vanishing, thereby proving that the corresponding state living on the AdS boundary is pure. 

In the remaining part of this Section, we will provide a brief review of some relevant results obtained in~\cite{Hubeny} and then extend them to the FGG issue. In this paper, the authors evaluate the holographic entanglement entropy for bag of gold metrics obtained by sewing a dS region on the left hand side of a Schwarzschild/AdS metric separated by a wall describing a type-(d) trajectory\footnote{Notice that here we are referring to Freivogel et al.'s classification of wall trajectories.}, i.e. a static configuration enabling both vacua to share the same timelike killing vector field. The main feature of the bag of gold metrics ends up being that $S_{\rm HEE}$ results equal for, both, the domain $A$ living on the AdS boundary and its complement $B$, no matter what is the extension of the domain $A$. However, since their difference should define the entanglement entropy of the whole boundary theory, $\delta S_{A}$, this would also mean that the boundary state associated to an AdS metric has to be pure. Such result would be in contradiction with their setup, since they claim that the process of gluing a dS portion to the Schwarzschild/AdS right hand side of the maximally-extended causal diagram should correspond to having traced over the AdS boundary on the left hand side of the black hole event horizon. Thereby, even prior to having added the dS part, the only boundary left is already in a mixed state\footnote{For a mixed state, $S_{\rho_{\Sigma}}\neq0$ is expected.}.  

The picture reported below, shows how the extremal surface probes deeper and deeper regions with increasing $\theta_{\infty}$. The maximum depth within the bulk is labeled $r_{o}$.  

\begin{figure}[h!] 
\begin{center} 
\includegraphics[scale=0.4]{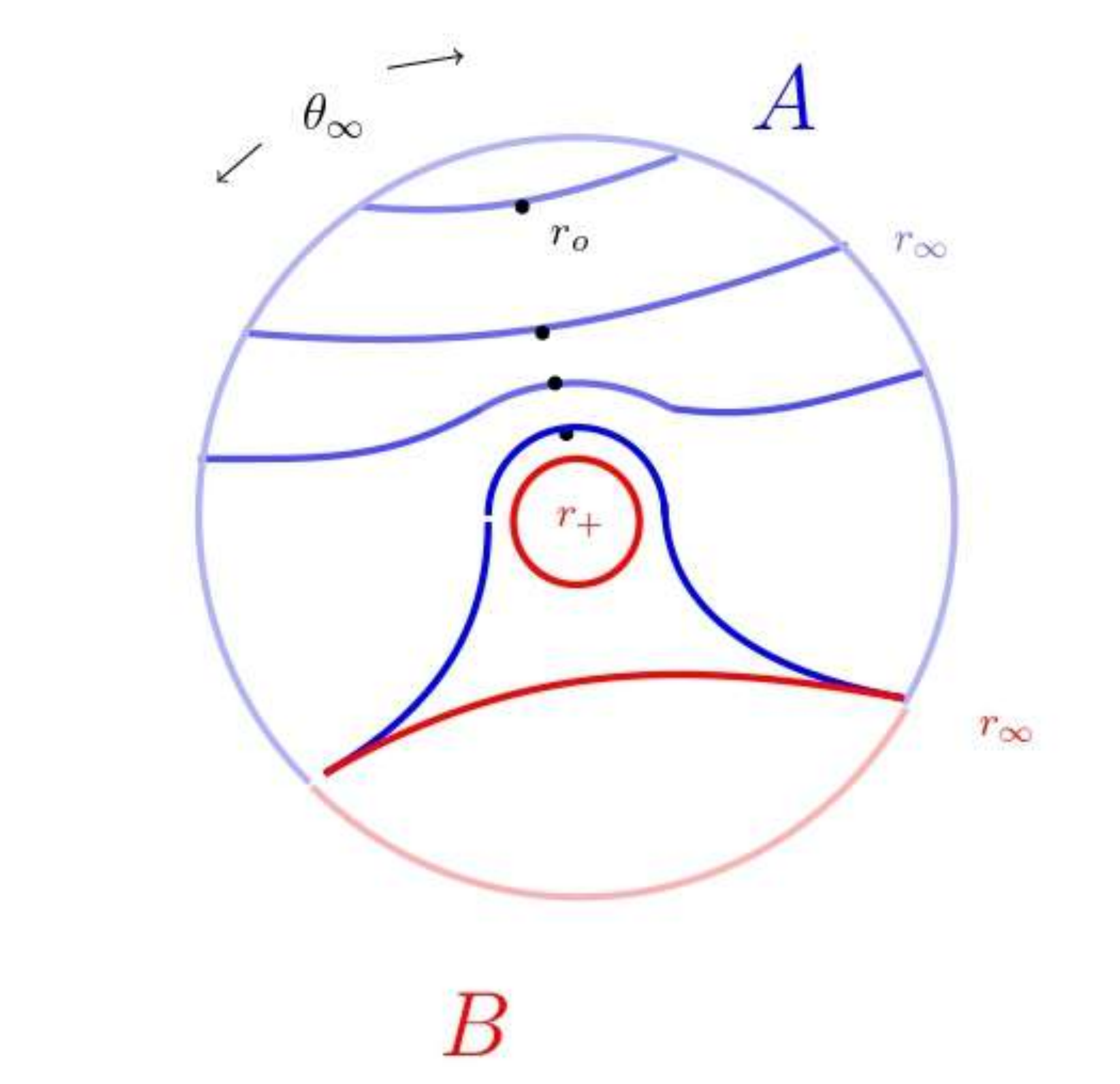} 
\caption{\footnotesize{This picture refers to the two-sided AdS black hole. The outer circle is the AdS boundary, located at $r_{\infty}$. The choice of the domain of $A$ (light blue arc) corresponds to a certain $\theta_{\infty}$. Its complement $B$ is depicted in red. Dark red lines indicate the disconnected minimal surfaces from the point of view of an observer in $B$. $r_{+}$ indicates the black hole event horizon, providing the (finite) thermal contribution to the holographic entanglement entropy evaluation. The black dots indicate the maximum depth reached by the extremal surface wihtin the bulk of the Schwarzschild/AdS spacetime.}} 
\end{center} 
\end{figure} 

\begin{figure}[h!]
\begin{center} 
\includegraphics[scale=0.4]{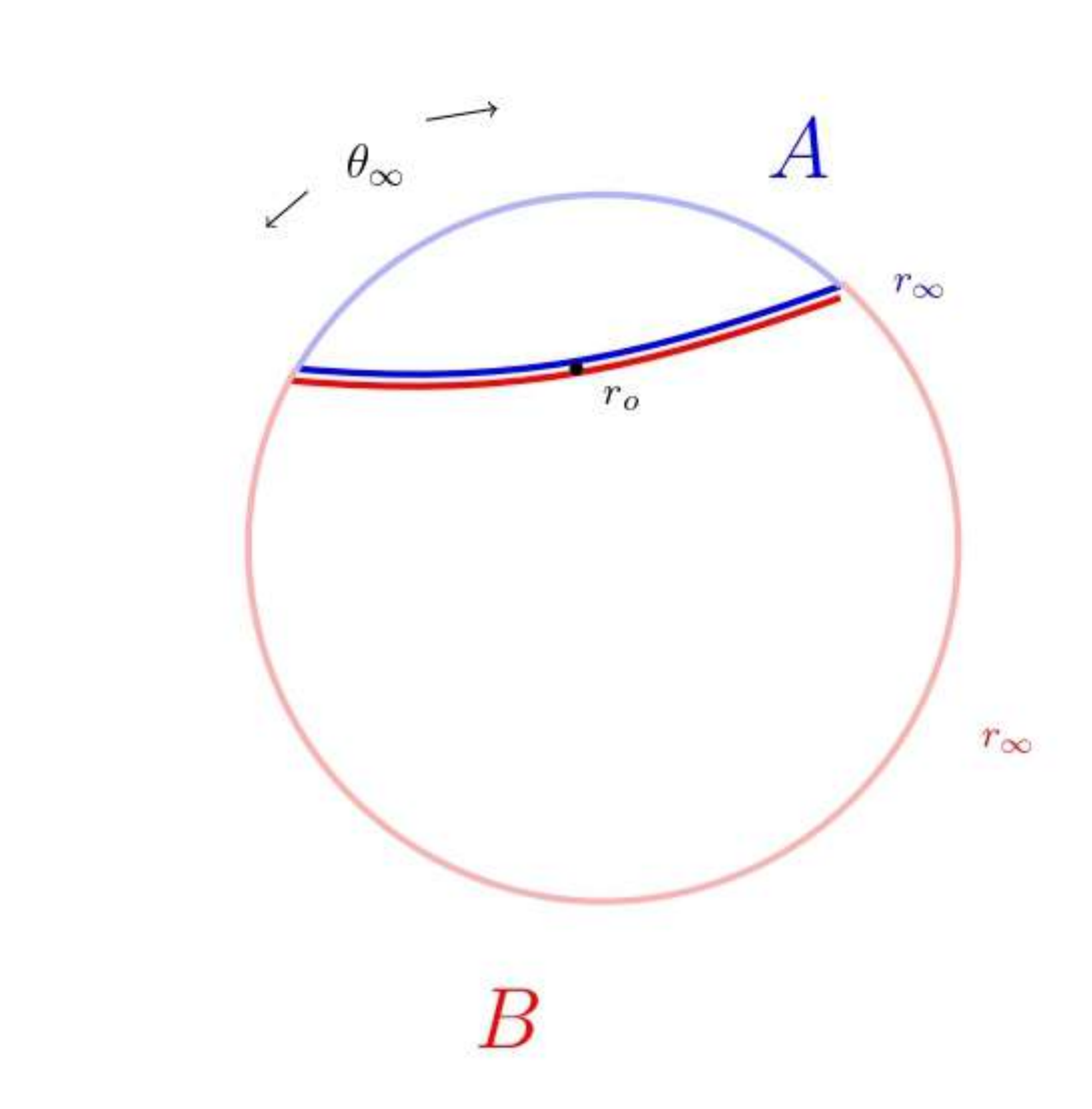} 
\caption{\footnotesize{This configuration is associated to the case in which both domains $A$ and $B$ lie on the same AdS right boundary. This setup is also assuming the presence of a dS patched up on the left hand side of the causal diagram replacing the other AdS boundary that would lie on the opposite side of the wedge. Please note that the picture is the same as before, but, this time, with no black hole horizon depicted.}} 
\end{center} 
\end{figure} 
\begin{figure}[h!] 
\begin{center}
\includegraphics[scale=0.9]{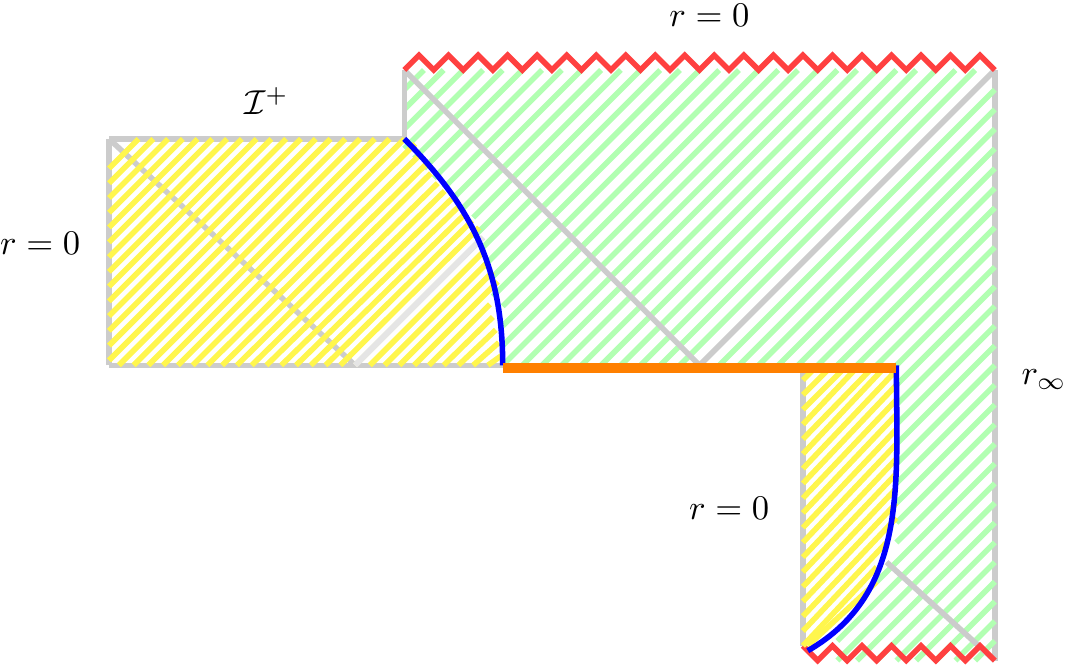}
\caption{\footnotesize{A composed Penrose diagram for an AdS/CFT illustration of the full bubble trajectory (in blue), starting at $r=0$ on trajectory (a) on the right handed geometry corresponding to Schwarzschild/AdS up to a maximum point, followed by the quantum tunnelling through the wormhole to trajectory (b) towards the $\mathcal{I}_+$ on the left part of the diagram corresponding to the relevant parts of dS spacetime. For fixed times just before and just after the tunnelling line we have a single boundary case (the AdS boundary on the right since $r=0$ on the dS part) which implies a pure state on the CFT boundary and no mixed states.}} 
\end{center}
\end{figure}

These surfaces have $\partial A$ whose centre lies at $\theta_{\infty}=0$. The bulk of the given metric, however, is also characterised by a black hole event horizon at $r_{+}$, which is responsible for the folding experiened by the blue lines in the plot. At some point, the blue extremal surface develops a neck which is too thin to ensure this being the minimal surface with boundary at $r_{\infty}$ equal to $\partial A$. Therefore, the need to turn to the complementary description (i.e. that of $B$), which, instead, experiences two disconnected extremal surfaces, one bounded at $r_{\infty}$, while the other is wrapping the black hole event horizon. Notice that the value of $\theta_{\infty}$ at which the two domains exchange is related to $r_{+}$ and also on the dimensionality of the spacetimes being considered. The need for adding the surface wrapping the black hole follows from homology constraints\footnote{The homology constraint takes into account the difference in the topology of the two boundary domains. To understand this, notice that the domain of $B$ is centered at $\theta_{\infty}=\pi$ rather than $\theta_{\infty}=0$, which is what happens for its complement $A$, instead.}. 

On the other hand, we adopt a different perspective, claiming, instead, that the bag of gold metric has a unique boundary, namely the AdS one on the right hand side of the wedge, and therefore it has to be a pure state. From this follows. Furthermore, focusing on the type-(a) trajectory, the patching in between the two metrics occurs on the right hand side of the causal wedge, therefore, there there is no need to add an extremal surface wrapping the black hole, which is only needed when regions $A$ and $B$ lie on different AdS boundaries by the homology constraint. In this case, therefore, the bulk surface $\gamma_{A}$ is such that, for any value of $\theta_{\infty}$ , $S_{A}=S_{B}$. 

In conclusion, holography provides strong arguments in favour of up-tunnelling processes.

\section{Summary and Conclusions}
\label{sec:Conclusions}

In this article we have revisited the questions regarding transitions among dS and Minkowski vacua from bubble nucleation. These are important from several points of view, since they can give information about initial conditions for inflation and also for the dynamics and population of the string landscape. We followed the Hamiltonian approach initiated by Fischler, Morgan and Polchinski~\cite{Fischler:1989se} and extended their results in several directions. We explicitly computed the transition rates for dS to dS including both down- and up-tunnelling agreeing with earlier results by Coleman and De Luccia~\cite{Coleman:1980aw} and Brown and Teitelboim~\cite{Brown:1988kg}. However, we argue that to properly compute the Minkowski to dS transition with the initial state being the zero energy state of the boundary Hamiltonian, the most appropriate way is to consider the direct calculation of zero mass Schwarzschild to dS instead of taking the limit of zero vacuum energy for the lowest dS vacuum. Fortunately, the calculation can be performed explicitly  and the transition agrees with the original conclusions of Farhi, Guth and Guven~\cite{Farhi:1989yr} and Fischler, Morgan and Pochinski~\cite{Fischler:1989se} in that it is possible to create  false vacuum bubbles starting from flat spacetime. 

Reproducing detailed balance between up- and down-tunnelling is a consistency check of our results. Obtaining the ratio of transitions in terms of the corresponding entropies for the two initial  states brought us to address the issue of entropy associated to Minkowski spacetime from the massless limit of Schwarzschild (which gives a unique vacuum state with zero entropy) as opposed to  the zero cosmological constant limit of dS space (which gives the entropy associated with the entire Hilbert space that can be built on Minkowski space).

We further  extended all previous treatments of vacuum transitions by explicitly computing the wave function away from the turning points and then we were able to analyse the rates more generally with the wave function exhibiting oscillatory behaviour as linear combinations of independent solutions of the Wheeler-DeWitt equation. This provides a more accurate picture of the relative probabilities of finding different classical configurations for the universe. Also this discussion shows that (in the case of dS to dS) detailed balance is only recovered (with the identification of dS entropy with the horizon area) if a certain constant is not set to zero to get the Vilenkin wave function (which in general is the sub-dominant term). 
 
More concretely, we can summarise our approach as follows. The vacuum transitions from a background $\mathcal{B}$ to a nucleated state $\mathcal{N}$ are determined in terms of relative probabilities
\be
\label{eq:ProbConclusion}
\mathcal{P}\left(\mathcal{B}\rightarrow \mathcal{N}\right)=\frac{|\Psi_\mathcal{N}|^2}{|\Psi_\mathcal{B}|^2} \,,
\ee
where $\Psi_{\mathcal{B},\mathcal{N}}$ are solutions of the Wheeler-DeWitt equation $\mathcal{H} \Psi=0$ with themselves being interpreted in terms of the probability to create the corresponding state from `nothing'. In the WKB approximation the wave functions take the general form 
\be
\Psi=a e^I + b e^{-I} \,,
\ee
where $I = i S$ evaluated on a classical solution, see Sec.~\ref{sec:TunnellingProbability} for more details. As usual, since the Wheeler-DeWitt equation is essentially a generalisation of the time independent  Schr\"odinger equation, it is not clear how to interpret these transitions as happening in time and we have to be content with the interpretation of transition probabilities as relative probabilities of the different configurations. In fact we are computing ratios of probability densities. Observe that, in Euclidean calculations \textit{à la} CDL, the expression in Eq.~\eqref{eq:ProbConclusion} is usually interpreted as the tunnelling rate per unit volume and unit time of the spacetime $\mathcal{N}$ from a spacetime $\mathcal{B}$. Addressing  dS to dS transitions in this way we can extract the following results:
\begin{itemize}
\item{} $\mathcal{P}$ being a relative probability is not restricted to be smaller than one. Being the ratio of squares of wave functions $\mathcal{P} $ may have well defined interpretations independent of measures as well as any ambiguity in the definition of `nothing'.
\item{} Usually the background is simply the standard wave function of the universe as defined either by Hartle-Hawking~\cite{Hartle:1983ai} or Vilenkin~\cite{Vilenkin:1982de, Vilenkin:1984wp} criteria. It is then determined in standard minisuperspace techniques. The nucleated wave function, being a composite state has to be computed going beyond minisuperspace approximation, including not only $t$ but also $r$ dependent metrics.
\item{} Boundary conditions determine which term in the expression for $\Psi$ above dominates. For coherent or real boundary conditions (generalising the ones used by Hartle-Hawking for the background wave function), the positive exponential dominates. In this case $\mathcal{P}$ is such that the denominator is the standard Hartle-Hawking probability with the  positive exponential and since the positive exponential in the denominator dominates, the total ratio reproduces the results of CDL and BT for both down and up-tunnelling of dS to dS transitions which are exponentially suppressed. Detailed balance is satisfied as expected since the boundary conditions are those of an equilibrium system with both outgoing and incoming waves. 
\item{} For boundary conditions imposing only outgoing waves in the classical region, corresponding to an expanding universe (generalising the Vilenkin boundary conditions beyond the background) the dominant term in the wave function is the negative exponential ($a=0$). In this case for a dS to dS transition the negative exponential in the denominator dominates and $\mathcal{P}$ is a positive exponential. Detailed balance is not satisfied, as expected given the fact that this is not an equilibrium situation since it  only has outgoing waves. At this moment we do not have a proper physical interpretation of these positive exponentials in terms of transition amplitudes, except for the original relative probabilities interpretation.
 \end{itemize}

On the question of nucleating dS spaces from Minkowski we found the following:
 \begin{itemize}
 \item{} Explicit calculation of the relative probability for being in a nucleated state of dS joined to a Minkowski compared to being in the Minkowski vacuum, obtained from the zero mass limit of Schwarzschild to dS transition (see Eq.~\eqref{eq:P1}), gives:
 \be
 \mathcal{P}(\mathcal{M}\rightarrow {\rm{dS}}) = \exp\left[\frac{\eta\pi}{G H^2} \left(1 - \frac{\kappa^{4}}{(H^{2}+\kappa^{2})^{2}}\right)\right] \,,
 \label{MdSrate}\ee
 with $\Lambda = 3 H^2 M_p^2$ being the dS cosmological constant and $\eta=\pm 1$. For $\eta=1$ and  $\kappa \ll H$ this is essentially the Hartle-Hawking probability except that now this has the interpretation of being the relative probability of finding a random state in the Hilbert space of dS compared to that of finding the Minkowski vacuum. The fact that this is positive is a reflection of the fact that it is more probable to be in a randomly selected state out of the large number of dS states for a given cosmological constant than to be in the Minkowski vacuum.
 \item{} As a consistency check, the same calculation (describing Minkowski in terms of zero-mass Schwarzschild) is done for the standard down-tunnelling Coleman-De Luccia amplitude, obtaining exactly the same result as Coleman-De Luccia.
  \item{} The rate satisfies detailed balance.
  \be
 \mathcal{P}(\mathcal{M}\rightarrow {\rm{dS}}) =  \mathcal{P}({{\rm{dS}}\rightarrow \mathcal{M}})\, \exp \left[\frac{\pi}{GH^{2}}\right] =\mathcal{P}({{\rm{dS}}\rightarrow \mathcal{M}})\, \exp\left(S_{\rm dS}\right)
  \ee
  with $S_{\rm dS}$ the dS entropy and the entropy of the Minkowski state (taken to be the zero eigenstate of the boundary Hamiltonian) is identified as zero.
  \item{} Imposing outgoing boundary conditions would lead in this case to an exponentially suppressed transition ($\eta=-1$) contrary to the dS to dS case. Still having a non-equilibrium situation detailed balance is not satisfied in this case.
 \item{} As mentioned above, the same techniques were used to include arbitrary dS to dS transitions and the fact that they reproduce the  known results for both up- and down-tunnelling gives confidence that the Minkowski to dS transition happens.
 \item{} Several standard criticisms against the Minkowski to dS transitions are addressed. In particular consistency with AdS/CFT.
 \end{itemize}

In order to quantify as much as possible, at the end of Sec.~\ref{sec:SummaryHamiltonian} we provided explicit expressions for the different Minkowski and dS transitions with the intention to record which transitions are allowed and which are more probable for a given starting point. The general results depend on the two main boundary conditions giving rise to the $\eta=\pm 1$ in the dominant exponentials but also on the relative sizes of $H_{\rm O}, H_{\rm I}$ and $\kappa$. It is interesting to notice that with Hartle-Hawking boundary conditions up-tunnelling to smaller values of the cosmological constant is preferred starting from both Minkowski and dS despite the fact that the transition amplitudes are positive exponentials in one case and negative in the other. The opposite conclusion happens with the outgoing waves boundary condition. 

We would like to stress that irrespective of the original motivation for creating a universe in the  laboratory, this framework offers an interesting perspective for  early universe cosmology. Nucleating a dS universe in a Minkowski background may be considered as  being similar to the nucleation of universes from `nothing' as proposed originally by Vilenkin and Hartle-Hawking. The zero black hole mass limit has only one single turning point and then only one classical region as in the dS case. The transition is then fully quantum mechanical closer to the creation from `nothing'. However, the fact that baby universes are created from white hole configurations seems tantalising. It is possible to have situations with two classical regions and turning points that can actually be considered as tunnelling. The Farhi, Guth, Guven scenario assumes that out of many bubbles created following trajectories (a) (see Fig.~\ref{fig:EffectivePotential}) most of them expand, reach the turning point and return to re-collapse, only  very few of them will tunnel to a trajectory (b) and disappear from the observers in the original black hole geometry. The tunnelling transition towards the inflating universe through the wormhole is fully quantum mechanical. Since the black hole is eternal,  it avoids the Borde, Guth and Vilenkin theorem~\cite{Borde:2001nh} regarding the need for a beginning of the universe but further study on this issue is needed. 

Our results may also be relevant to the study of the string landscape in several directions. First it indicates that dS vacua may be actually considered as a resonance (see for instance~\cite{Freivogel:2004rd, Freivogel:2005vv, Maltz:2016iaw}) that may appear after a transition from Minkowski space and later decay back to lower or higher vacuum energies, including Minkowski again. Second, despite the fact that there is no control on the measure of the landscape, we may have an indication of which transitions are more probable by considering relative probabilities. But in order to properly address landscape issues our approach needs to be extended to extra dimensions including  a proper 10D study of the transitions with different geometries such as the products of 4D Minkowski and dS spacetimes times a compact manifold. Also, a detailed discussion to address the challenges raised in~\cite{Johnson:2008vn} about populating the landscape needs the consideration of multi-field scalar potentials with a hierarchy of barriers and runaway directions. We hope to come back to these and other related questions in the future.

Finally let us reiterate that in this paper we have given a fresh perspective on the decades old discussions arising from applying flat space arguments about tunnelling to transitions amongst space-times {\it à la} Coleman-De Luccia, by formulating these questions in terms of solutions to the Wheeler-DeWitt equation, thus eliminating issues about unitarity, purity of states, negative mode issues\footnote{For recent discussions see for instance~\cite{Bramberger, Gregory, Bramberger2} and references therein.}, and degenerate metrics, that plague the Euclidean calculations. 

\section*{Acknowledgements}
We would like to especially thank Joe Polchinski for many interesting exchanges and encouragement during the early stages of this project and, unfortunately, the last stages of his life. We dedicate this article to him with much gratitude and admiration. We also thank useful communications with Mohsen Alishaia, Stefano Ansoldi, Thomas Bachlechner, Cliff Burgess, Sebasti\'an C\'espedes, Joe Conlon, Alan Guth, Veronika Hubeny, Anshuman Maharana, Juan Maldacena, Kyriakos Papadodimas, Mukund Rangamani, Leo Pando-Zayas and Steve Shenker.

\end{document}